\documentclass[11pt,a4paper]{article}
\usepackage{jheppub}
\usepackage{soul}
\usepackage[utf8x]{inputenc}
\usepackage{graphicx}
\usepackage{subfigure}
\usepackage{bm}
\usepackage{xspace}
\usepackage{relsize}
\usepackage{placeins}
\usepackage[bottom]{footmisc}
\newcommand{\beq}{\begin{equation}}
\newcommand{\eeq}{\end{equation}}
\newcommand{\bea}{\begin{eqnarray}}
\newcommand{\eea}{\end{eqnarray}}

\newcommand{\HEPfit}{\texttt{HEPfit}\xspace}
\usepackage{lmodern}

\usepackage{color}
\usepackage{float}
\usepackage[usenames,dvipsnames,svgnames,table]{xcolor}
\usepackage{hyperref}
\usepackage[all]{hypcap}
\hypersetup{
    bookmarks=true,         
    unicode=false,          
    pdftoolbar=true,        
    pdfmenubar=true,        
    pdffitwindow=false,     
    pdfstartview={FitH},    
    pdftitle={Heavy Higgs Searches: Flavour Matters.},    
    pdfauthor={Stefania Gori, Christophe Grojean, Aurelio Juste and Ayan Paul},     
    pdfsubject={hep-ph, hep-ex},   
    pdfcreator={TeXShop},   
    pdfproducer={MacTex, PDFLaTeX}, 
    pdfkeywords={2HDM} {Heavy Higgs} {Flavour Violation}, 
    pdfnewwindow=true,      
    colorlinks=true,       
    linkcolor=blue,          
    citecolor=violet,        
    filecolor=magenta,      
    urlcolor=cyan,           
    linktocpage=true
}

\begin{document}

\preprint{
\begin{flushright}
DESY 17-103
\end{flushright}}

\title{Heavy Higgs Searches: Flavour Matters}

\author[a]{Stefania Gori,}
\author[b,c]{Christophe Grojean,\footnote{On leave from Instituci\'o Catalana de Recerca i Estudis Avan\c cats, 08010 Barcelona, Spain}}
\author[d,e]{Aurelio Juste}
\author[f]{and Ayan Paul}

\affiliation[a]{Department of Physics, University of Cincinnati, Cincinnati, Ohio 45221, USA}
\affiliation[b]{DESY, Notkestrasse 85, D-22607 Hamburg, Germany}
\affiliation[c]{Institut f\"ur Physik, Humboldt-Universit\"at zu Berlin, D-12489 Berlin, Germany}
\affiliation[d]{Institut de F\'isica d'Altes Energies (IFAE), E-08193 Bellaterra, Barcelona, Spain}
\affiliation[e]{Instituci\'o Catalana de Recerca i Estudis Avan\c{c}ats (ICREA), E-08010 Barcelona, Spain}
\affiliation[f]{INFN, Sezione di Roma, Piazzale A. Moro 2, I-00185 Roma, Italy}

\emailAdd{gorisa@ucmail.uc.edu}
\emailAdd{christophe.grojean@desy.de}
\emailAdd{juste@ifae.es}
\emailAdd{ayan.paul@roma1.infn.it}

\abstract{
We point out that the stringent lower bounds on the masses of additional electrically neutral and charged Higgs bosons crucially depend on the flavour structure of their Yukawa interactions. We show that these bounds can easily be evaded by the introduction of flavour-changing neutral currents in the Higgs sector. As an illustration,
we study the phenomenology of a two Higgs doublet model with a Yukawa texture singling out the third family of quarks and leptons.  We combine constraints from low-energy flavour physics measurements, LHC measurements of the 125 GeV Higgs boson rates, and LHC searches for new heavy Higgs bosons. We propose novel LHC searches that could be performed in the coming years to unravel the existence of these new Higgs bosons. 
}

\maketitle

\section{Introduction}

Progress in physics often comes through understanding whether an intriguing observation is a relevant question worthy of further exploration.
Particle physics is at a unique moment of its history and more than ever there is a need to identify the right questions: on one hand  the Standard Model (SM) beautifully describes with an astonishing accuracy the vast majority of the data collected at various colliders; on the other hand, there are indisputable experimental evidences that this model has to be amended to account for instance of dark matter, to explain the matter-antimatter asymmetry, and also to be able to describe gravitational phenomena in the quantum regime. In addition, the values of the parameters defining the SM raise a number of questions: why is the Higgs boson mass so small compared to the scale of quantum gravity while it is subject to large quantum corrections? why is the electron mass 200  times smaller than the muon mass and 3,500 times smaller than the tau mass, while no quantum number distinguishes electrons from muons and taus? 

The first question  either requires  new particles, new forces or new space-time structure, or large accidental cancelations in a multiverse. The second question is a less obvious guide toward the understanding of what should come next. Indeed, in the SM, an electron is different from a muon simply because it has a different Yukawa coupling to the Higgs boson. These Yukawa couplings are fundamental parameters of the SM and any values of these couplings are perfectly legitimate and not subject to large quantum corrections due to the chiral symmetry. But is it the correct description of nature? While the SM picture is compatible with the wealth of flavour measurements, the fact that the masses of the quarks and leptons truly originate from minimal Yukawa interactions with a minimal Higgs sector is still to be demonstrated. In particular, one needs to understand if the Yukawa pattern is completely arbitrary or if it is governed by some yet hidden structure(s). Additionally, the Yukawa couplings to first and second generations still need to be measured.

Within the minimal set-up of the SM, two particular empirical structures  emerge: (i)~the couplings of fermions to the Higgs boson are proportional to the fermion masses and (ii)~these couplings are flavour diagonal. Were there  new vector-like fermions mixing with the SM fermions, or were there more than one Higgs vacuum energies, these two emerging structures would be distorted and the Higgs interactions could trigger sizable flavour-violating processes. That is why the $\sim$$2\sigma$ excess in the decay $h \to \tau\mu$ observed by the CMS Collaboration using LHC Run 1 data~\cite{Khachatryan:2015kon} raised much interest, although it has not been confirmed by subsequent analyses~\cite{Aad:2016blu, CMS:2016qvi, CMS-PAS-HIG-17-001}.
That is also why the search for $t \to ch$ decays is of high priority~\cite{Aad:2014dya, Khachatryan:2014jya, Aad:2015pja, Khachatryan:2016atv,FCNC_ATLAS13}.
The confirmation of an excess or a null result would give invaluable information on the (non-)minimality of the Higgs sector and on the origin of flavour.

Conversely, the existence of sizable flavour-violating interactions in the Higgs interactions can have large impact on the phenomenology of additional Higgs bosons. This is the central question we want to explore in this paper. In particular, we demonstrate, by a concrete example, how the standard  lower bounds on the masses of additional neutral and charged Higgs bosons can be easily evaded by departing from Type~I and II two Higgs doublet models (2HDM) and by allowing for a small amount of flavour-violating interactions between the 125\,GeV Higgs boson and the SM fermions at a level perfectly compatible with all flavour constraints. The flavour-violating window opens  new rich phenomenological opportunities to which traditional searches for additional Higgs bosons remain blind.
A full exploration of this new flavour territory is beyond the scope of this paper.
Instead, we will limit ourselves to illustrating our points with a study of a 2HDM that does not fulfill the Glashow--Weinberg criteria~\cite{Glashow:1976nt} and can thus accommodate Higgs-mediated tree-level flavour-changing interactions. The most general form of this so-called Type III  2HDM has still a very large number of free parameters.\footnote{A study of the flavour and LEP experimental signatures of Type III 2HDMs can be found in Ref.~\cite{Atwood:1996vj}.} We will appeal to a specific texture in which the right-handed top quark is a singular player such that interesting signals can be expected in the forthcoming searches involving top quarks. The particular texture we will resort to was reconsidered recently by Ref.~\cite{Chiang:2015cba} to address the strong CP problem with an axion model free of the usual domain wall problem, making an interesting connection amongst three different topics (strong CP violation, dark matter abundance and flavour physics). Other scenarios have brought forward Higgs flavour-violating interactions to address some other experimental anomalies, notably the persistently troublesome value of the muon magnetic moment~\cite{Omura:2015nja,Arroyo-Urena:2015uoa} or $B\to D^{(*)}\tau\nu$~\cite{Celis:2012dk,Crivellin:2012ye,Kim:2015zla,Iguro:2017ysu}, $B \to K^{(*)}\mu\mu$~\cite{Crivellin:2015mga}, or to explain the large mass hierarchy between the three generation quarks and leptons \cite{Altmannshofer:2015esa,Ghosh:2015gpa,Altmannshofer:2016zrn}. We leave these considerations aside, and focus our attention on how the presence of flavour-violating interactions impact the search for heavy Higgs bosons, and how new search strategies have to be designed to seize this novel opportunity.

Our paper is organized as follows. In Section~\ref{sec:model}, we present our extended Higgs model.
In Section~\ref{sec:constraints}, we review the various experimental constraints from Higgs measurements and heavy meson decays. In Section~\ref{sec:signatures}, we advocate new collider signatures not yet explored at the LHC, as, for example, the associated production of a neutral or charged heavy Higgs boson followed by a subsequent flavour-violating decay, $H\to tc$ or $H^\pm \to cb$.
Finally, we present our conclusions in Section~\ref{sec:conclusions}. Some useful formulae and details of the constraints on the model are collected in the appendices.

\section{A top-philic flavour-violating two Higgs doublet model}\label{Sec:model}
\label{sec:model}
A Type III 2HDM has generalized flavour-diagonal and off-diagonal Higgs couplings. In its most general form, the model is not consistent with experimental data on flavour-changing neutral current (FCNC) processes, since it predicts Higgs-mediated tree-level flavour transitions. However, as we will demonstrate, there exist specific flavour structures not based on $Z_2$ symmetries that are only weakly constrained by existing measurements (see also Refs.~\cite{Cheng:1987rs,Hou:1991un,Branco:1996bq,Pich:2009sp,Altmannshofer:2012ar,Altunkaynak:2015twa,Altmannshofer:2015esa,Ghosh:2015gpa,Altmannshofer:2016zrn,Arroyo:2013tna,Crivellin:2015mga,Botella:2015hoa,Bertuzzo:2015ada,Bauer:2015kzy,Bauer:2015fxa,Aloni:2015wvn,Cline:2015lqp,Bizot:2015qqo,Wang:2016rvz,Dery:2016fyj,Herrero-Garcia:2016uab,Gori:2017qwg,Gomez:2017dhl} for additional interesting models).

The most general renormalizable Yukawa sector involving two Higgs doublets, $H_u$ and $H_d$ respectively of hypercharge 1/2 and -1/2, and respecting the full local symmetry of the SM can be written as (with $\tilde{H}_i = i\sigma^2 H^*_i$)
\begin{equation}
- \mathcal{L} =   Y^u_{ij} \bar{u}_{i} H_u Q_j + \hat\epsilon^{u\dagger}_{ij} \bar{u}_{i} \tilde{H}_d Q_j
- Y^d_{ij} \bar{d}_{i} H_d Q_j + \hat\epsilon^{d\dagger}_{ij} \bar{d}_{i} \tilde{H}_u Q_j
- Y^\ell_{ij} \bar{e}_{i} H_d L_j + \hat\epsilon^{\ell\dagger}_{ij} \bar{e}_{i} \tilde{H}_u L_j +h.c.
\end{equation}
Traditional Type~II models correspond to vanishing non-holomorphic couplings, $\hat\epsilon^{u,d,\ell}=0$, while Type~I models have $Y^{d,\ell}=\hat\epsilon^u=0$. It is straightforward to write these Yukawa interactions in the mass-eigenstate basis of the Higgs bosons:
\begin{eqnarray}
H_u & = & \left( \begin{array}{c} \cos \beta\, H^+ \\ v \sin \beta + \frac{1}{\sqrt{2}} \phi^0_u  \end{array} \right)\ \ \textrm{with} \ \ \phi^0_u= \cos \alpha\, h + \sin \alpha\, H - i \cos \beta\, A,\\
H_d & = & \left( \begin{array}{c}  v \cos \beta + \frac{1}{\sqrt{2}} \phi^0_d \\
\sin \beta\, H^- \\ \end{array} \right)
\ \ \textrm{with} \ \ \phi^0_d=- \sin \alpha\, h + \cos \alpha\, H - i \sin \beta\, A, 
\end{eqnarray}
with $v\sim 174$\,GeV. Here $\alpha$ is the angle that defines the rotation from the interaction basis to the mass eigenstate basis of the CP even neutral Higgs boson fields. The angle $\beta$ determines the rotation from the Higgs basis~\cite{Lavoura:1994fv} to the interaction basis. 
In full generality, the fermion masses are diagonalized by bi-unitary transformations:
\begin{eqnarray}
\label{eq: rotationU}
U_R  \, v\left( Y^u \sin\beta + \hat\epsilon^{u\dagger} \cos\beta \right)  U^\dagger_L = \textrm{diag}(m_u, m_c, m_t) \equiv {m}^u,\nonumber\\
D_R  \, v\left( Y^d \cos\beta + \hat\epsilon^{d\dagger} \sin\beta \right)  D^\dagger_L =  \textrm{diag}(m_d, m_s, m_b)\equiv {m}^d,\nonumber\\
E_R  \, v\left(Y^\ell \cos\beta + \hat\epsilon^{\ell\dagger} \sin\beta \right)  E^\dagger_L =  \textrm{diag}(m_e, m_\mu, m_\tau)\equiv {m}^\ell.
\end{eqnarray}
In the mass eigenstate basis, the interaction Lagragian between the fermions and the Higgs bosons becomes
\begin{eqnarray}
\mathcal{L}  & = & 
 \left( \frac{  \phi^0_u }{\sqrt{2} v \sin\beta}  \left(-  {m}^u + v \cos\beta \, {\epsilon}^{u\dagger} \right)_{ij} 
- \frac{{\phi^0_d}^*}{\sqrt{2}} {\epsilon}^{u\dagger}_{ij}  \right) \bar{u}_{R\, i}\, u_{L\, j} +h.c.\nonumber\\
& & + \left( \frac{ \phi^0_d }{\sqrt{2}v \cos \beta}  \left(- {m}^d + v \sin\beta \, {\epsilon}^{d\dagger} \right)_{ij} 
- \frac{{\phi^0_u}^*}{\sqrt{2}} {\epsilon}^{d\dagger}_{ij}  \right) \bar{d}_{R\, i}\, d_{L\, j} +h.c.\nonumber\\
& & + \left( \frac{ \phi^0_d }{\sqrt{2}v \cos \beta}  \left(-{m}^\ell + v \sin\beta \, {\epsilon}^{\ell\dagger} \right)_{ij} 
- \frac{{\phi^0_u}^*}{\sqrt{2}} {\epsilon}^{\ell\dagger}_{ij}  \right) \bar{e}_{R\, i}\, e_{L\, j} +h.c.\nonumber\\
&& + \left( \frac{{m}^u V}{v \sin \beta}  - (\tan\beta+\textrm{cotan} \beta)  {\epsilon}^{u\dagger}  V \right)_{ij} 
\cos \beta \, H^+  \bar{u}_{R\, i}\, d_{L\, j} +h.c.\nonumber\\
&& + \left( \frac{{m}^d V^\dagger }{v \cos \beta}  -(\tan\beta+\textrm{cotan} \beta) {\epsilon}^{d\dagger} V^\dagger  \right)_{ij}  \sin\beta \, H^-  \bar{d}_{R\, i}\, u_{L\, j} +h.c.\nonumber\\
&& + \left( \frac{{m}^\ell }{v \cos \beta}  -(\tan\beta+\textrm{cotan} \beta) {\epsilon}^{\ell\dagger}  \right)_{ij} 
\sin\beta \, H^-  \bar{e}_{R\, i}\, \nu_{j} +h.c.
\end{eqnarray}
where ${\epsilon}^{u\dagger} = U_R \hat\epsilon^{u\dagger} U_L^\dagger$, 
${\epsilon}^{d\dagger} = D_R \hat\epsilon^{d\dagger} D_L^\dagger$,
${\epsilon}^{\ell\dagger} = E_R\hat\epsilon^{d\dagger} E_L^\dagger$
 and $V=U_L^{\vphantom{\dagger}} D_L^\dagger$ is the CKM matrix.
 It is well known~\cite{Gunion:2002zf} that in the decoupling limit, $\cos(\beta-\alpha)\to 0$, the flavour- and CP-violating coupling of the light Higgs $h$, vanish but not those of $H$ and $A$.

Tree level flavour violating couplings of the first two light generations of up type quarks in the Yukawa sector are tightly constrained by measurements of flavour violating processes. However, interesting phenomenology can be generated by allowing for flavour violating Yukawa couplings of the second and the third generations of fermions in the up sector in the light of recent experimental developments. For this we introduce a mixing between the right-handed charm and top quarks which can possibly be maximal since it involves only the right handed fermions. A similar argument can be placed for the lepton sector to motivate for possible interesting observations of flavour violating dynamics while suppressing flavour violation involving leptons from the first two families. The right handed mixing matrix for the up quark sector and the lepton sector are then given by
 \beq
U_R\equiv \left(\begin{array}{ccc}
1 & 0 & 0 \\
0 & \cos\frac{\rho_u}{2} & \sin\frac{\rho_u}{2}   \\
0 & -\sin\frac{\rho_u}{2} & \cos\frac{\rho_u}{2}
\end{array}\right),~~E_R\equiv \left(\begin{array}{ccc}
1 & 0 & 0 \\
0 & \cos\frac{\rho_\ell}{2} & \sin\frac{\rho_\ell}{2}   \\
0 & -\sin\frac{\rho_\ell}{2} & \cos\frac{\rho_\ell}{2}
\end{array}\right).
\eeq

For concreteness, we will further consider the set-up proposed by Chiang et al. in Ref.~\cite{Chiang:2015cba} that assigns a particular role to the top quark. This follows from a variant axion model endowed with a Peccei--Quinn symmetry acting only on the right-handed top. In this set-up, it follows that
\begin{equation}
Y^u = \left( \begin{array}{ccc} \cdot & \cdot & \cdot \\  \cdot & \cdot & \cdot \\ 0 & 0 & 0 \end{array} \right) \ \ \textrm{and} \ \ \hat\epsilon^{u\dagger} = \left( \begin{array}{ccc} 0 & 0 & 0\\ 0 & 0 & 0\\ \cdot & \cdot & \cdot  \end{array} \right)
\end{equation}
Upon performing the rotation defined in Eq.~(\ref{eq: rotationU}), we obtain the non-holomorphic couplings of the up-quarks and of the leptons:
\begin{equation}\label{eq:epsilons}
{\epsilon}^u=
\left(\begin{array}{ccc}
\frac{m_u}{v\cos\beta} & 0 & 0\\
0                                  & \frac{m_c}{v\cos\beta}\frac{1+\cos\rho_u}{2} & -\frac{m_c\sin\rho_u}{2v\cos\beta}\\
0                                  & -\frac{m_t\sin\rho_u}{2v\cos\beta} &  \frac{m_t}{v\cos\beta}\frac{1-\cos\rho_u}{2}
\end{array}\right)
,\;
{\epsilon}^{\ell}=
\left(\begin{array}{ccc}
0				   & 0 & 0\\
0                                  & \frac{m_\mu}{v\sin\beta}\frac{1-\cos\rho_\ell}{2} & \frac{m_\mu\sin\rho_\ell}{2v\sin\beta}\\
0                                  & \frac{m_\tau\sin\rho_\ell}{2v\sin\beta} &  \frac{m_\tau}{v\sin\beta}\frac{1+\cos\rho_\ell}{2}
\end{array}\right).
\end{equation}

For the down quark sector, most flavour bounds can be respected by limiting the Yukawa couplings of Type II and hence we set $\epsilon^d=0$. The explicit forms of the Higgs-fermion couplings are listed in Appendix~\ref{app:HEC}.
It is useful to note that the flavour-violating light Higgs couplings are proportional to the combination\begin{equation}
a=(\tan\beta+\cot\beta)\cos(\beta-\alpha)\,,
\label{eq:def_a}
\end{equation}
vanishing, therefore, in the $\cos(\beta-\alpha)\to 0$ limit.

For a matter of simplicity, in our phenomenological analysis of Sections~\ref{sec:constraints} and~\ref{sec:signatures} we will assume $\rho_u=\rho_\ell \equiv\rho$. Relaxing this condition would lead to flavour-violating effects in the lepton sector which would be, to a certain extent, independent of the quark sector.  Most notable, $h\to \tau \mu$ decays would be governed by $\rho_l$ while $t \to c h$ decays would be governed by $\rho_u$ rendering them theoretically uncorrelated. In addition, the scaling of the $h\tau\tau$ coupling could, effectively, be different from the $htt$ coupling in our model. While this could be an interesting study, we will not focus on it in this work. However, we will briefly comment on what changes when allowing different mixing angles in the up-quark and lepton sectors.

\section{Constraints on the model parameters}
\label{sec:constraints}

In the model, due to the mixing with the second Higgs doublet, the couplings of the 125\,GeV Higgs boson will be modified from what is predicted in the SM.
We first analyze the constraints from Higgs coupling measurements in both flavour-conserving and flavour-violating processes (Sections~\ref{Sec:HiggsConstraints} and~\ref{Sec:HiggsflavourConstraints}, respectively). These bring about constrains on $\beta -\alpha$, $\tan\beta$, and $\rho$. In addition, the model allows for significant flavour violation, and therefore, its parameter space can be constrained by the measurement of several low-energy flavour transitions, such as $b\to s \gamma$, $B\to\tau\nu$, $R_D$, and $R_{D^*}$. We study the most significant constraints from low-energy flavour-violating processes on the parameter space of the model (Section~\ref{Sec:flavourConstraints}). These processes add to the constraints on $\beta -\alpha$, $\tan\beta$, and $\rho$, and also constraint the mass of the charged Higgs ($m_{H^\pm}$), and, to a less extent, the mass of the neutral heavy Higgs bosons ($m_H,m_A$). We comment on additional weaker flavour constraints in Section~\ref{sec:otherConstraints}. Finally, we combine all relevant constraints in Section~\ref{Sec:combination} and we demonstrate that heavy Higgs bosons as light as 200\,GeV can be compatible with all constraints.

For combining the different constraints we use \HEPfit~\cite{HEPfit}, a code for the combination of indirect and direct constraints on High Energy Physics models. We use a Bayesian framework based on a Markov Chain Monte Carlo routine implemented within \HEPfit. A custom version of the code is used since the code also allows for user-defined models.\footnote{The code can be made available upon request.} We vary the parameters in the range: $0.1\le\tan\beta\le15$, $-\pi\le\rho\le\pi$, and $0\lesssim (\beta - \alpha)\lesssim \pi$ which are assigned flat priors. We restrict our investigation to not too large values of $\tan\beta$, in anticipation of avoiding constraints from low-energy flavour-violating processes. The fourth parameter of concern, the charged Higgs boson mass $m_{H^\pm}$, is varied in the range 200\,GeV to 1200\,GeV.

\subsection{Higgs couplings measurements}
\label{Sec:HiggsConstraints}
\begin{table}[b!]
\begin{center}
{\footnotesize
\begin{tabular}{|c|c|c|}
\hline
&\hspace{0.15in}Mean\hspace{0.15in}	&\hspace{0.15in}RMS$\;\;\;\;$\\
\hline
$\kappa_{gZ}$ 					&1.090     &0.110 	\\
$\lambda_{Zg}$ 				&1.285	&0.215	\\
$\lambda_{tg}$ 					&1.795	&0.285	\\
$\lambda_{WZ}$ 				&0.885	&0.095	\\
$\left|\lambda_{\gamma Z}\right|$ 	&0.895	&0.105	\\
$\left|\lambda_{\tau Z}\right|$ 		&0.855	&0.125	\\
$\left|\lambda_{bZ}\right|$ 		&0.565	&0.175	\\
\hline
\end{tabular}
\qquad
\begin{tabular}{|c|ccccccc|}
\hline
& $\kappa_{gZ}$	&$\lambda_{Zg}$	&$\lambda_{tg}$	&$\lambda_{WZ}$	&$\left|\lambda_{\gamma Z}\right|$	&$\left|\lambda_{\tau Z}\right|$	&$\left|\lambda_{bZ}\right|$ \\
\hline
$\kappa_{gZ}$ 					&\;1.00   &-0.03  &-0.24     &-0.62    &-0.57     &-0.38     &-0.34\\
$\lambda_{Zg}$				&-0.03   &\;1.00  &\;0.51    &-0.59    &-0.51     &-0.62     &-0.54\\
$\lambda_{tg}$ 					&-0.24   &\;0.51  &\;1.00    &-0.21    &-0.23     &-0.28     &-0.35\\
$\lambda_{WZ}$				&-0.62   &-0.59   &-0.21    &\;1.00    &\;0.66    &\;0.55    &\;0.55\\
$\left|\lambda_{\gamma Z}\right|$	&-0.57   &-0.51   &-0.23    &\;0.66    &\;1.00    &\;0.58    &\;0.51\\
$\left|\lambda_{\tau Z}\right|$ 		&-0.38   &-0.62   &-0.28    &\;0.55    &\;0.58    &\;1.00    &\;0.49\\
$\left|\lambda_{bZ}\right|$ 		&-0.34   &-0.54   &-0.35    &\;0.55    &\;0.51    &\;0.49    &\;1.00\\
\hline
\end{tabular}
}
\caption{Higgs effective couplings in the $\kappa - \lambda$ framework from Ref.~\cite{Khachatryan:2016vau}. The root mean square (RMS) values are symmetrized in our fit procedure and are given in the table on the left. The table on the right contains the correlation matrix amongst the seven free parameters defined in Eq.~(\ref{eq:SevenParFit}).}
\end{center}
\label{tab:HC}
\end{table}%
The measurement of the effective couplings of the Higgs boson to SM particles puts strong constraints on the parameter space of the model, which predicts significant deviations of these couplings from the corresponding SM values at sizable values of the mixing angle, $\rho$, and $\cos(\beta-\alpha)\neq 0$. 
To study such constraints we use the results from the combination of measurements by both ATLAS and CMS collaborations in Run 1~\cite{Khachatryan:2016vau}. Although more recent Run~2 ATLAS and CMS analyses are available~\cite{ATLAS-CONF-2017-045,ATLAS-CONF-2017-043,ATLAS-CONF-2017-047,ATLAS-CONF-2016-112,ATLAS-CONF-2017-041,ATLAS-CONF-2016-080,ATLAS-CONF-2016-058,ATLAS-CONF-2016-068,Aaboud:2017ojs,CMS-PAS-HIG-16-040,CMS-PAS-HIG-16-041,CMS-PAS-HIG-16-021,CMS-PAS-HIG-16-043,CMS-PAS-HIG-16-038,CMS-PAS-HIG-17-004,CMS-PAS-HIG-17-003},
we decide to use the Run 1 combination, since the fits to and correlations amongst the ratios of the effective couplings are available, and this provides a more coherent way of comparing model parameters with experimental results.

In Table~\ref{tab:HC} we summarize the fit values and uncertainties for the effective couplings under the so-called ``$\kappa - \lambda$ framework" (coupling modifier ratio parameterization), along with the corresponding correlation matrix. These values have been obtained under the assumption that there is no exotic or invisible decay of the 125\,GeV Higgs boson, in accordance with what is predicted by the model under consideration. Note that the asymmetric errors have been symmetrized since we use a symmetric Gaussian multivariate distribution to quantify the likelihood constructed from the fit to the experimental measurements. The details of the constraints set by the Higgs coupling data along with all the relevant formula can be found in Appendix~\ref{app:HEC}.\footnote{We have neglected the charged Higgs contribution to the gluon fusion production and the radiative decays, such that the Higgs coupling measurements directly constrain $\beta-\alpha$, $\tan \beta$ and $\rho$ independently of the value of the heavy Higgs masses. The charged Higgs contributions are, in fact, negligible in the entire parameter space.}

\subsection{Constraints from the 125\,GeV Higgs flavour-violating couplings}\label{Sec:HiggsflavourConstraints}
\begin{figure}[t!]
\begin{center}
\subfigure{\includegraphics[trim = 5mm 0mm 10mm 5mm, clip, width=.5\textwidth]{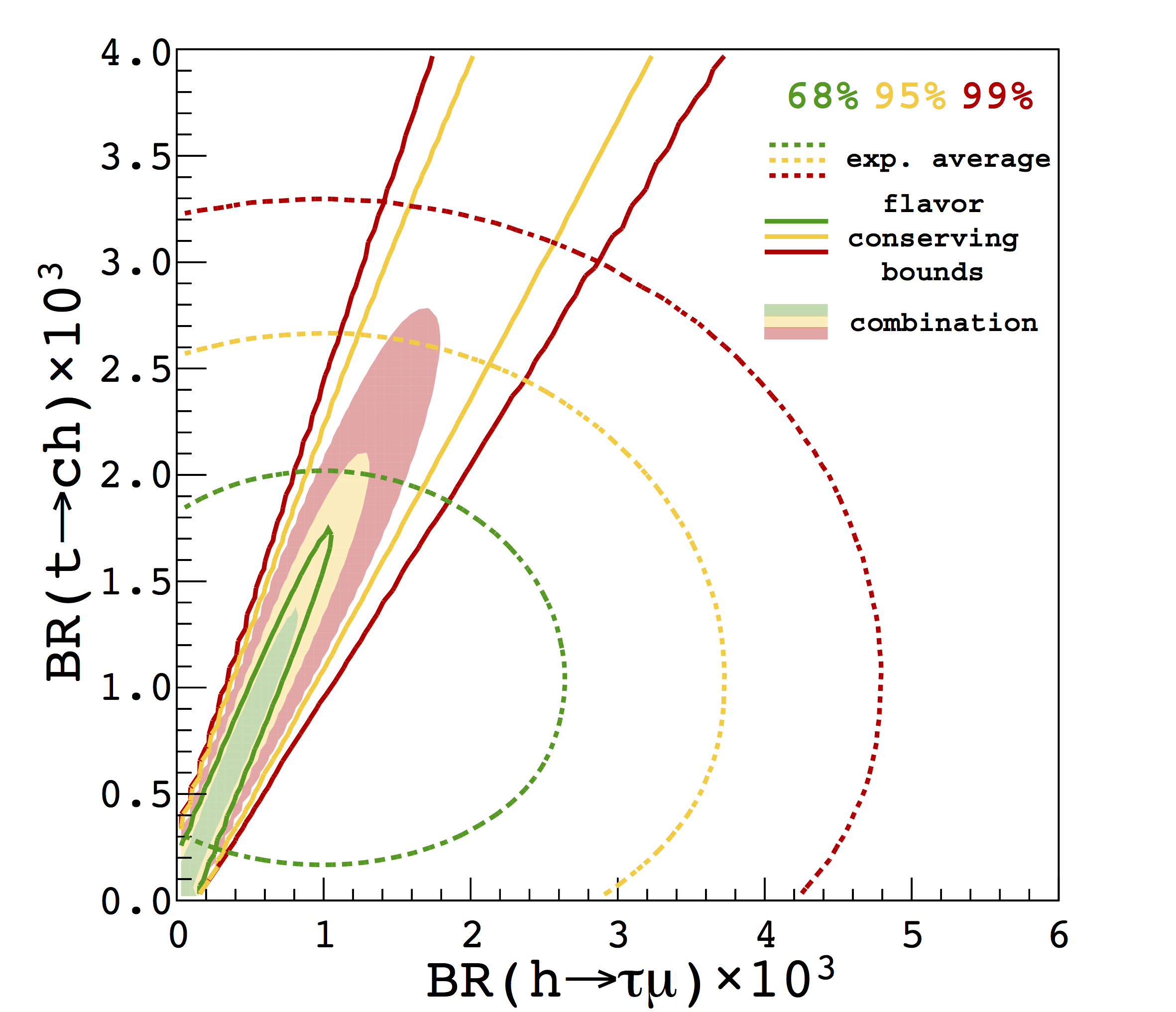}}
\caption{The average of the experimental results for $h\to\tau\mu$ and $t\to ch$ (dashed lines) compared to the region allowed by the constraints on the Higgs effective couplings given in Table~\ref{tab:HC} (solid lines). The shaded region is the combination of the two. The green, yellow and red lines (shaded areas) mark the 68\%, 95\%, and 99\% regions, respectively.}
\label{fig:Htaumutch}
\end{center}
\end{figure}
Measurements of the top-quark flavour-violating decays have been performed by both the ATLAS and CMS collaborations \cite{Aad:2015pja,Khachatryan:2016atv,FCNC_ATLAS13}. While for the branching ratio measurements of $h\to\tau\mu$ both CMS~\cite{Khachatryan:2015kon,CMS-PAS-HIG-17-001} and ATLAS~\cite{Aad:2016blu} provide central values and RMS allowing us to build a likelihood profile, for $t\to ch$ we use only the ATLAS results~\cite{Aad:2015pja,FCNC_ATLAS13} to put constraints on the model under consideration since the results from CMS~\cite{Khachatryan:2016atv} are cited only as  upper bounds and have lower sensitivity than the latest ATLAS result. 
The numbers that are used in the fit are collected in Table~\ref{tab:HFNC}, where we show 
a naive average of the $\textrm{BR}(h\to\tau\mu)$ and $\textrm{BR}(t\to ch)$ measurements. 

We plot these averages in Fig.~\ref{fig:Htaumutch} (dotted lines) in the $\textrm{BR}(h\to\tau\mu)-\textrm{BR}(t\to ch)$ plane.\footnote{The details of the analysis leading to this fit are reported in Appendix~\ref{app:FVD}.} The solid lines represent the boundaries of the 68\%, 95\%, and  99\% regions (green, yellow, and red line, respectively) of the prediction for $\textrm{BR}(h\to\tau\mu)$ and $\textrm{BR}(t\to ch)$ coming from the fit to the effective couplings in Table~\ref{tab:HC} only. The shaded region in Fig.~\ref{fig:Htaumutch} depicts how the measurements of the two flavour-violating processes constrain this prediction. From Fig.~\ref{fig:Htaumutch} it is clear that the experimental measurements of $\textrm{BR}(h\to\tau\mu)$ and $\textrm{BR}(t\to ch)$ constrain the parameter space of the model in addition to what is already constrained by the Higgs effective couplings data. The combination of the flavour-preserving and flavour-violating Higgs data is presented in Appendix~\ref{app:FVD}. The parameter space of our model is further constrained by low-energy flavour observables, as discussed in the next section.

\begin{table}[t!]
\begin{center}
\begin{tabular}{|c|c|c|}
\hline
Experiment&    $\textrm{BR}(h\to\tau\mu)$	&$\textrm{BR}(t\to ch)$\\
\hline
ATLAS 8\,TeV 20.3\,fb$^{-1}$	&($0.53 \pm 0.51$)\%~\cite{Aad:2016blu}			&($0.22 \pm 0.14$)\%~\cite{Aad:2015pja}\\
CMS 8\,TeV 19.7\,fb$^{-1}$		&($0.84^{+0.39}_{-0.37}$)\%~\cite{Khachatryan:2015kon}	& $<0.40\%$ @ 95\% CL$^\dagger$~\cite{Khachatryan:2016atv}\\
ATLAS 13\,TeV 36.1\,fb$^{-1}$		& -- 	 									&($0.069^{+0.075}_{-0.054}$)\%~\cite{FCNC_ATLAS13}\\
CMS 13\,TeV 35.9\,fb$^{-1}$		&($0.00 \pm 0.12$)\%~\cite{CMS-PAS-HIG-17-001}	&--\\
\hline
Average					&($0.10 \pm 0.11$)\%							&($0.109\pm0.061$)\%\\
\hline
\end{tabular}
\caption{The experimental results for $h\to \tau\mu$ and $t\to c h$ searches by the ATLAS and CMS collaborations, along with our simple weighted averages of $\textrm{BR}(h\to\tau\mu)$ and $\textrm{BR}(t\to ch)$. The weighted averages are computed considering only actual measurements with uncertainties. ($^\dagger$Not used in the fit.)}
\end{center}
\label{tab:HFNC}
\end{table}%

\subsection{Low energy flavour constraints}\label{Sec:flavourConstraints}

Having flavour off-diagonal couplings, the several Higgs states can leave very distinctive signatures in FCNC and tree-level charged-current processes in low-energy mesonic decays. To validate the parameter space of our interest, it is important to look at the possible constraints coming from these decays both in inclusive and exclusive channels, as well as from neutral meson oscillations.
As we will discuss, flavour constraints favor the part of the parameter space that involves a relatively low value of $\tan\beta$. While the Higgs couplings data and the flavour-violating processes discussed in Sections~\ref{Sec:HiggsConstraints} and~\ref{Sec:HiggsflavourConstraints} put bounds on only $\tan\beta$, $\rho$ and $\beta-\alpha$, the low-energy flavour-violating processes that we consider here also put constraints on the mass of the charged Higgs boson. As we will comment in Section~\ref{sec:otherConstraints}, constraints on the mass of the heavy neutral Higgs boson are much milder.

In a Type II 2HDM, the bound on $m_{H^\pm}$ coming from the measurement of ${\rm BR}(b\to s \gamma)$ is very stringent~\cite{Belle:2016ufb,Misiak:2017bgg} and is in the 570--800\,GeV range at $95\%$ confidence level\footnote{The bound on the charged Higgs boson mass depends quite sensitively on the method applied for its determination. For a detailed discussion we refer the reader to Ref.~\cite{Misiak:2017bgg}.}, independent of the value of $\tan\beta$, for $\tan\beta \gtrsim2$. However, this constraint can get weakened in a Type III 2HDM~\cite{Crivellin:2012ye} because of a possible destructive interference with the SM contribution for certain ranges of $\epsilon^u_{23}$ and $\epsilon^u_{32}$ allowing for much lower values of $m_{H^\pm}$ to remain compatible with measurements. In terms of the parameters in this model, a non-zero $\rho$ can bring about contributions that interfere destructively with the SM contributions allowing for compatibility with experimental measurements for relatively low values of $m_{H^\pm}$. The $\tan\beta$ dependence of the new physics (NP) contributions from our model is also quite different from a Type II 2HDM, since both $u_L^i d_R^j$ and $u_R^i d_L^j$ charged Higgs couplings are $\tan\beta$-enhanced for $\rho\neq 0$ (see Eqs.~(\ref{eq:HpmuLdR})--(\ref{eq:HpmuRdL})). 

In addition to $b\to s \gamma$, the charged-current process $B\to\tau\nu$ can get contributions due to a flavour-violating charged current mediated by the charged Higgs boson. Finally, the observables $R_D$ and $R_{D^*}$ involving a $b\to c \tau\nu$ transition can also set constraints on our parameter space. In our analysis, we do not try to explain the long-standing discrepancy between the SM predictions and the experimental measurements of these observables, but we simply include the two observables in a global fit.

We collect the measurements and SM predictions of the low-energy flavour-violating  processes that we use in this study in Table~\ref{tab:flavour}. More details about the decay modes $b\to s\gamma$, $B\to\tau\nu$ and the observables $R_D$ and $R_{D^*}$ can be found in Appendix~\ref{app:FL} where we also discuss the subtleties involved in performing the necessary calculation in the particular model under consideration.

We note that additional flavour observables will be modified by the charged Higgs exchange. Examples are leptonic meson decays, $D^+_{(s)}\to\mu^+\nu$, $D^+_{(s)}\to\tau^+\nu$, $K^+\to\mu^+\nu/\pi^+\to\mu^+\nu$, $K^+(\pi^+)\to e^+\nu/\pi^+\to e^+\nu$, and tau decays, $\tau\to K\nu/\tau\to\pi\nu$. However, the constraints coming from the measurement of these transitions are always much milder than the ones arising from $b\to s \gamma$ and $B\to\tau\nu$. For this reason, we do not introduce these additional observables in our global fit.

\begin{table}[t!]
\begin{center}
\begin{tabular}{|c|c|c|}
\hline
Process					&    Measurement				& SM Prediction\\
\hline
${\rm BR}(b \to s \gamma)	$	&$(3.32 \pm 0.15)\times 10^{-4}$	&$(3.36 \pm 0.23)\times 10^{-4}$\\
${\rm BR}(B \to \tau \nu)$		&$(1.06 \pm 0.19)\times 10^{-4}$	&$(0.807 \pm 0.061)\times 10^{-4}$\\
$R_D$					&$0.403 \pm 0.47$	 			&$0.299 \pm 0.003$\\
$R_{D*}$					&$0.310 \pm 0.17$				&$0.257 \pm 0.003$\\
\hline
\end{tabular}
\caption{The experimental measurements for the flavour changing observables used in the fit. The statistical and systematic uncertainties on the $R_D$ and $R_{D*}$ measurements have been summed in quadrature. These two measurement have a correlation coefficient of $-0.23$.}
\end{center}
\label{tab:flavour}
\end{table}%

\subsection{Other indirect constraints}
\label{sec:otherConstraints}

Given the flavour structure of the model, one could expect that the amount of flavour violation encoded by the parameter $\rho$ can be severely constrained by additional flavour observables beyond those that we have discussed in Section~\ref{Sec:flavourConstraints}. While we do not present a detailed analysis of additional constraints, in this section we discuss why FCNC observables do not affect our current study. In particular, we will discuss why neutral Higgs mediated processes are only weakly constrained, leading to very mild bounds on the heavy Higgs mass, $m_H$, in the regions of $\tan\beta,\rho$ allowed by the charged current constraints discussed in the previous section.

In generic Type III 2HDMs, amongst the tree-level FCNC processes mediated by the several neutral Higgs states, important constraints can arise from $\Delta F = 1$ leptonic meson decays like $B_{s,d}\to\mu^+\mu^-$, $K_L\to\mu^+\mu^-$, and $\bar{D}^0\to\mu^+\mu^-$. Our model, however, does not predict tree-level flavour-violating neutral Higgs couplings to $\bar b s(d)$, $\bar s d$, and to $\bar u c$, as can be seen from the structure of the couplings in Eq.~(\ref{Higgs-vertices-decoupling}), since $\epsilon^d_{23,32}=\epsilon^d_{13,12}=\epsilon^d_{12,21}=0$, and $\epsilon^u_{12,21}=0$. 
A similar argument applies to $\Delta F=2$ processes like $B_s$--$\bar{B}_s$, $B_d$--$\bar{B}_d$ and $K^0$--$\bar{K}^0$ mixings, for which tree-level contributions are absent for the same reason. In our model, additional loop-induced contributions to $\Delta F=2$ processes arise. These contributions are mediated by the charged Higgs boson and are proportional to $\epsilon^u_{ij}$. Being loop contributions, even for charged Higgs bosons with a mass just above the LEP bound, the constraints on the absolute value of the relevant elements of this matrix are of ${\cal O}(1)$, 
values that can be obtained only for values of $\tan\beta$ much larger than the ones allowed by $b\to s \gamma$. 

Neutral Higgs mediated tree-level contributions to lepton flavour-violating (LFV) decays like $\tau^-\to\mu^-\mu^+\mu^-$ and $\tau^-\to e^-\mu^+\mu^-$ arise in our model. However, the contributions are rather small, since they are proportional to the lepton mass of the respective generation, and the experimental bounds are relatively weak.
This translates into weak bounds on $\epsilon^\ell_{23,32}$, $\epsilon^\ell_{13,31}$ which do not affect our analysis. 
Better measured LFV processes like $\mu^-\to e^-e^+e^-$ do not receive tree-level NP contributions since $\epsilon^\ell_{12,21}=0$ in our model.
The bounds from radiative LFV processes like $\tau^-\to \mu^-\gamma$ and $\tau^-\to e^-\gamma$ are significantly weaker.

Additional low energy observables will be affected by the exchange of neutral Higgs bosons. There is a long-standing discrepancy between the SM prediction and the measurement of the muon magnetic moment, $a_\mu=(g-2)/2$.
In our model, loop contributions mediated by the neutral (and charged) Higgs boson exchange are generated. As shown by the couplings in Eqs.~(\ref{eq:Hll})--(\ref{Higgs-leptons-vertices-decoupling}), a very large value of $\epsilon_{22}^\ell(\tan\beta+\cot\beta)$ is needed to obtain a relatively sizable NP effect. Since $\epsilon_{22}^\ell\propto m_\mu/v$, $a_\mu$ receives only a small NP effect in the region at low values of $\tan\beta$, as required by the measurement of $b\to s \gamma$. 

Finally, also electroweak precision observables like the $S$, $T$ oblique parameters can set important constraints on the parameter space of a 2HDM. These electroweak parameters were studied in a generalized Type III 2HDM e.g. in Refs.~\cite{Haber:2010bw,Funk:2011ad}. In these studies, it has been shown that the $T$ parameter is generically the most important constraint,
and its bound depends on the mass splitting of the charged Higgs boson and the most massive neutral Higgs boson.  Since, in our investigation we do not fix the relative mass of the heavy Higgs states and they can possibly be quasi-degenerate, the $T$ parameter will not put significant constraints on the parameter space under consideration.

\subsection{Combining constraints}
\label{Sec:combination}

\begin{figure}[h!]
\begin{center}
\subfigure{\includegraphics[trim = 10mm 0mm 10mm 0mm, clip, width=.45\textwidth]{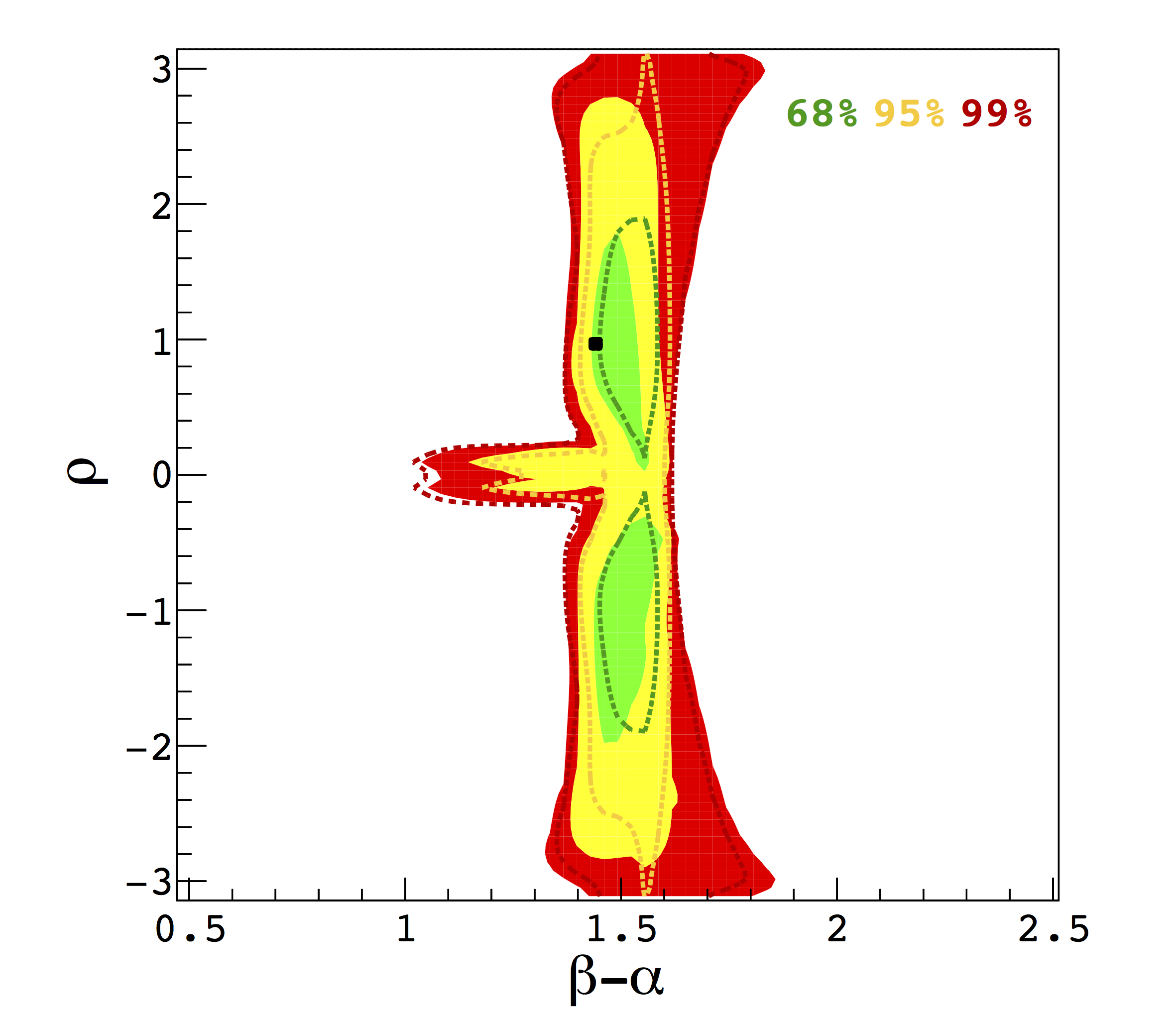}}\hspace{0.25in}
\subfigure{\includegraphics[trim = 10mm 0mm 10mm 0mm, clip, width=.45\textwidth]{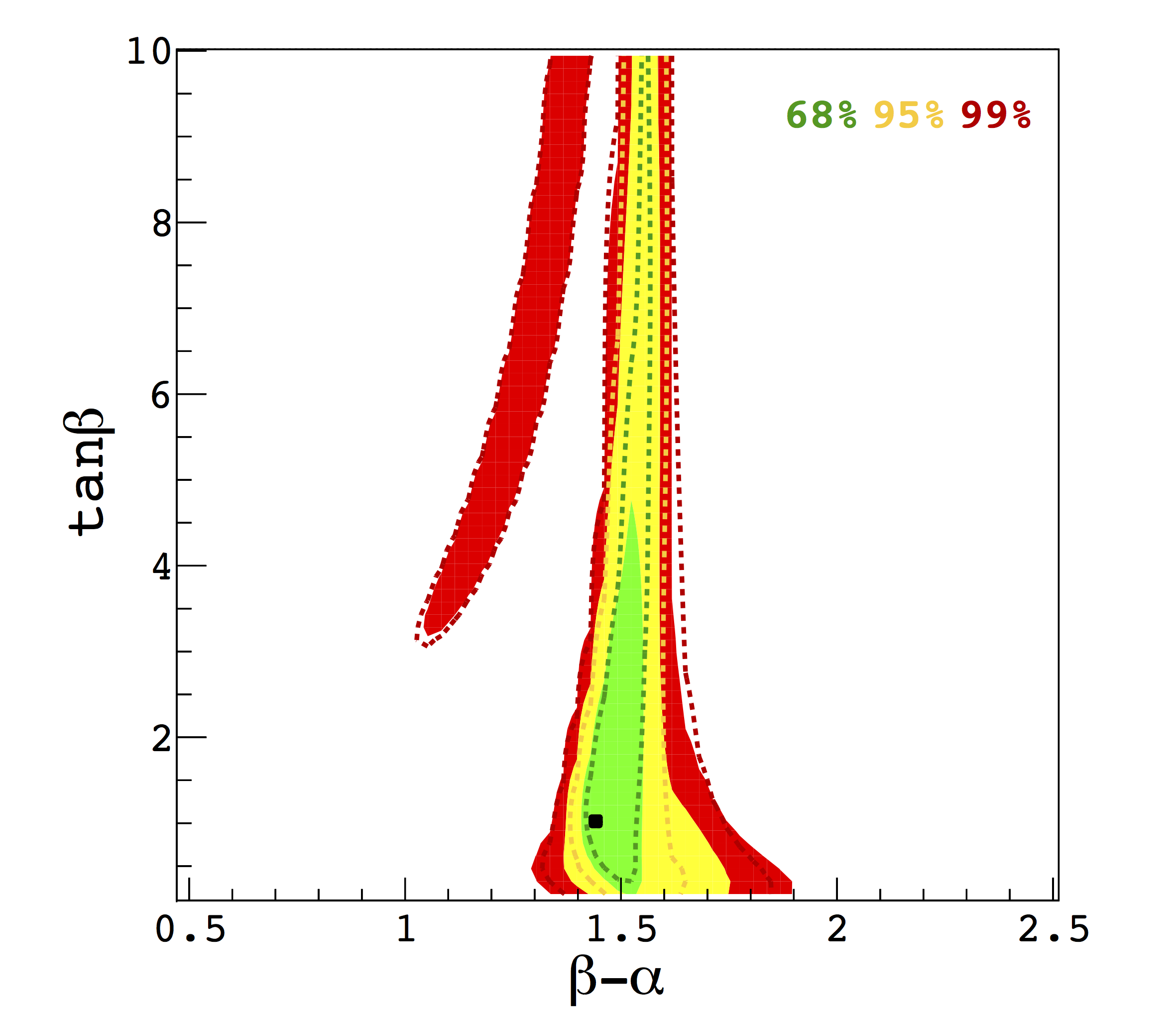}}\\
\subfigure{\includegraphics[trim = 10mm 0mm 10mm 0mm, clip, width=.45\textwidth]{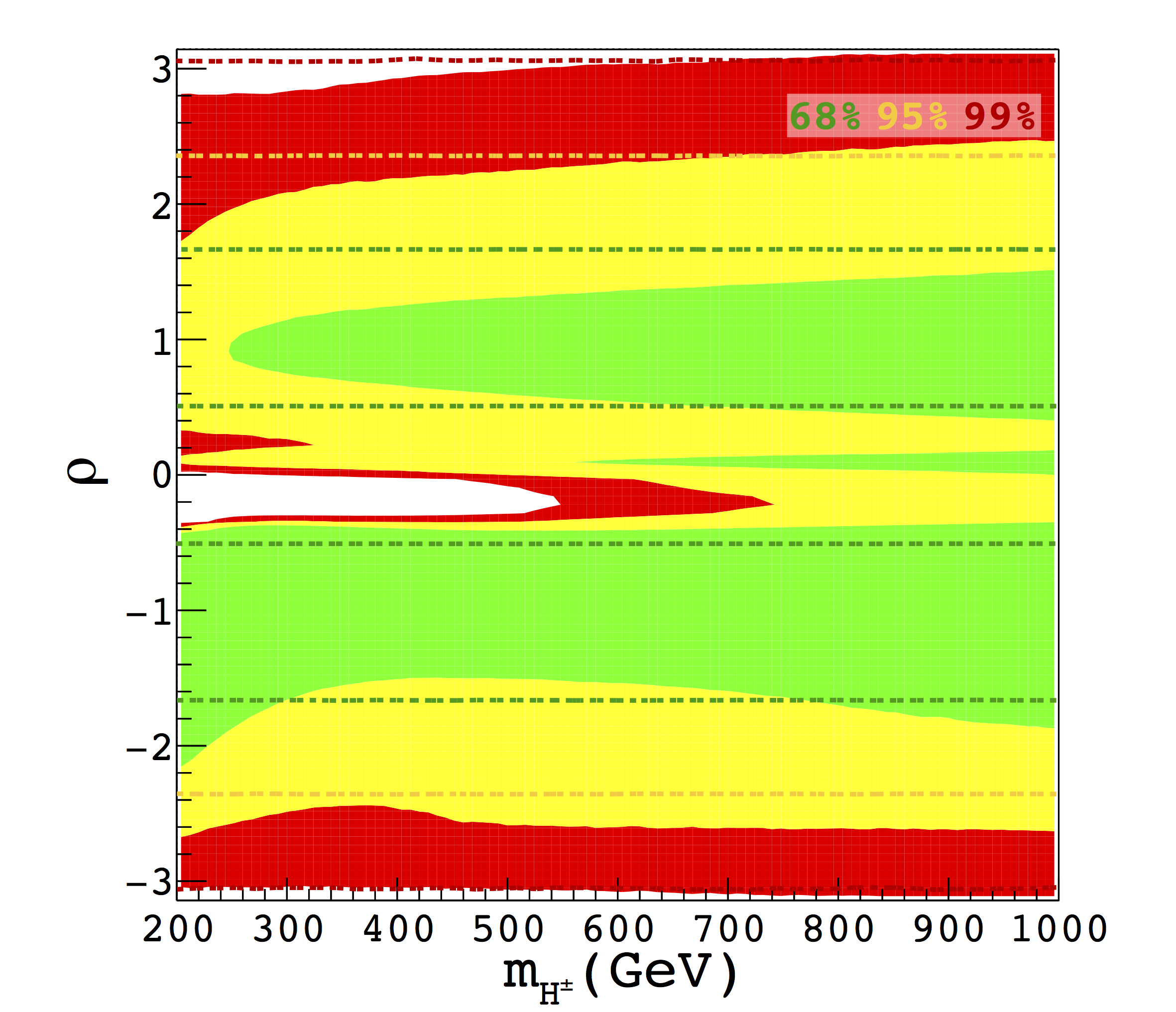}}\hspace{0.25in}
\subfigure{\includegraphics[trim = 10mm 0mm 10mm 0mm, clip, width=.45\textwidth]{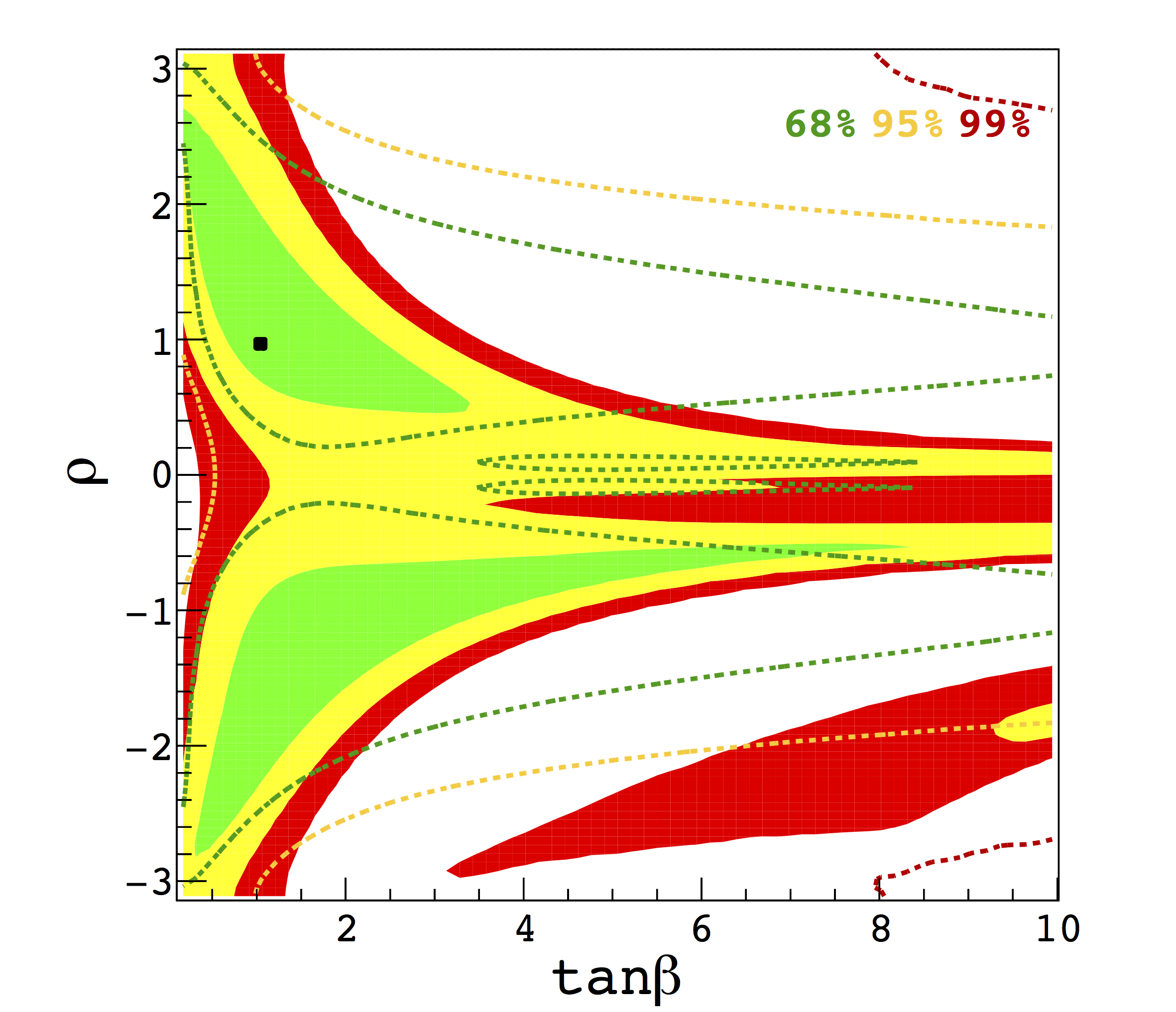}}
\caption{2D marginalized posterior distributions of the relevant model parameters combining all constraints from Higgs couplings data (see Table~\ref{tab:HC}), $h \to \tau \mu$, $t \to c h$ (see Table~\ref{tab:HFNC}) and flavour measurements (see Table~\ref{tab:flavour}). The green, yellow, and red regions are the 68\%, 95\%, and  99\% regions respectively. The dashed contours  represent the 68\%, 95\%, and  99\% contours from constraints coming only from the Higgs flavour-conserving and flavour-violating measurements. The black dot in each plot marks the benchmark point we will use for the discussion of the collider phenomenology in Sec. \ref{sec:signatures}, and they correspond to $\rho = 1$, $\tan\beta = 1$, and $\beta - \alpha = 1.4454$ ($\cos(\beta - \alpha) = 0.125$).
}
\label{fig:comb}
\end{center}
\end{figure}

As a final step to determine possible benchmark points that are allowed by all the constraints, we perform a combined fit using the Higgs effective coupling results tabulated in Table~\ref{tab:HC}, the Higgs and top flavour-violating decay results collected in Table~\ref{tab:HFNC}, and the charged-current process measurements listed in Table~\ref{tab:flavour}.

We show the results of our combination in Fig.~\ref{fig:comb}, as a function of the  free parameters, $\beta-\alpha,~\rho,~\tan\beta,~m_{H^\pm}$.
As we have deduced before, the primary constraints come from measurements of Higgs coupling, of the branching ratios of $t \to c h$, and of $b \to s \gamma$. The favored values of $\beta - \alpha$ is clustered at around $\pi/2$, corresponding to the decoupling or alignment limit. 
The dashed contours in Fig.~\ref{fig:comb} represent the 68\%, 95\%, and  99\% contours from constraints coming only from the Higgs flavour-conserving and flavour-violating measurements discussed in Sections~\ref{Sec:HiggsConstraints} and \ref{Sec:HiggsflavourConstraints} (see also Fig.~\ref{fig:HiggsFlav} in Appendix~\ref{app:FVD}). As can be seen from the right panels of Fig.~\ref{fig:comb}, the addition of the flavour violating low-energy observables, and in particular of $b \to s \gamma$, brings about a preference for  lower values of $\tan\beta$ as well as a bit smaller values of $|\rho|$. The dramatic change of the constraints to the parameter space can be seen comparing dashed and solid lines in the lower right panel of Fig.~\ref{fig:comb} in the ($\rho-\tan\beta$) plane.  
As shown by the lower left panel of the figure, even very light charged Higgs bosons are allowed, for a wide range of values for $\rho$.

To complete this analysis, we choose a benchmark point that we will use in the next section to discuss the collider phenomenology of our model. The point of our choice is marked with a black dot in the panels in Fig.~\ref{fig:comb}, and it is in the 68\% probability region in all the 2D marginalized posterior distributions and corresponds to $\rho = 1$, $\tan\beta = 1$, and $\beta - \alpha = 1.4454$ ($\cos(\beta - \alpha) = 0.125$). 
In what follows we shall see that this point in the parameter space brings about some interesting phenomenological implications for  heavy Higgs boson production and decay. As we will show, varying the point in the favored region of parameter space will not qualitatively affect the heavy Higgs boson phenomenology.

\section{Collider signatures}
\label{sec:signatures}

\begin{figure}[t!]
\begin{center}
\subfigure{\includegraphics[width=0.2\textwidth]{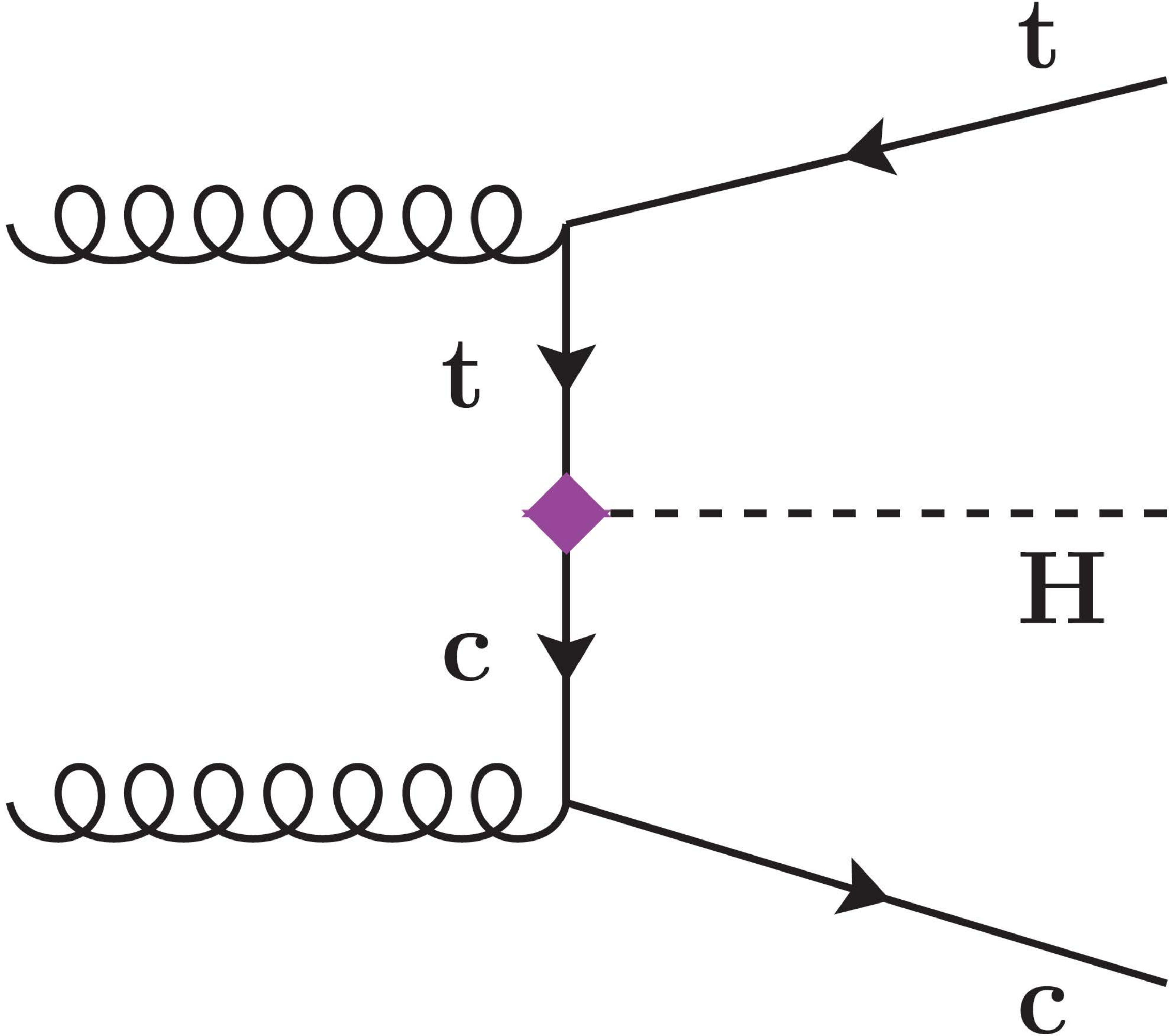}}~~~~~~
\subfigure{\includegraphics[width=0.2\textwidth]{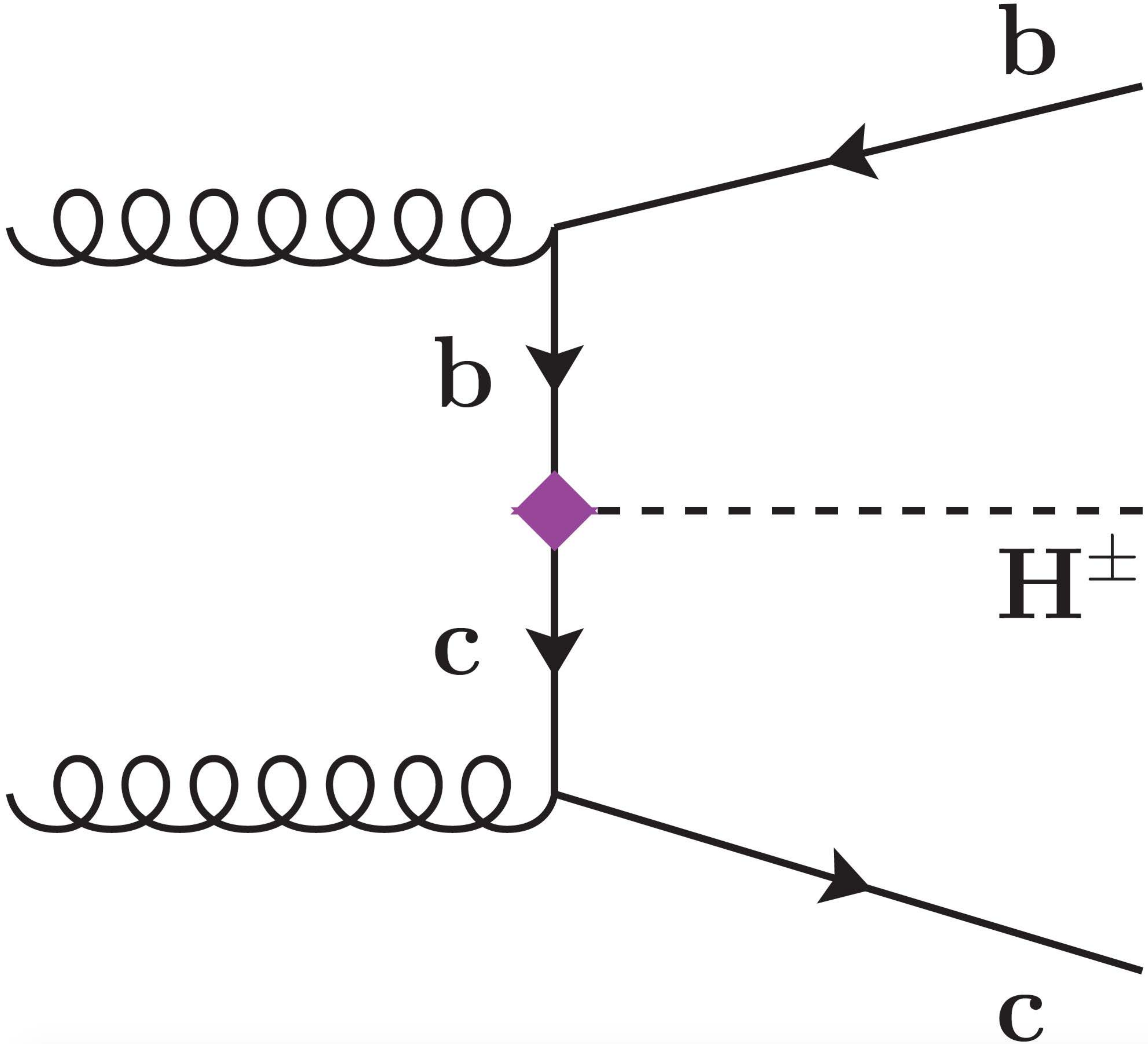}}~~~~~~~~~~~~
\subfigure{\includegraphics[width=0.2\textwidth]{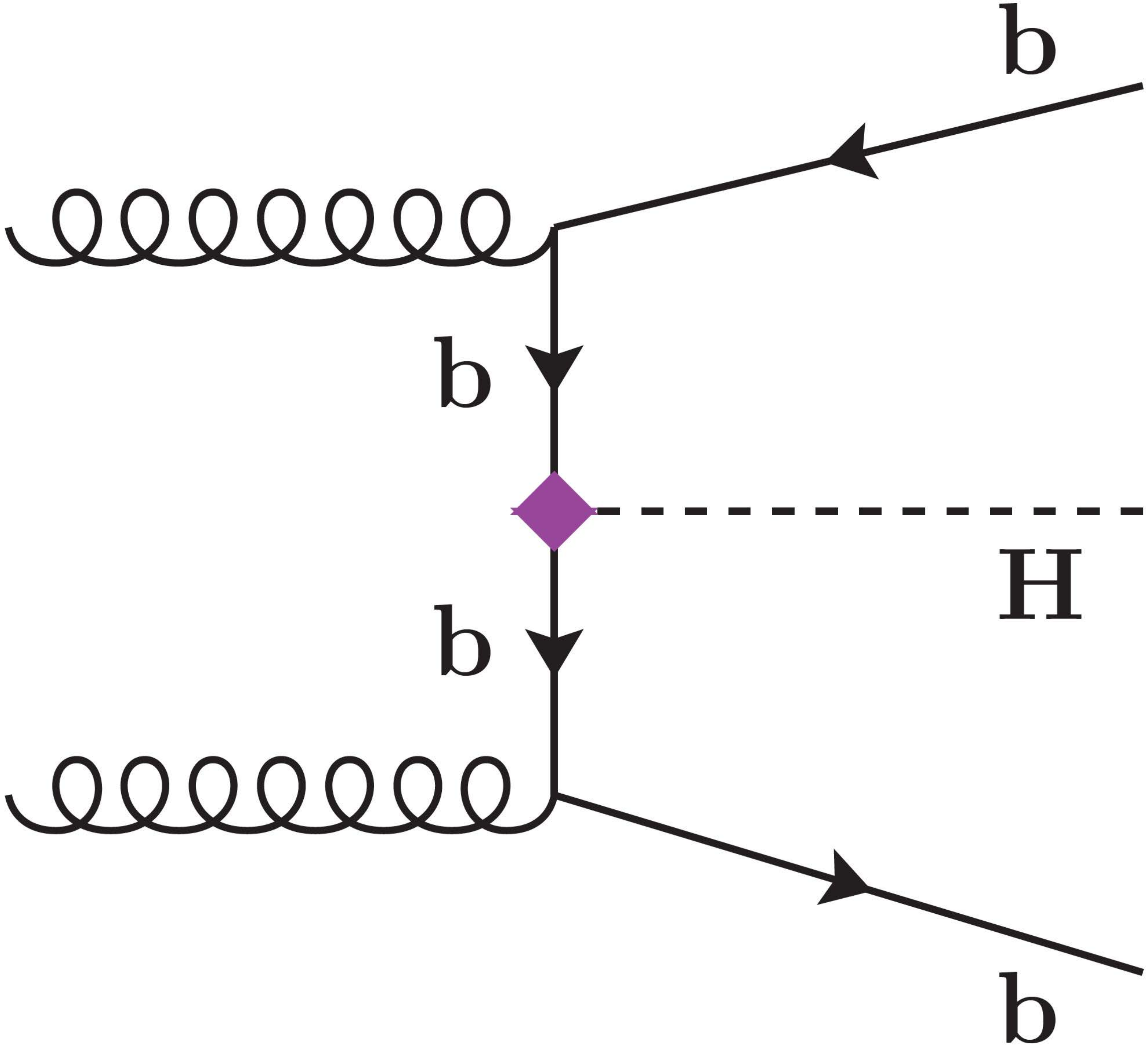}}~~~~~~
\subfigure{\includegraphics[width=0.2\textwidth]{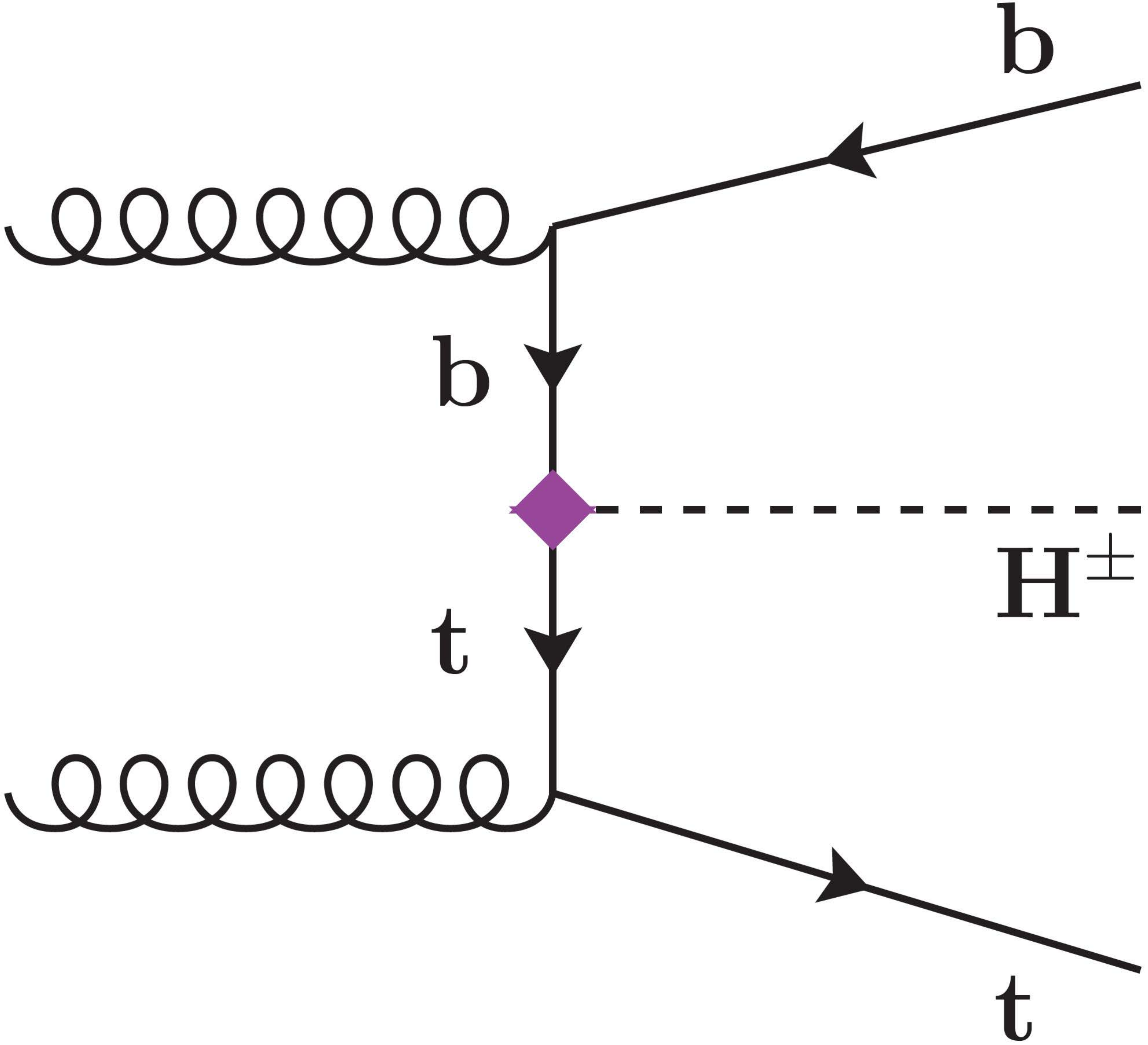}}
\caption{Some of the most important Feynman diagrams contributing to heavy Higgs production in our model (first two panels), as compared to a Type II 2HDM (last two panels). We do not report the gluon fusion diagram since it is common to the several types of 2HDMs.}
\label{fig:FD}
\end{center}
\end{figure}

One of the primary goals of our study is to shed light on possible interesting collider signatures that have not been searched for yet at the LHC.
As we will show in this section, over a broad region of the allowed parameter space of the model, the phenomenology
of the heavy Higgs bosons, $H,~A,~H^\pm$, differs significantly from what is typically assumed in direct searches at the LHC. Particularly, the heavy Higgs bosons can have large branching ratios for flavour-violating decays, similarly to what is predicted by different flavour structures such as e.g. the flavourful 2HDM (F2HDM) of Refs.~\cite{Altmannshofer:2015esa,Ghosh:2015gpa,Altmannshofer:2016zrn}. Furthermore, also some of the main production mechanisms will be different than the ones predicted by a Type II 2HDM (see Fig.~\ref{fig:FD} for the most important Feynman diagrams contributing to the heavy Higgs production in our model, as compared to a Type II 2HDM. In the figure, we do not report the gluon fusion diagram since it is common to the several types of 2HDMs). For definiteness, and following the discussion in Section~\ref{sec:constraints}, we consider a benchmark scenario with $\tan\beta=1$ and $\cos(\beta-\alpha)=0.125$, and
study how the phenomenology depends on the assumed value of $\rho$ and the heavy Higgs boson masses. The choice of a different benchmark with $\cos(\beta-\alpha)\neq 0$ would not alter the phenomenology qualitatively.

The predicted cross sections and branching ratios were obtained using {\tt FeynRules} v2.3~\cite{Alloul:2013bka,Degrande:2014vpa} and {\tt MadGraph\;5}\;v2.3.3~\cite{Alwall:2014hca}.
In order to obtain next-to-leading-order (NLO) cross sections from {\tt MadGraph}, we implemented our own model in {\tt FeynRules} based on the NLO implementation of the 2HDM~\cite{Degrande:2014qga}
and modifying it to a 3, 4 or 5 massless flavour theory as and when necessary. The default {\tt MadGraph} run cards were used with the NNPDF2.3 PDF sets~\cite{Ball:2014uwa} with
matched order. To study the gluon fusion production cross sections of the neutral Higgs boson we used {\tt HIGLU}~\cite{Spira:1996if}, since this code allows for the independent extraction of the top-quark and $b$-quark
loop contributions and their interference. We then rescaled these cross sections with the modified couplings from our model. 

\subsection{Phenomenology of the neutral heavy Higgs boson}

\begin{figure}[t!]
\begin{center}
\subfigure{\includegraphics[width=0.5\textwidth]{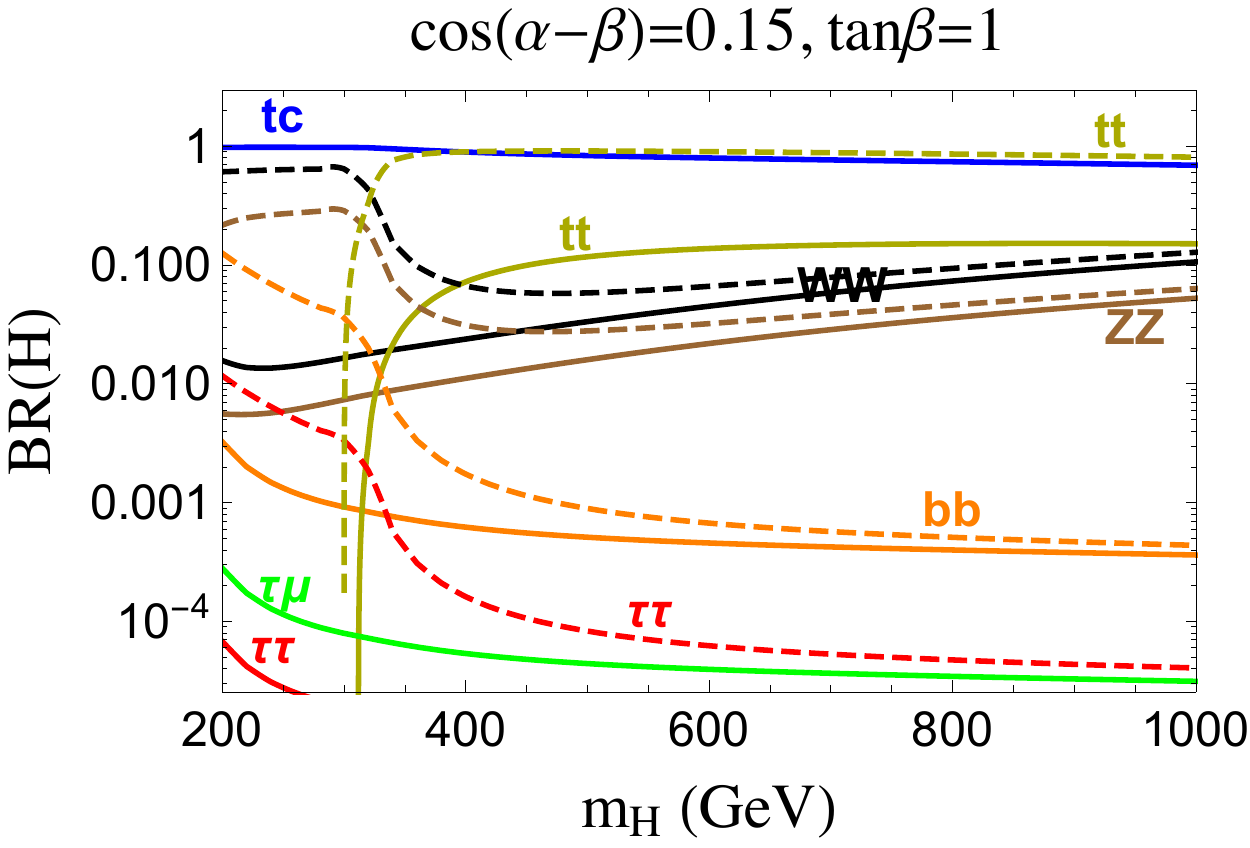}}
\subfigure{\includegraphics[width=0.49\textwidth]{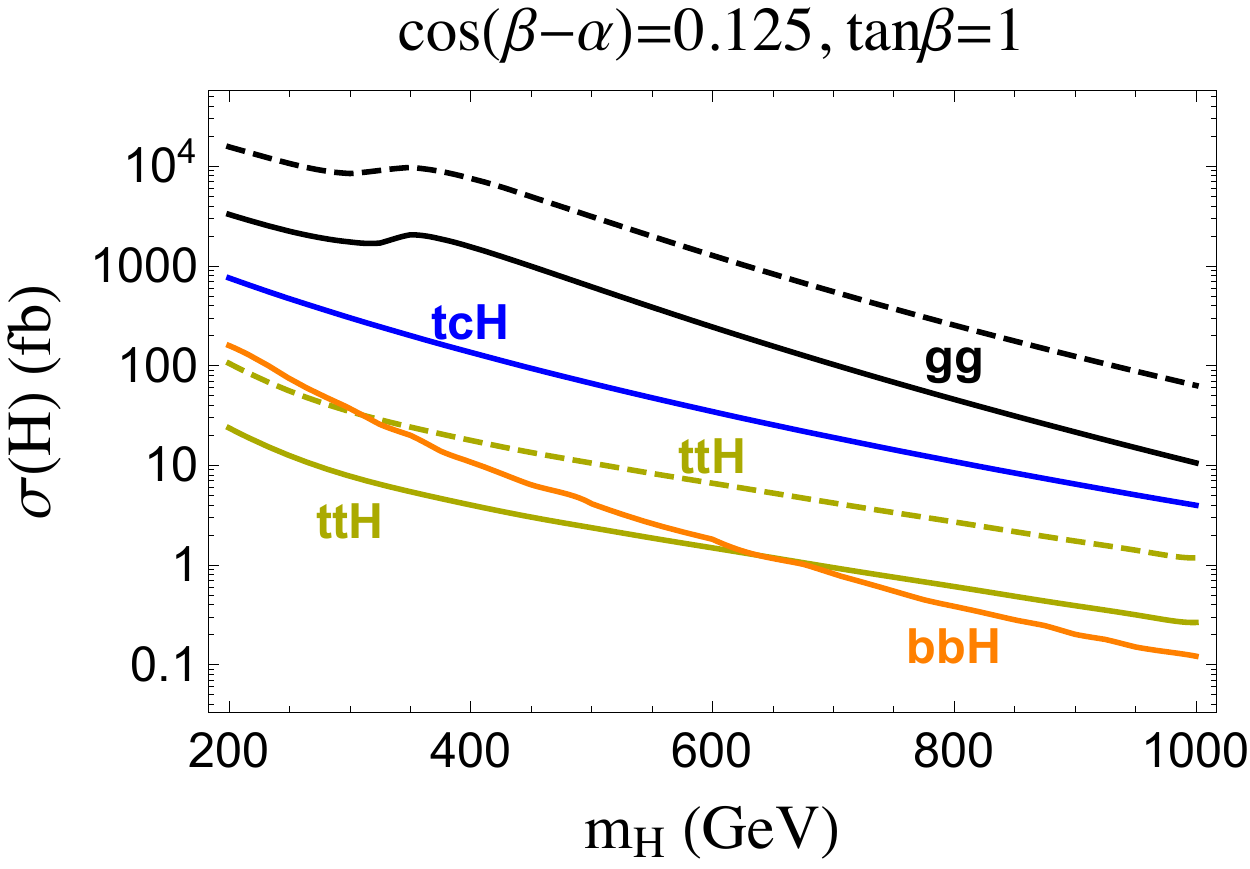}}
\caption{Branching ratios (left panel) and production cross sections at the 13 TeV LHC (right panel) for the neutral Higgs boson as a function of its mass, $m_{H}$, having fixed $\cos(\beta-\alpha)=0.125$, $\tan\beta=1$. Solid lines correspond to $\rho=1$, dashed lines to the flavour conserving case, $\rho=0$. Note that the solid and dashed lines for the $b\bar{b}H$ process overlap in the right panel.}
\label{fig:BRsAndSigma}
\end{center}
\end{figure}

As shown in Section~\ref{sec:model}, the couplings of the neutral heavy Higgs bosons to fermions depend strongly on the value of $\rho$.
In the case of $\rho=0$, the couplings are purely flavour conserving, whereas for $\rho=\pi/2$ the flavour violating couplings can be even larger than the corresponding flavour-conserving ones.

To illustrate the effect of a sizable value of $\rho$ (which is nevertheless consistent with the experimental constraints from Higgs and flavour transition measurements), in Fig.~\ref{fig:BRsAndSigma} we present the branching ratios (left panel) and production cross sections at the 13 TeV LHC (right panel) for the CP-even neutral Higgs boson as a function of its mass, having fixed $\rho=1$ (solid lines in the plots).\footnote{The phenomenology of the pseudoscalar, $A$, is analogous to that of the CP-even heavy scalar, $H$, with the exception of the decays to $WW$ and $ZZ$ that are loop suppressed in the case of the pseudoscalar~\cite{Bernreuther:2009ts}.} For comparison, we also show the corresponding curves in the flavour conserving case $\rho=0$ (dashed lines in the figure). In this limit, the Higgs couplings to third generation show a Type IV structure, namely Type II-like for the bottom and top couplings and Type I-like for the tau coupling.

\begin{figure}[h!]
\begin{center}
\subfigure{\includegraphics[trim = 10mm 0mm 10mm 0mm, clip, width=0.45\textwidth]{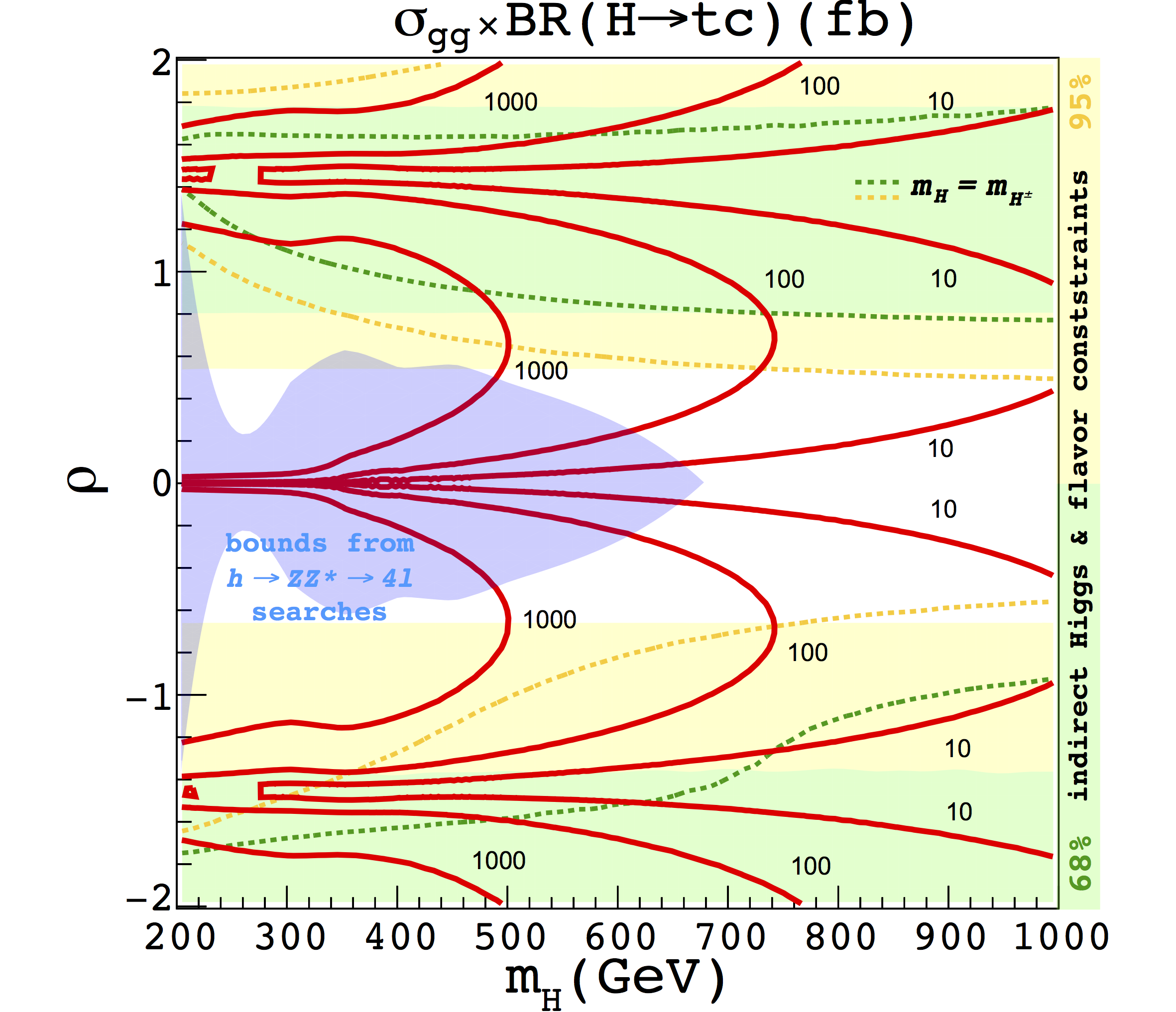}}\hspace{0.25in}
\subfigure{\includegraphics[trim = 10mm 0mm 10mm 0mm, clip, width=0.45\textwidth]{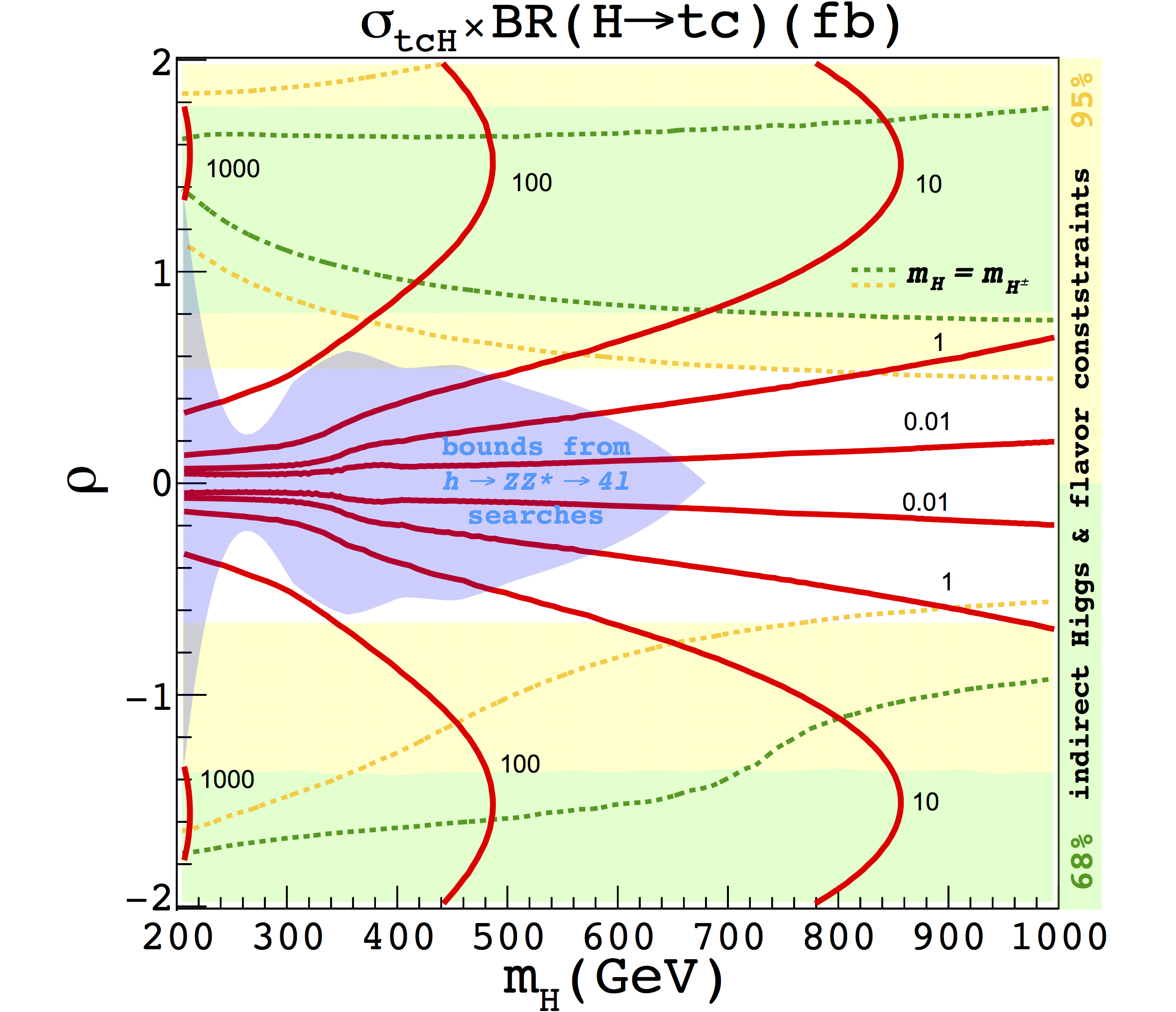}}\\
\subfigure{\includegraphics[trim = 10mm 0mm 10mm 0mm, clip, width=0.45\textwidth]{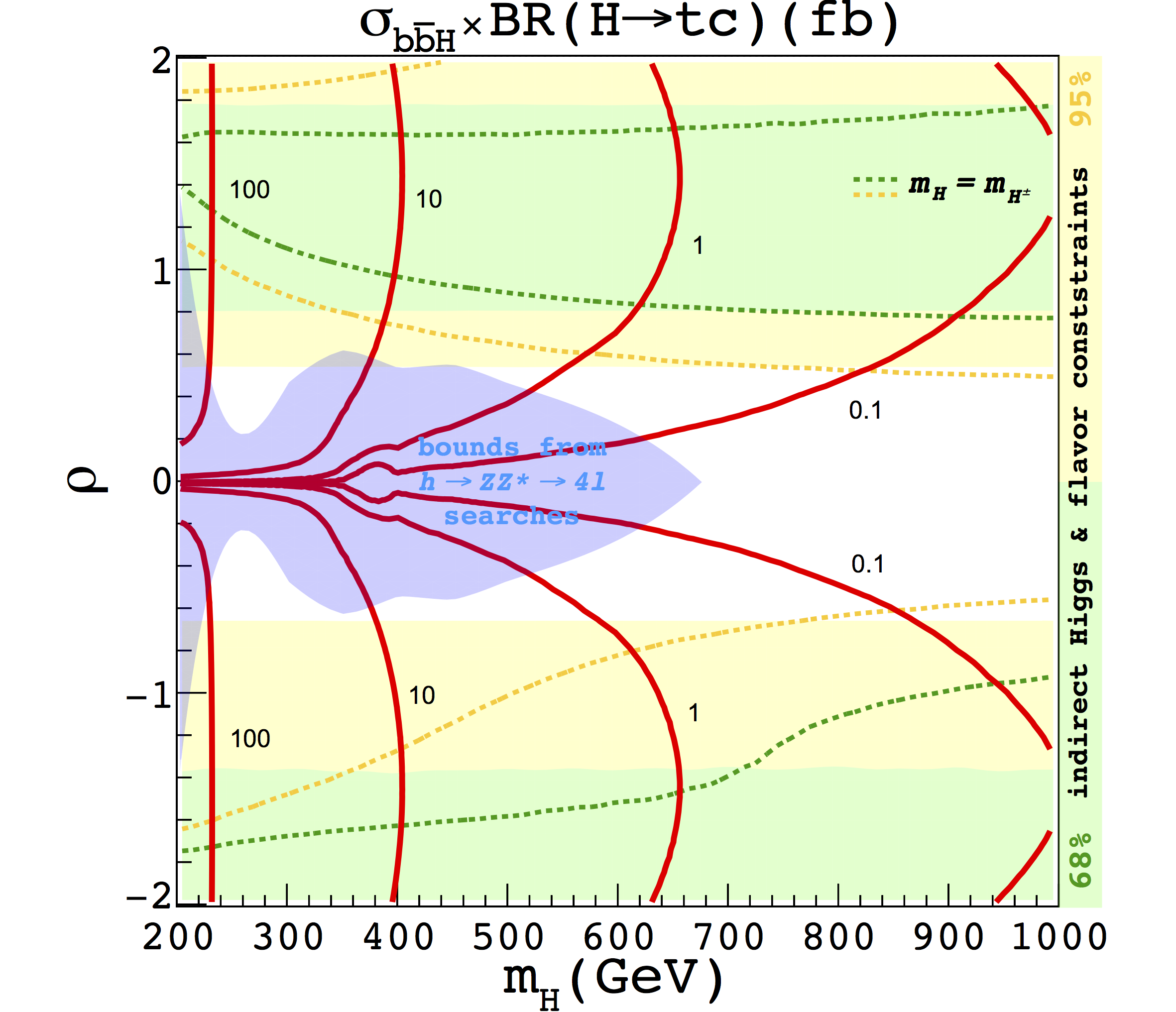}}\hspace{0.25in}
\subfigure{\includegraphics[trim = 10mm 0mm 10mm 0mm, clip, width=0.45\textwidth]{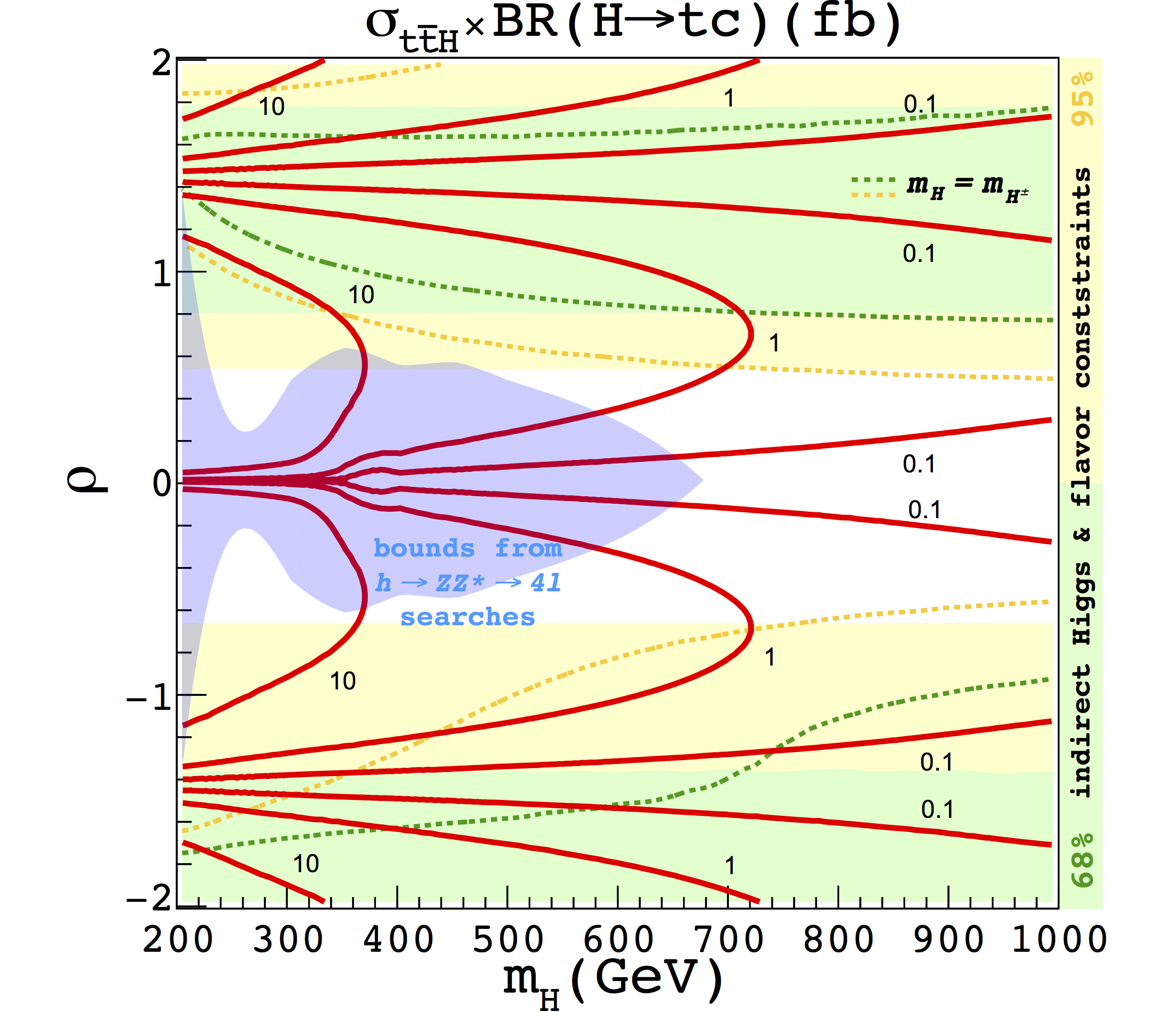}}
\caption{13 TeV cross section times branching ratio for the novel signatures of the neutral Higgs bosons, as a function of $\rho$ and of the neutral Higgs boson mass, $m_{H}$, having fixed $\cos(\beta-\alpha)=0.125$ and $\tan\beta=1$.
The shaded blue region is the region excluded at 95\% CL by LHC searches for $pp\to H\to ZZ^*\to 4\ell$ at $\sqrt{s}=13$\,TeV~\cite{ATLAS-CONF-2017-058,CMS:2016ilx}. The green and yellow regions are favored at the 68\% and 95\% CL by the constraints listed in Tables~\ref{tab:HC},~\ref{tab:HFNC}, and~\ref{tab:flavour} but setting $\tan\beta = 1$ and $\cos(\beta-\alpha) = 0.125$, and marginalizing over $m_{H^\pm}$. The dashed lines correspond to the same regions but assuming $m_{H}=m_{H^\pm}$.}
\label{fig:RatesNewSignals}
\end{center}
\end{figure}

As we will also see in Fig.~\ref{fig:RatesNewSignals}, sizable values of $\rho$ result in $H\to tc$ being the main decay mode.
The branching ratio into $t\bar{t}$ is suppressed by about a factor of ten, when compared to the well-studied Type II 2HDM at low $\tan\beta$ (see the dashed yellow line in the figure). The choice of $\tan\beta=1$ leads to only very small branching ratios into $b\bar{b}$ and $\tau^+\tau^-$, as it can be seen from the expression of the couplings in Appendix~\ref{app:HEC}. Particularly, the $\tau^+\tau^-$ decay mode is quite suppressed if compared to the Type IV 2HDM prediction (see solid vs. dashed red lines in the figure).
As our benchmark point requires $\cos(\beta-\alpha)=0.125$ and thus requires some mixing between the two Higgs doublets,
the branching ratios into $W^+W^-$ and $ZZ$ are non-zero but small, in the range of $\sim$0.5\%--10\%, depending on mass. Finally, the branching ratio for the flavour-violating decay $H\to\tau\mu$ is generically larger than the one for $H\to\tau\tau$, even if it is, in any case, quite small (see e.g. Refs.~\cite{Buschmann:2016uzg,Sher:2016rhh,Primulando:2016eod} for models predicting sizable lepton-flavour-violating branching ratios).

The main production mechanisms are gluon fusion, followed by associated production with a top quark and a $c$-quark ($tcH$). This novel production mode has a sizable cross section due to the sizable $Htc$ coupling (see Eq.~(\ref{eq:Htc})).
The productions in association with top quarks ($t\bar{t}H$) and $b$-quarks ($b\bar{b}H$) are relatively small, both having cross sections of the order of tens of fb for a heavy Higgs boson with a mass of a few hundred GeV. Particularly, the $t\bar{t}H$ cross section is rather suppressed if compared to a Type II 2HDM (solid vs. dashed yellow lines in the figure); the $b\bar{b}H$ cross section is the same as in a Type II 2HDM, since the down quark sector of our framework has the same coupling structure as a Type II 2HDM.

In Fig.~\ref{fig:RatesNewSignals}, we show the cross sections times branching ratio, $\sigma \times {\rm BR}$, for the novel signatures of the neutral Higgs bosons,
as a function of $\rho$ and of the neutral Higgs boson mass, $m_{H}$ at the 13 TeV LHC. In particular, we show $\sigma \times {\rm BR}$ for the processes $pp\to H (\to tc)$ (upper left panel),
$pp\to tc H (\to tc)$ (upper right panel), $pp\to b\bar{b} H (\to tc)$ (lower left panel), and $pp\to t\bar{t} H (\to tc)$ (lower right panel). The shaded blue region in the figure is the region of parameter space excluded at 95\% CL by the present LHC searches. The primary bounds arise from the searches for $pp\to H\to ZZ\to 4\ell$~\cite{ATLAS-CONF-2017-058,CMS:2016ilx}, while searches
for $pp\to H\to WW$~\cite{ATLAS-CONF-2016-074,ATLAS-CONF-2017-051}, $pp\to H\to t\bar{t}$~\cite{ATLAS-CONF-2016-073},\footnote{Here we consider only those searches that take into
account the interference effect between the non-resonant $t\bar{t}$ background and the signal, which significantly distorts the spectrum, creating a peak-dip structure \cite{Dicus:1994bm,Craig:2015jba,Jung:2015gta,Gori:2016zto,Carena:2016npr,Djouadi:2016ack}.}
$pp\to H\to b\bar{b}$~\cite{CMS-PAS-HIG-16-025}, and $pp\to H\to \tau^+\tau^-$~\cite{ATLAS-CONF-2017-050,CMS-PAS-HIG-16-037}
do not set relevant constraints on the parameter space, and are not shown in the figure.
Figure~\ref{fig:RatesNewSignals} also shows the 68\% and 95\% favored regions that are obtained from combining all the constraints listed in Tables~\ref{tab:HC},~\ref{tab:HFNC} and~\ref{tab:flavour}, setting $\tan\beta = 1$ and $\cos(\beta-\alpha) = 0.125$, and marginalizing over the value of $m_{H^\pm}$ (shaded green and yellow regions, respectively). The dashed lines correspond to the same regions but assuming $m_{H}=m_{H^\pm}$. 

The processes with the largest signal rate at sizable values of $\rho$ ($\sim$1) are $pp\to H (\to tc)$ and $pp\to tc H (\to tc)$, with cross sections that can reach $\sim 3$ pb and $\sim 1$ pb, respectively, in the region of parameter space not yet probed by the present LHC heavy Higgs searches (blue shaded region) and in agreement with
Higgs and flavour measurements (green and yellow shaded regions).
These two processes are also complementary in their coverage of $\rho$, with the former process peaking at $|\rho| \sim 0.6$ while
the latter peaks at  $|\rho| \sim 1.5$, for the  values of the other parameters, $\tan\beta=1$ and $\cos (\beta-\alpha)=0.125$, chosen in our benchmark. The additional production modes, $b\bar bH$ and $t\bar tH$, lead to cross sections that are at least one order of magnitude smaller than $tc H$ and $pp\to H$, and, therefore, are not the first smoking gun of our framework.

Experimentally the $pp\to H (\to tc)$ process is quite challenging
because of the relatively small signal rate in comparison with the large expected backgrounds, mainly coming from the $t\bar{t}$+jets and $W$+jets SM processes.
Also, for heavy Higgs bosons with masses of few hundred~GeV, the
resulting top quarks would typically not be boosted, which experimentally implies focusing on leptonic top-quark decays in order to be able to trigger on the events.
A similar signature is probed by searches for $pp \to H^+ (\to tb)$~\cite{Aad:2015typ} and $pp \to W^\prime (\to tb)$~\cite{Aad:2014xea,Chatrchyan:2014koa,CMS-PAS-B2G-17-010},
which have some difficulty in probing low values of mass ($<500$\,GeV), even with two $b$-tagged jets in the event, a key requirement to suppress backgrounds
from $W$+jets production. In the case of our signal of interest, one of those bottom jets would be replaced by a charm jet, which is harder to discriminate from light-flavour jets,
although the LHC experiments are making progress on optimizing the performance of charm-tagging algorithms~\cite{ATL-PHYS-PUB-2015-001}.
It will be important to assess the prospects for probing the process $pp\to H (\to tc)$ in the coming years of the LHC using these new developments, even if we expect that
reaching the required level of sensitivity would be rather challenging.

On the other hand, the process $pp\to tcH (\to tc)$, with two top quarks and two additional charm jets in the final state,
provides more experimental  handles to suppress the backgrounds. There are two experimental signatures that are potentially promising.
(1)~In the case of only one top quark decaying leptonically, the final state is characterized by one lepton, some missing transverse momentum ($E_{\rm T}^{\rm miss}$)
due to the presence of a neutrino, and at least six jets, of which two are bottom jets and two are charm jets. In this case the main background
is $t\bar{t}$+jets production, and in particular $t\bar{t}$+$b\bar{b}$ and $t\bar{t}$+$c\bar{c}$, which are subject to large uncertainties both
theoretical and experimental. However, the LHC experiments have developed sophisticated search strategies exploiting large-statistics
subsidiary data samples for the purpose of constraining background uncertainties. Examples are searches for $tbH^\pm (\to tb)$~\cite{ATLAS-CONF-2016-089}
and $b\bar{b}H (\to t\bar{t})$~\cite{ATLAS-CONF-2016-104}, although they are not optimized for this signal.
A natural evolution of those searches to target this signal would include exploiting kinematic reconstruction of the heavy Higgs boson mass,
and possibly using explicit charm tagging as a way to suppress the $t\bar{t}$+$b\bar{b}$ background. It will be very interesting to assess  the reach of this possible search. 
(2)~An additional and even more promising avenue to probe the $pp\to tcH (\to tc)$ signal is through the final-state signature of same-charge dilepton plus bottom and charm jets,
since half of the events have two same-charge top quarks. In this case SM backgrounds are heavily suppressed compared to the lepton-plus-jets
final state discussed previously, although instrumental backgrounds from $t\bar{t}$+jets with charge-misreconstruction or non-prompt/fake leptons need
to be precisely estimated using data-driven techniques.
Examples of this kind of searches, currently not tailored for the $pp\to tc H (\to tc)$ signal, but which could be optimized accordingly, are searches for
pair-production of vector-like quarks~\cite{ATLAS-CONF-2016-032} or anomalous four-top-quark production, e.g. via $pp\to t\bar{t}H(\to t\bar{t})$~\cite{ATLAS-CONF-2016-032,Sirunyan:2017uyt} (see also Refs.~\cite{Craig:2015jba,Gori:2016zto,Craig:2016ygr}).

Finally, searches for same-charge dilepton or trilepton signatures are also very promising to probe the $pp\to tcH (\to t\bar{t})$ process, 
with three top quarks in the final state, provided the $H\to t \bar{t}$ branching ratio is sufficiently large.\footnote{The $pp\to tcA (\to tc)$ and $pp\to tcA (\to t\bar{t})$ 
processes are also predicted in the context of flavon models~\cite{Bauer:2016rxs}. We thank Martin Bauer for pointing this out.}
However, we do not discuss this channel in great detail here since the flavour violation in the model under consideration leaves more dominant 
signatures in the decay modes discussed above.
\subsection{Phenomenology of the charged Higgs boson}

\begin{figure}[t!]
\begin{center}
\subfigure{\includegraphics[width=0.49\textwidth]{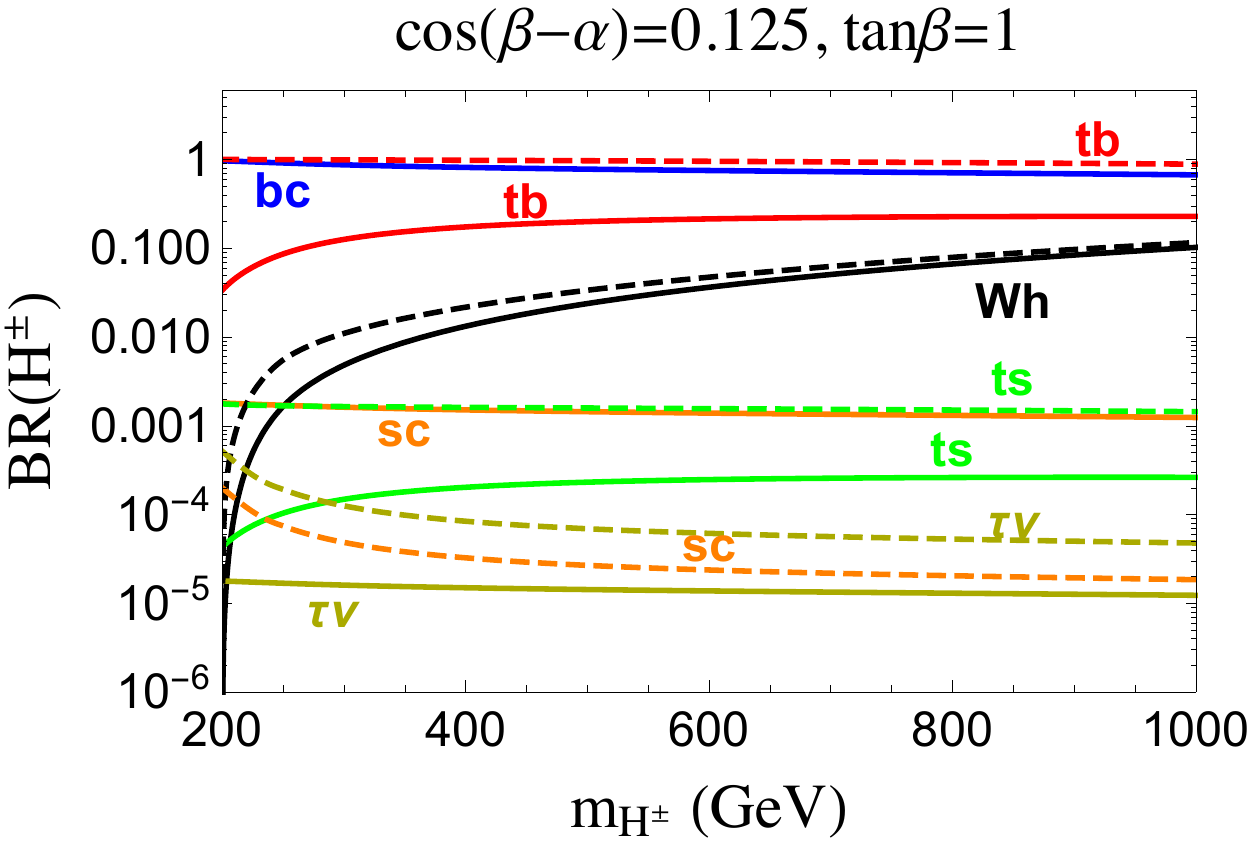}}
\subfigure{\includegraphics[width=0.49\textwidth]{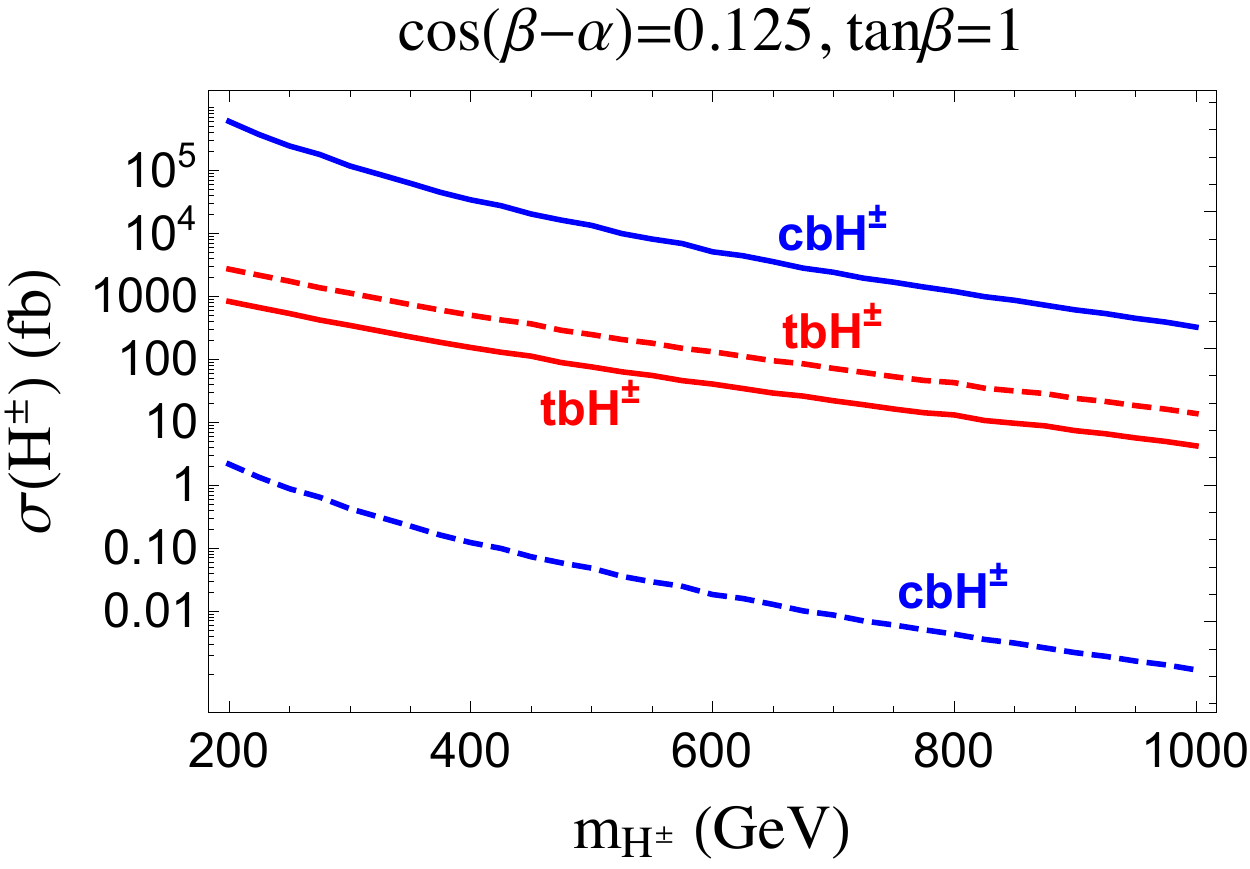}}
\caption{Branching ratios (left panel) and production cross sections at 13 TeV LHC (right panel) for the charged Higgs boson as a function of its mass $m_{H^\pm}$,  having fixed $\cos(\beta-\alpha)=0.125$, $\tan\beta=1$. Solid lines correspond to $\rho = 1$, dashed lines to the flavour conserving case, $\rho = 0$.}
\label{fig:BRsAndSigmaHpm1}
\end{center}
\end{figure}

The phenomenology of the charged Higgs boson also changes dramatically at non-zero values of $\rho$, if compared to the better studied Type I-II 2HDMs.
In Fig.~\ref{fig:BRsAndSigmaHpm1} we present the branching ratios (left panel) and production cross sections (right panel) for the charged Higgs boson as a function of its mass, assuming 
$\cos(\beta-\alpha)=0.125$, $\tan\beta=1$, and $\rho=1$ (solid lines) or $\rho=0$ (dashed lines). Note that, contrary to the neutral Higgs case, in the charged Higgs case the $\rho\to0$ limit correspond to formally finite but numerically small deviations from Type II Yukawa couplings in the quark sector. In the lepton sector, the couplings do not match any of the Type I-IV 2HDMs. For the $\rho=1$ benchmark, the $H^\pm \to cb$ decay mode becomes dominant, with a branching ratio well above $\sim$70\%, followed by $H^\pm \to tb$ with a branching
ratio below $\sim$20\% and quite smaller than the corresponding prediction in a Type II 2HDM (see solid vs. dashed red lines in the figure). This exact ratio of branching ratios is due to the specific benchmark chosen. Particularly the ratio of the $tb$ and $cb$ couplings can be approximated by $\sim (1+\tan^2\beta)\sin\rho/[1-\tan^2\beta+(1+\tan^2\beta)\cos\rho]$, showing that the hierarchy between the $tb$ and $cb$ decay modes can also be reversed, even if typically the $cb$ coupling is larger than the $tb$ one. Because of the non-zero value of $\cos(\beta-\alpha)$ the branching ratio $H^\pm \to Wh$ is also sizable. Furthermore the very well studied $\tau\nu$ mode is very suppressed and has branching ratios at the level of few $\times 10^{-5}$. This branching ratio is smaller than the corresponding one in a Type II 2HDM and in the $\rho=0$ case (solid vs. dashed yellow lines in the figure).

For our benchmark, because of the enhancement of the $H^\pm cb$ coupling relative to the $H^\pm tb$ coupling, the main production mechanism is $pp \to cbH^\pm$, with a cross
section orders of magnitude above that for $pp \to tbH^\pm$, and much larger than what is predicted by a Type II 2HDM (solid vs. dashed blue lines in the figure). The $H^\pm tb$ cross section is, instead, a factor of a few smaller than in a Type II 2HDM (solid vs. dashed red lines in the figure). As a result, in our scenario the usual signal process searched for at the LHC, $pp\to tbH^\pm (\to tb)$ \cite{ATLAS-CONF-2016-089,Khachatryan:2015qxa}, is quite suppressed,
while there are other signatures that are much more promising and remain relatively unexplored.

\begin{figure}[H]
\begin{center}
\subfigure{\includegraphics[trim = 10mm 0mm 10mm 0mm, clip, width=0.45\textwidth]{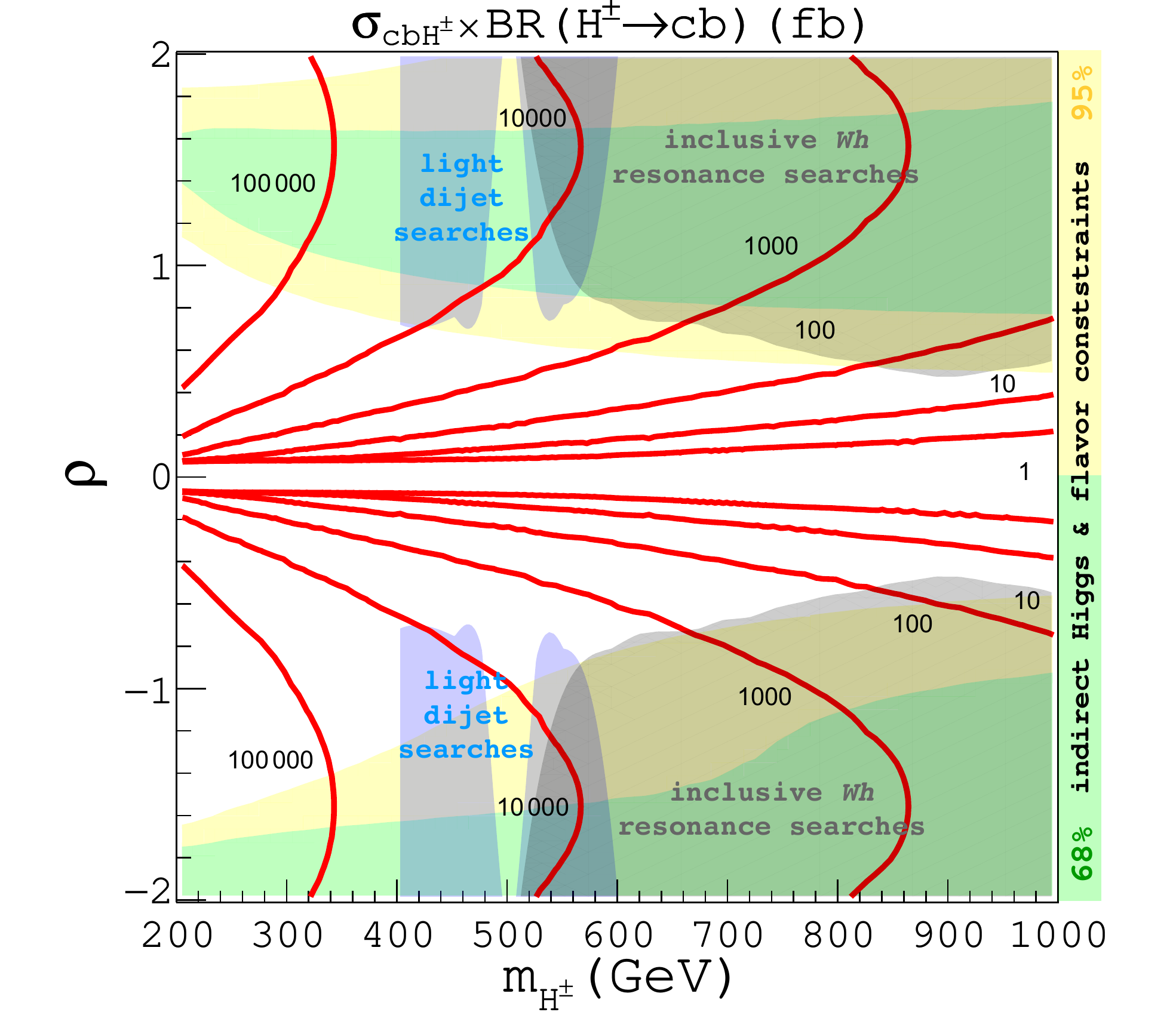}}\hspace{0.25in}
\subfigure{\includegraphics[trim = 10mm 0mm 10mm 0mm, clip, width=0.45\textwidth]{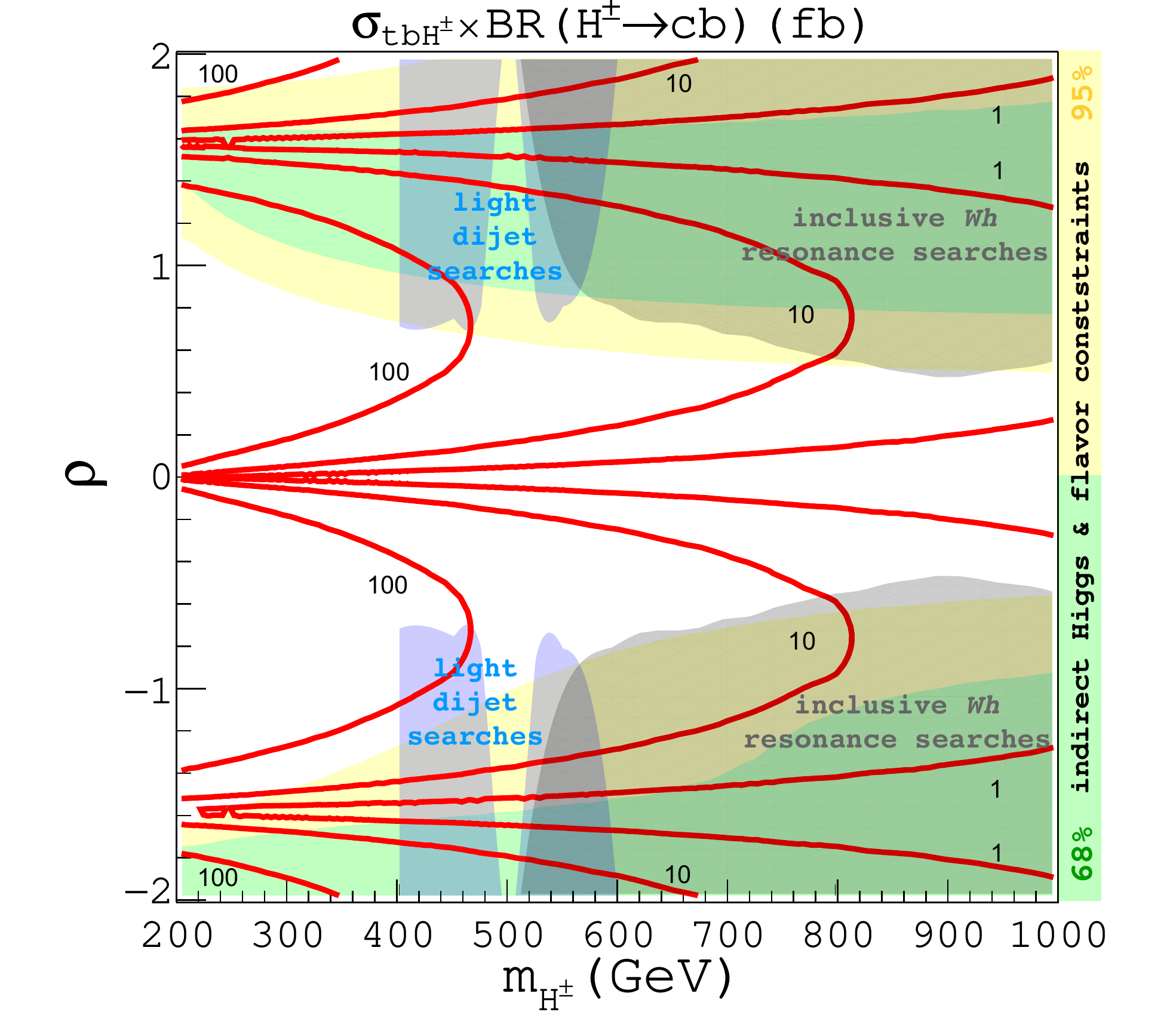}}\\
\subfigure{\includegraphics[trim = 10mm 0mm 10mm 0mm, clip, width=0.45\textwidth]{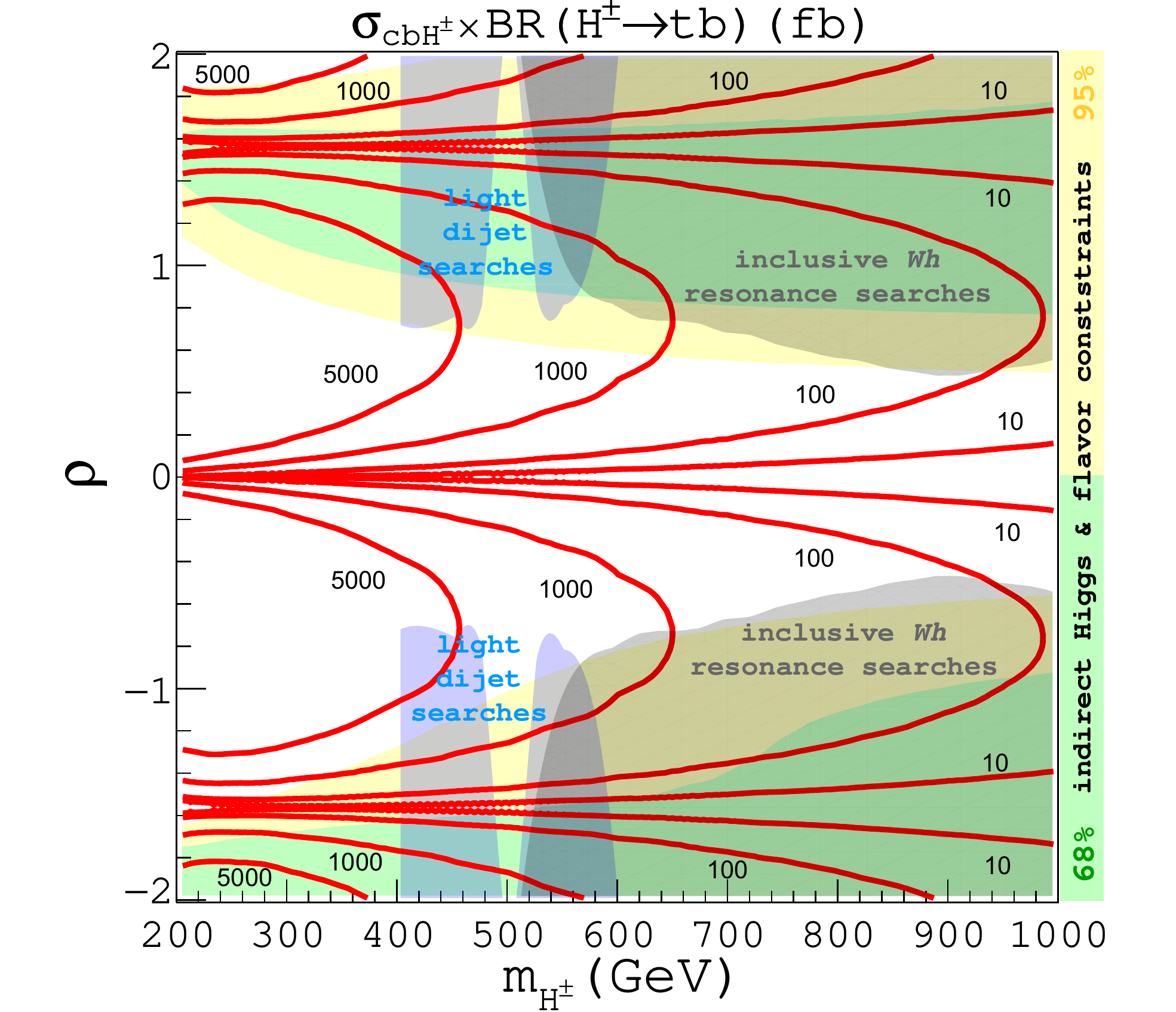}}\hspace{0.25in}
\subfigure{\includegraphics[trim = 10mm 0mm 10mm 0mm, clip, width=0.45\textwidth]{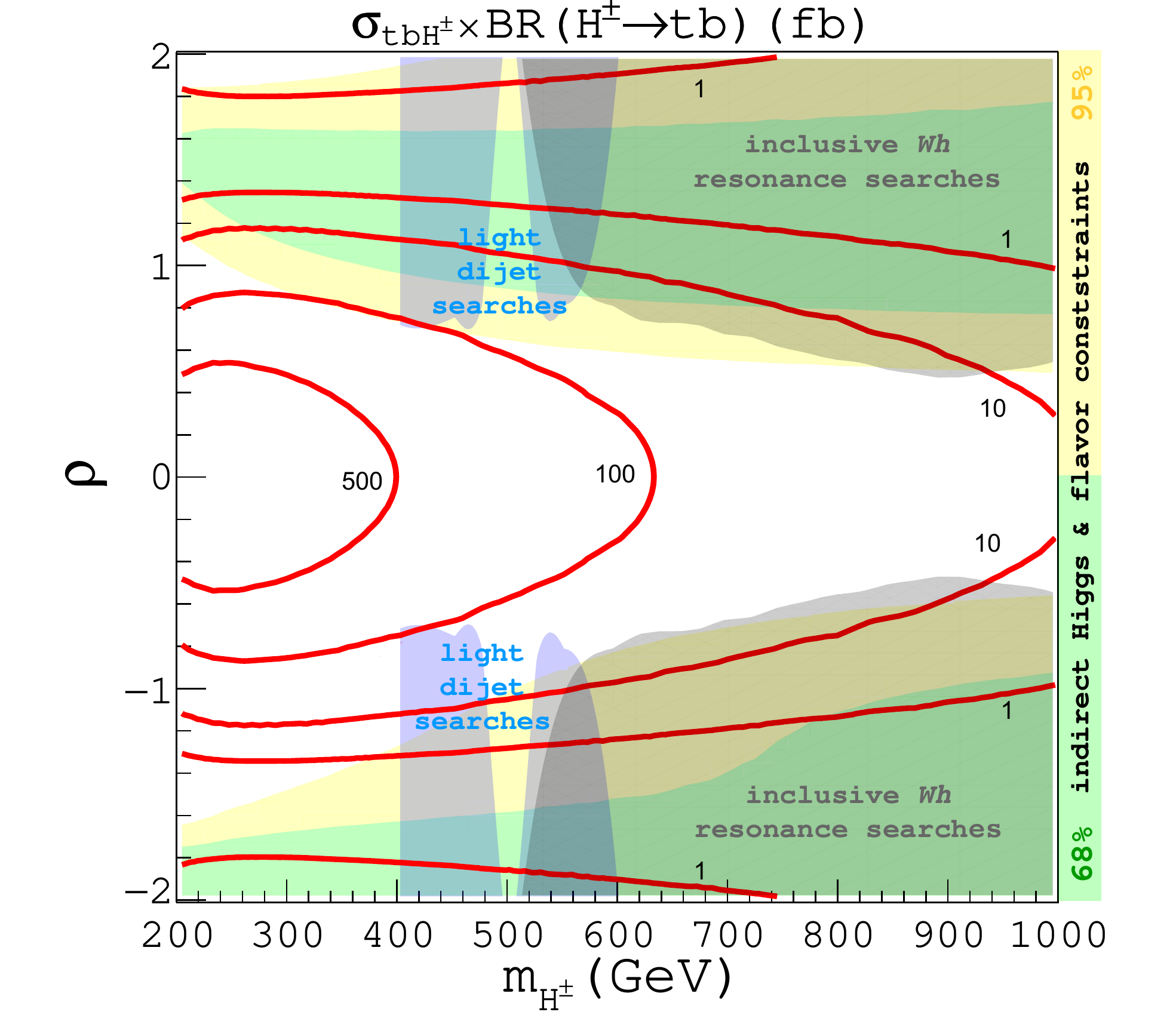}}\\
\subfigure{\includegraphics[trim = 10mm 0mm 10mm 0mm, clip, width=0.45\textwidth]{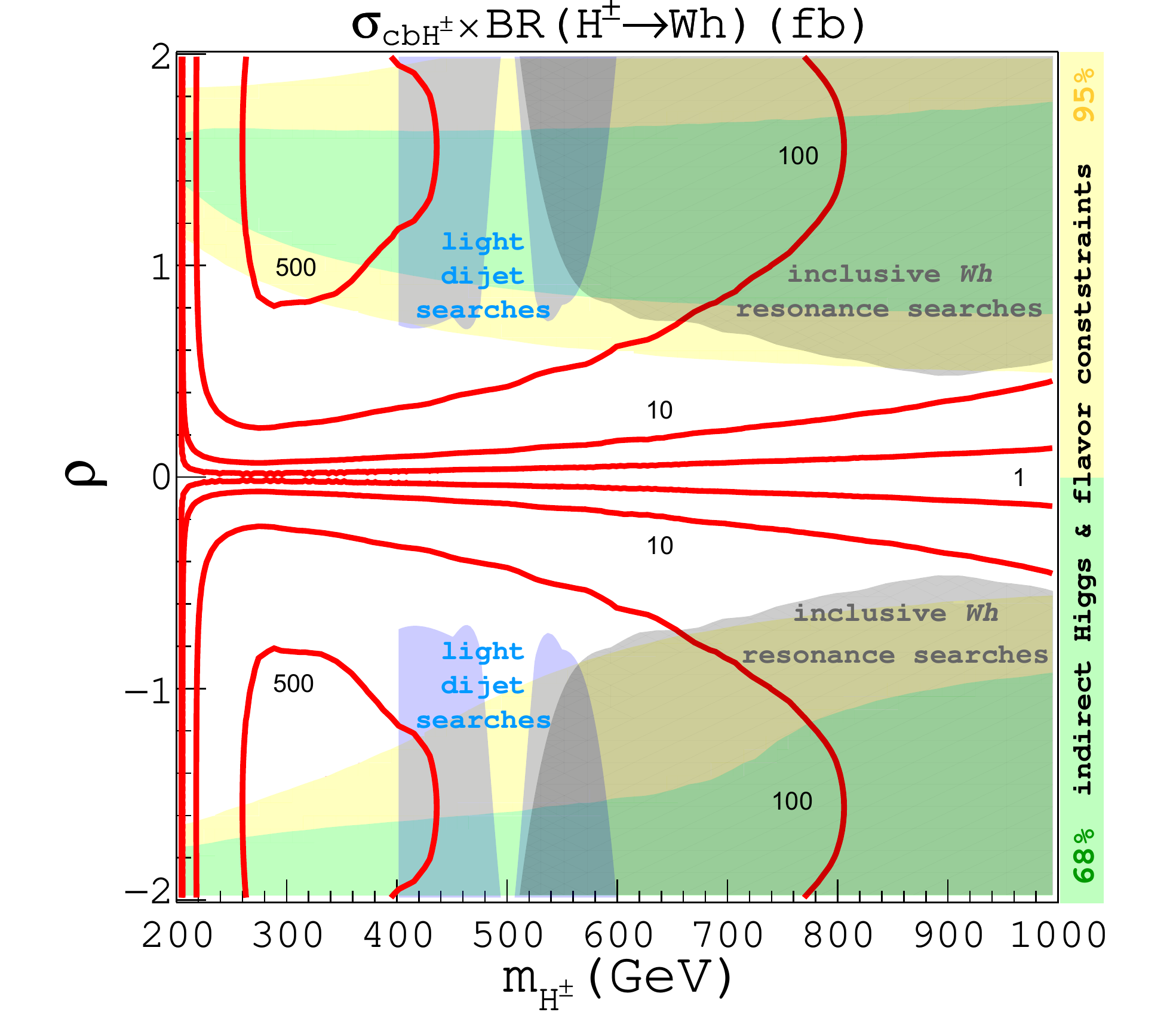}}\hspace{0.25in}
\subfigure{\includegraphics[trim = 10mm 0mm 10mm 0mm, clip, width=0.45\textwidth]{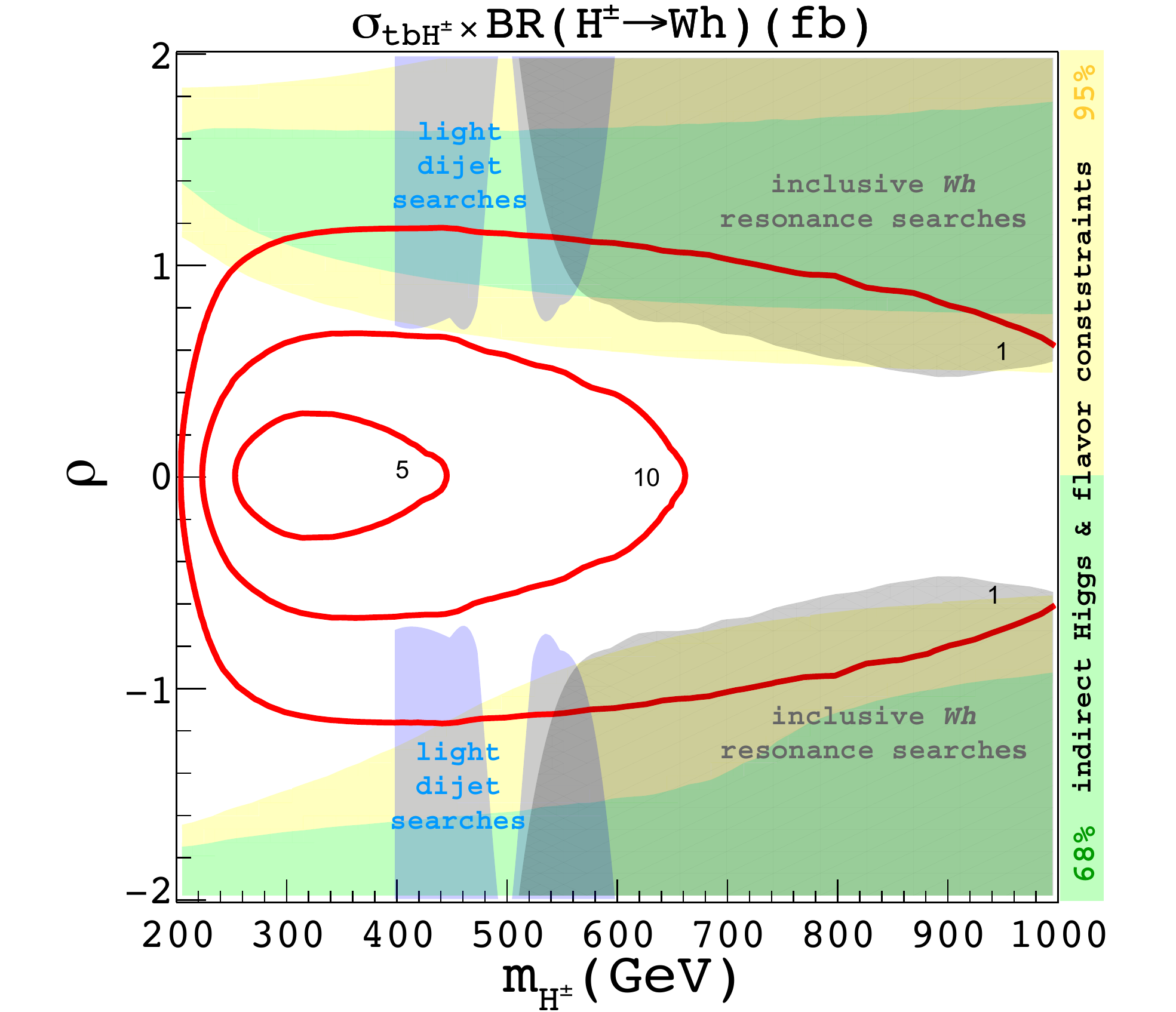}}
\caption{13 TeV cross section times branching ratio for the novel signatures of the charged Higgs boson, as a function of $\rho$ and charged Higgs boson mass $m_{H^\pm}$, having fixed $\cos(\beta-\alpha)=0.125$ and $\tan\beta=1$.
The shaded blue region is the region excluded at 95\% CL by LHC light dijet searches. The shaded gray region is probed by inclusive $Wh$ resonance searches. The green and yellow regions are the 68\% and  95\% favored regions respectively that are obtained from combining all the constraints listed in Tables~\ref{tab:HC},~\ref{tab:HFNC}, and~\ref{tab:flavour}, setting $\tan\beta = 1$ and $\cos(\beta-\alpha) = 0.125$. }
\label{fig:RatesNewSignalsHpm}
\end{center}
\end{figure}

In Fig.~\ref{fig:RatesNewSignalsHpm} we show $\sigma \times {\rm BR}$ for several interesting $H^\pm$ at the 13 TeV LHC, as a function of $\rho$ and of the charged Higgs boson mass, $m_{H^\pm}$.  The 68\% and 95\% favored regions that are obtained from combining all the constraints discussed in the previous sections and listed in Tables~\ref{tab:HC},~\ref{tab:HFNC} and~\ref{tab:flavour} are shaded in green and yellow, respectively. Over most of the parameter space, except when $\rho$ is small, the dominant process is $pp\to cbH^\pm (\to cb)$, with  $\sigma \times {\rm BR}$  ranging from $\mathcal O(500~{\rm{pb}})$ at  $m_{H^\pm}\sim 200$~GeV to $\mathcal O(1~{\rm{pb}})$ at $m_{H^\pm}\sim 800$~GeV. This process can, in principle, be probed by inclusive searches for dijet resonances at the LHC.  Currently a few searches probe resonance masses below
1\,TeV (see Refs.~\cite{ATLAS-CONF-2016-030,ATLAS-CONF-2016-031,ATLAS:2016bvn,ATLAS:2016jcu,CMS-PAS-EXO-16-056,CMS-PAS-EXO-17-001} for LHC
\noindent 
searches at $\sqrt{s}=13$\,TeV), which is the mass range of interest in this study. In particular, among these searches, the most stringent constraint on our model is set by Ref.~\cite{ATLAS-CONF-2016-030}, and is displayed as a shaded blue region in Fig.~\ref{fig:RatesNewSignalsHpm}. To obtain this region, we recast the ATLAS analysis~\cite{ATLAS-CONF-2016-030}. We use {\tt MadGraph} in the four-flavour scheme to compute cross sections and generate parton-level events at $\sqrt{s}=13$\,TeV for the process $pp\to cbH^\pm (\to cb)$, for different values of the charged Higgs mass. Subsequently, we require  the two leading jets to have a
$p_{\rm T}$ larger than 185\,GeV and 85\,GeV, respectively, and $|\eta|<2.8$, as required by the ATLAS analysis.
The requirement for $|y^*|<0.3$ or $|y^*|<0.6$, depending on the signal region, is also added ($y^*$ is the difference in rapidity of the two leading jets, $y^*\equiv (y_1-y_2)/2$).

 Finally, we extract the bound on the parameter space, by comparing the cross section of our model after cuts, with the excluded cross section. As can be appreciated, most of the mass region is still allowed even at high values of $\rho$.
However, these analyses only use a fraction of the Run 2 dataset and further optimizations for this signal should be possible, including making $b$-(and c-)tagging requirements.
Therefore, we tentatively conclude that dedicated future searches for resonances in multijet final states would be quite important to probe a significant
fraction of the parameter space in this scenario.

The process with the next highest rate is $pp\to cbH^\pm (\to tb)$, with a $\sigma \times {\rm BR}$ of several pb below a mass of $\sim$600\,GeV.
Since there is a high probability that the associated $b$-quark and $c$-quark fall outside the ATLAS and CMS detector acceptance (only a few percents of events have both charm and
$b$-jets satisfying the standard $p_{\rm T}>20$\,GeV and $|\eta|<2.5$ requirements)
the main signature is a $tb$ resonance, for which there are dedicated searches~\cite{Aad:2014xea,Aad:2015typ,Chatrchyan:2014koa,CMS-PAS-B2G-17-010}.
In practice, however, current searches~\cite{CMS-PAS-B2G-17-010} at $\sqrt{s}=13$\,TeV are able to probe $\sigma \times {\rm BR} \sim 1$ pb at a mass of 1\,TeV,
about one order of magnitude above the expected $\sigma \times {\rm BR}$ in our scenario. 
With sufficient integrated luminosity and further optimisations, these searches may eventually be able
to probe our scenario at masses lower than 1\,TeV, although it is not clear if masses as low as 200\,GeV can be reached.

Other signals have smaller cross sections, with $\sigma \times {\rm BR}$ exceeding 0.5 pb only at low mass. This includes not only the standard LHC search mode, $pp\to tbH^\pm (\to tb)$,
but also new processes such as $pp\to cbH^\pm (\to W^\pm h)$ and $pp\to tbH^\pm (\to cb)$.
In the case of $pp\to tbH^\pm (\to cb)$, the final state signature is the same as in LHC searches for $t\bar{t} \to Wb H^\pm(\to cb) b$~\cite{CMS-PAS-HIG-16-030}, which so far have assumed a
light charged Higgs boson in top-quark decays and, thus, are currently not optimized to probe this signal. Our model offers, therefore, a motivation to search for $H^\pm \to cb$ resonances
even above the top threshold.\footnote{See also Ref.~\cite{Altmannshofer:2016zrn} for another model predicting this signature.}

In the case of $pp\to cbH^\pm (\to W^\pm h)$, since the associated $b$-quark and $c$-quark
are often not reconstructed, the signal is similar to the one probed in searches for the SM Higgs boson in the $Wh$ production mode, except for the presence of a $Wh$ resonance.
Existing searches for $W^\prime \to Wh(\to b\bar{b})$ resonances at $\sqrt{s}=13$\,TeV~\cite{ATLAS-CONF-2017-055,Sirunyan:2017wto} 
currently target resonance masses above at least 500\,GeV,
although masses down to 300\,GeV were also probed by Run 1 searches~\cite{Aad:2015yza}. The most significant constraint on the parameter space of our model is set by the search in Ref.~\cite{ATLAS-CONF-2017-055}, which leads to the bound shaded in gray in Fig.~\ref{fig:RatesNewSignalsHpm}. Due to the large $cbH^\pm$ production cross section,\footnote{Note that the $cbH^\pm$ production cross section in our model can be much larger than the $tbH^\pm$ production in a Type II 2HDM, even at very large values of $\tan\beta$, and can lead to a much larger number of charged Higgs bosons produced at the LHC.} already sizable regions of parameter space are probed, leading us to conclude that there will be very good prospects for $Wh$ searches to probe our model, in the case of $\cos(\alpha-\beta)\neq 0$. Modifications of the current searches can also be envisioned, 
particularly targeting the $cbH^\pm$ associated production through the requirement of an associated heavy-flavour jet, which may lead to further improvements in sensitivity.

In the model used in this work the branching fraction of $H^\pm \to c\bar{s}$ is negligibly small compared to the other decay modes in contrast with come other models where $H^\pm$ has significant flavour violating couplings with the second generation of quarks. A search by CMS for this decay channel can be found in Ref.~\cite{Khachatryan:2015uua}.

\subsection{Comparison with other top-philic flavour-violating 2HDMs}

As we have demonstrated in this section, 2HDM frameworks that single out the third generation of quarks and leptons have a very different phenomenology, if compared to 2HDMs with natural flavour conservation. Particularly, it is very easy to evade the constraints from the present LHC direct searches for new Higgs bosons and, at the same time, predict sizable cross sections for signatures not yet searched for at the LHC. Different scenarios will, however, predict a different pattern of branching ratios and cross sections. In this section, we discuss how to disentangle experimentally between the framework discussed in this paper and other scenarios, in particular the F2HDM proposed in Refs.~\cite{Altmannshofer:2015esa,Ghosh:2015gpa} that generates the third generation fermion masses mainly through the SM Higgs mechanism and the 1st and 2nd generation masses through the Higgs mechanism of an additional Higgs doublet. 

Both frameworks predict sizable branching ratios for $H,A$ flavour-changing decays. However, in the F2HDM the flavour-violating couplings between 2nd and 3rd generations scale as $m_{2^{\rm{nd}}}/v\tan\beta$, where $m_{2^{\rm{nd}}}$ is the mass of the second generation fermions, and have therefore a similar scaling as the flavour-diagonal couplings to second generations. In our framework, instead, the $2\leftrightarrow 3$ flavour-violating couplings in the up and lepton sectors can be quite larger than the coupling to second generations, since $c_{23}\sim (m_{3^{\rm{rd}}}/v) (\cot\beta+\tan\beta)\sin\rho$, where $m_{3^{\rm{rd}}}$ is the mass of the third generation fermions (see Eq. (\ref{Eq:A11})). 
Similarly, flavour-violating couplings involving the first generation quarks and leptons are suppressed by the first generation masses in the F2HDM; in the minimal realization of our framework where the third generation up quark and lepton mixes only with the second generation, there are no neutral Higgs couplings to first and second generations. If we consider a richer flavour structure that mixes the first generation quarks and leptons with the second and third (see the discussion in Section~\ref{sec:model}), the two flavour-violating couplings will scale as $c_{13}\propto m_{3^{\rm{rd}}}/v$ and $c_{12}\propto m_{2^{\rm{nd}}}/v$, 
eventually generating larger flavour-changing effects that, however, will be constrained by low energy flavour observables like $D$--$\bar D$ mixing.

This different scaling of the several couplings results in a quite different phenomenology for the heavy Higgs bosons. In particular, searches for di-muon and di-jet resonances will be more relevant to test the neutral Higgs bosons arising in the F2HDM, than those of the framework discussed in this paper, where the only relevant bounds 
arise from searches for $H\to ZZ$ (see Fig.~\ref{fig:RatesNewSignals}). 
Similarly, also the phenomenology of the charged Higgs boson presents both differences and similarities with the one predicted by the F2HDM. In both 2HDMs the flavour-violating decay $H^\pm \to bc$ has sizable branching ratios. However, contrary to the F2HDM, in our framework this decay mode is typically the dominant one together with the $tb$ decay mode, and has a larger branching ratio than the $cs$ decay mode. Additionally, the charged Higgs boson is typically more copiously produced at the LHC in our framework than in the F2HDM, due to its larger flavour-violating $cb$ coupling, that scales as $m_t/v$.

Another class of models that warrants a mention here is the one proposed by Branco--Grimus--Lavoura (BGL) in Refs.~\cite{Branco:1996bq,Botella:2009pq,Botella:2011ne}. In these models the flavour-changing couplings of the neutral Higgs sector are connected directly with the quark and lepton mixing matrices. Recently, these models have been used to study signatures like $h\to\tau\mu$ and $t\to ch$ also considered in our work. It has been shown that these models can saturate the experimental bounds on these channels~\cite{Botella:2015hoa}. However, in these models the flavour off-diagonal couplings of the neutral Higgs are proportional to the products of the off-diagonal elements of the mass mixing matrices which, in the quark sector, are suppressed by powers of the Wolfenstein parameter. Hence, the flavour-violating effects can never be as large as those allowed in the model we use in this study, where we have shown that the parameter space of the model is actually significantly constrained by the experimental measurements of $h\to\tau\mu$ and $t\to ch$. 

Flavour violation in the top sector can also be enhanced while suppressing the one in the down sector by assuming the Cheng--Sher ansatz~\cite{Cheng:1987rs}. The Yukawa coupling structure is assumed to have the form
$
Y_{ij} = \lambda_{ij}\frac{\sqrt{\,2m_im_j}}{v},
$
where $i,j$ denote the family indices and $\lambda_{ij}$ can be of $O(1)$. This makes the flavour violating effects proportional to the geometric mean of the masses and hence suppressing the FCNC effects in the first two families while allowing significant effects in the processes involving the third family, especially the top quark. The flavour violating effects can be significant in the top sector, a study of which can be found in Ref.~\cite{Kim:2015zla}. However, since their study focuses on the alignment limit, where the SM Higgs sector is decoupled from the NP Higgs sector, they do not predict any NP effects in channels like $t\to c h$ or $h \to \tau \mu$, in contrast with the model that we study in this work.

\section{Conclusions}
\label{sec:conclusions}

The discovery of the Higgs boson came with pride and prejudice.
Pride that a new dynamical principle, namely spontaneous symmetry breaking, envisioned by theorists about 50 years ago, found its incarnation in Nature.
Prejudice that the Higgs boson is not the last particle of the SM but rather the first particle en route towards physics beyond the Standard Model.
The Higgs boson remains so far the only particle that exists in a single species.
Of course, this could well be the result of large quantum corrections that push to the mass of elementary scalars to high value.
If they are present around the weak scale, extra Higgs bosons are very much expected to be accompanied with a plethora of new particles, be they supersymmetric or composite, that would balance the large corrections to the Higgs masses.   These new particles can affect the production and the decay of these extra Higgs bosons in a significant way and it is therefore difficult to put robust and model-independent bound on the heavy Higgs boson masses. Therefore the stringent bounds obtained in standard searches in minimal Type I and Type II 2HDMs should be taken with care.

The purpose of this paper has been to show that simple departures from minimal set-ups fool the standard direct and indirect constraints from Higgs coupling measurements and flavour data. Rich opportunities for discovery open up in unexplored channels. Not only neutral and charged Higgs bosons as light as 200\,GeV can exist, but they can be copiously and predominantly produced and decay in channels that have not be explored yet. For instance, we showed that a neutral heavy Higgs boson can be produced in association with a top and charm quarks and later decay into another top-charm pair, leading to a final state with same charge dilepton and bottom and charm jets that can be easily emerged from a not so dominant background. A heavy charged Higgs boson can be produced in association with a bottom and charm quarks and can decay into another bottom-charm pair with a total rate at or above 100\,pb. It will be very interesting to extend the present LHC program for searches of new Higgs bosons, to include the plethora of new signatures predicted by our model (see Table~\ref{tab:signatures}). For sure rather spectacular signatures are  expected and wait for the interest  of the experimental community to reveal the first direct evidence of new physics and to unravel the origin of flavour, which remains one of the deepest questions of high-energy physics.

\begin{table}[t!]
\begin{center}
\begin{tabular}{|c|c|c|c|}
\hline
$pp \to H \to tc$	 &  $pp \to tcH(\to tc)$ & $pp \to bc H^{\pm}(\to bc)$ & $pp \to bc H^{\pm}(\to Wh)$\\
\hline
1 charged lepton & 2 same-sign leptons & dijet resonance & $Wh$ resonance \\
$E_{\rm T}^{\rm miss}$ & 2 $b$-jets & $\geq$1 $b$-jet & $\geq$1 $b$/$c$-jet\\
1 $b$-jet  & $\geq$1 $c$-jet & $\geq$1 $c$-jet  & \\
1 $c$-jet &&&\\
\hline
\end{tabular}
\caption{Summary of the most promising signatures associated to the dominant (flavour-violating) production and decays of heavy neutral and charged Higgs bosons.}
\end{center}
\label{tab:signatures}
\end{table}%

\acknowledgments
A.P. would like to thank Laura Reina, Andreas Crivellin, Otto Eberherdt and Debtosh Chowdhury for useful discussions. S.G. acknowledges support from
the University of Cincinnati. S.G. is supported by the NSF CAREER grant PHY-1654502.
C.G. is supported by the European Commission through the Marie Curie Career Integration Grant 631962,  by the
Helmholtz Association through the recruitment initiative, and by the Collaborative Research Center SFB676 of the
Deutsche Forschungsgemeinschaft (DFG), Particles, Strings and the Early Universe.
A.J. is supported in part by the Spanish Ministerio de Econom\'ia y Competitividad  under projects
FPA2015-69260-C3-1-R and Centro de Excelencia Severo Ochoa SEV-2012-0234.
A.P. would like to acknowledge support from the ERC Ideas Starting Grant n.~279972 ``NPFlavour''.  S.G. is grateful to the hospitality of the Kavli Institute for Theoretical Physics in Santa Barbara, CA, supported in part
by the National Science Foundation under Grant No. NSF PHY11-25915, the
Aspen Center for Physics, supported by the National
Science Foundation Grant No. PHY-1066293, and the
Mainz Institute for Theoretical Physics (MITP) where some of the
research reported in this work was carried out.
Finally, the authors would like to thank the Galileo Galilei Institute in Florence, the Centro de Ciencias Pedro Pascual in Benasque, and the Mainz Institute for Theoretical Physics, where part of this project has been conducted during the workshops ``Gearing up for LHC13", ``Higgs Tasting" and ``The TeV scale: a threshold to new physics?", respectively.

\appendix

\section*{Appendix}
\section{Higgs couplings to fermions}
\label{app:Hferm}
 
The physical Higgs-quark and Higgs-lepton couplings ($H_k^0 = H,h,A$)~\cite{Crivellin:2013wna} are given by:\footnote{Due to the hermiticity of the Lagrangian the relation $\Gamma_{q^f_R q^i_L }^{H}\, =\, \Gamma_{q^i_L q^f_R }^{H\,\star}$ holds for Higgs boson couplings to the quarks and leptons.}
{\allowdisplaybreaks
\begin{eqnarray}
{\Gamma_{u^f_L u^i_R }^{H_k^0} } &=& x_u^k\left( \frac{m_{u_i }}{v_u}
\delta_{fi} - \epsilon_{fi}^{u}\cot\beta \right) + x_d^{k\star}
\epsilon_{fi}^{u}\,, \\[0.1cm]
{\Gamma_{d^f_L d^i_R }^{H_k^0 } } &=& x_d^k \left( \frac{m_{d_i
}}{v_d} \delta_{fi} - \epsilon_{fi}^{d}\tan\beta \right) +
x_u^{k\star}\epsilon_{fi}^{ d} \,, \\[0.1cm]
{\Gamma_{u^f_L d^i_R }^{H^\pm } } &=& \sin\beta\,\sum\limits_{j = 1}^3
{ V_{fj} \left( \frac{m_{d_i }}{v_d} \delta_{ji}-
  \epsilon^{d}_{ji}\left(\tan\beta + \cot\beta \right)\right)\,,}\label{eq:HpmuLdR}
\\[0.1cm]
{\Gamma_{d^f_L u^i_R }^{H^ \pm } } &=&\cos\beta\, \sum\limits_{j = 1}^3
{ V_{jf}^{\star} \left( \frac{m_{u_i }}{v_u} \delta_{ji}-
  \epsilon^{u}_{ji}\left(\tan\beta + \cot\beta \right) \right)\, }\label{eq:HpmuRdL}\,,\\[0.1cm]
 \label{Higgs-vertices-decoupling}
{\Gamma_{\ell^f_L \ell^i_R }^{H_k^0} } &=& x_d^k\left( \dfrac{m_{\ell_i }}{v_d}
\delta_{fi} - \epsilon_{fi}^{\ell}\tan\beta \right) + x_u^{k\star} \epsilon_{fi}^{\ell}\label{eq:Hll}\,,\\[0.1cm]
{\Gamma_{\nu_L \ell^i_R }^{H^\pm } } &=& \sin\beta\,\sum\limits_{j = 1}^3
{  \left( \dfrac{m_{\ell_i }}{v_d} \delta_{ji}-
  \epsilon^{\ell}_{ji}\left(\tan\beta + \cot\beta \right) \right)\,.}
      \label{Higgs-leptons-vertices-decoupling}
\end{eqnarray}
}
Here we have defined $v_d\equiv\langle\Phi_2\rangle=v\cos\beta$ and $v_u\equiv\langle\Phi_1\rangle=v\sin\beta$ with $v=174$\,GeV, and $V$ is the Cabibbo--Kobayashi--Maskawa (CKM) matrix. For the coupling of the charged Higgs boson with leptons, we sum over the neutrino flavour ($\nu_L\equiv\sum_f\nu_L^f$).
We have defined the coefficients $x_q^{k}$ as
\begin{equation}
\renewcommand{\arraystretch}{1.4}
\begin{array}{l}
x_u^k \, = \, \left(-\dfrac{1}{\sqrt{2}}\sin\alpha,\,-\dfrac{1}{\sqrt{2}}\cos\alpha,
\,\dfrac{i}{\sqrt{2}}\cos\beta\right) \,,\\[0.3cm]
x_d^k \, = \,\left(-\dfrac{1}{\sqrt{2}}\cos\alpha,\,\dfrac{1}{\sqrt{2}}\sin\alpha,
\,\dfrac{i}{\sqrt{2}}\sin\beta\right) \,.
 \end{array}
\end{equation}
Using the expressions in (\ref{eq:epsilons}) for $\epsilon_{ij}^u$, we find that the 125 GeV Higgs couplings to up quarks are given by 
{\allowdisplaybreaks
\begin{eqnarray}
c_f^h
=\frac{m_f}{\sqrt{2}v}\begin{cases}
\,\sin(\beta - \alpha) + \left(\cot\beta - \frac{1-\cos\rho_u}{2}(\tan\beta + \cot\beta)\right)\cos(\beta-\alpha) & (\mbox{for } f=t)
\,, \\
\,\sin(\beta-\alpha) - \left(\tan\beta - \frac{1-\cos\rho_u}{2}(\tan\beta + \cot\beta)\right)\cos(\beta-\alpha) &
(\mbox{for } f=c)
\,,\\
\,-\sin(\alpha) /\cos(\beta)& (\mbox{for $f=u$})~,
\end{cases}
\label{eq:kappa2h}
\end{eqnarray}
For completeness, we also report here the corresponding couplings of the heavy Higgs bosons, $H$ and $A$:

\begin{eqnarray}
c_f^H=\frac{m_f}{\sqrt{2} v}\begin{cases}
\,\cos(\beta - \alpha) - \left(\cot\beta - \frac{1-\cos\rho_u}{2}(\tan\beta + \cot\beta)\right)\sin(\beta -\alpha) & (\mbox{for } f=t)
\,, \\
\,\cos(\beta-\alpha) + \left(\tan\beta - \frac{1-\cos\rho_u}{2}(\tan\beta + \cot\beta)\right)\sin(\beta-\alpha) &
(\mbox{for } f=c)
\,,\\
\,\cos(\alpha) /\cos(\beta)& (\mbox{for $f=u$})~,
\end{cases}
\label{eq:kappa2H}
\end{eqnarray}
}
\begin{eqnarray}
c_f^A=\frac{m_f}{\sqrt{2} v}\begin{cases}
\, - \cot\beta + \frac{1-\cos\rho_u}{2}(\tan\beta + \cot\beta)& (\mbox{for } f=t)\,, \\
\, \tan\beta - \frac{1-\cos\rho_u}{2}(\tan\beta + \cot\beta) &
(\mbox{for } f=c)
\,,\\
\, \tan\beta & (\mbox{for $f=u$}) \,.
\end{cases}
\label{eq:kappa2A}
\end{eqnarray}
The flavour-violating couplings of the several Higgs bosons with charm and top quarks are given by
{\allowdisplaybreaks
\begin{eqnarray}\label{Eq:A11}
c_{23}^h&=&\frac{m_t}{2\sqrt{2}v}(\cot\beta+\tan\beta)\cos(\beta - \alpha)\sin\rho_u\,,\nonumber\\
c_{32}^h&=&\frac{m_c}{2\sqrt{2}v}(\cot\beta+\tan\beta)\cos(\beta - \alpha)\sin\rho_u\,,\nonumber\\\label{eq:Htc}
c_{23}^H&=&-\frac{m_t}{2\sqrt{2}v}(\cot\beta+\tan\beta)\sin(\beta - \alpha)\sin\rho_u\,,\nonumber\\
c_{32}^H&=&-\frac{m_c}{2\sqrt{2}v}(\cot\beta+\tan\beta)\sin(\beta - \alpha)\sin\rho_u\,,\nonumber\\
c_{23}^A&=&\frac{m_t}{2\sqrt{2}v}(\cot\beta+\tan\beta)\sin\rho_u\,,\nonumber\\
c_{32}^A&=&\frac{m_c}{2\sqrt{2}v}(\cot\beta+\tan\beta)\sin\rho_u\,.
\end{eqnarray}
}

The couplings of the Higgs bosons to leptons have a similar form, requiring an exchange of $\rho_u\rightarrow\rho_\ell$ and the $\tau$($\mu$) coupling being similar to the top(charm) coupling.
Finally, all couplings of the Higgs bosons to down-type quarks are as in a Type II 2HDM.

\section{Constraints from Higgs effective couplings}
\label{app:HEC}

 The Higgs effective couplings that parametrize the fit to the measurements of Higgs boson productions and decays can be written in terms of the reduced couplings, $\kappa$, as
    \begin{equation}
\kappa_{gZ} = \frac{\kappa_g\kappa_Z}{\kappa_h}\;\; \textrm{and} \;\;
\lambda_{ij}= \frac{\kappa_i}{\kappa_j},\;\; (i,j)=(Z,g),(t,g),(W,Z),(\gamma,Z),(\tau,Z),(b,Z).
\label{eq:SevenParFit}
\end{equation}
In all generality, in the absence of exotic Higgs decays, the scaling of the light neutral Higgs boson width is given by~\cite{Khachatryan:2016vau}:
\begin{eqnarray}
\kappa_h^2 &\simeq& 0.57 \kappa_b^2 + 0.22\kappa_W^2 + 0.09\kappa_g^2 + 0.06\kappa_t^2
            + 0.03\kappa_Z^2 + 0.03\kappa_c^2  \nonumber\\
            &+& 2.3\times 10^{-3}\kappa_\gamma^2
            + 1.6\times 10^{-3}\kappa_{Z\gamma}^2
            + 10^{-4}\kappa_s^2 + 2.2\times 10^{-4}\kappa_\mu^2,
\end{eqnarray}
where we are neglecting the decays to the light generations, $s,u,d,e$. In the case of no beyond the Standard model particles running in the loops for the Higgs to gluon, photon and $Z\gamma$ effective couplings, the scalings of the couplings of the Higgs to the gauge bosons are given by:
\begin{eqnarray}
\kappa_W &=& \kappa_Z =  \sin(\beta-\alpha),\nonumber\\
\kappa_{Z\gamma}^2 &=& 0.00348\kappa_t^2 + 1.121\kappa_W^2 - 0.1249\,\kappa_t\kappa_W,\nonumber\\
\kappa_g^2 &=& 1.06\kappa_t^2 + 0.01\kappa_b^2 - 0.07\,\kappa_b\kappa_t,\nonumber\\
\kappa_\gamma^2 &=& 1.59\kappa_W^2 + 0.07\kappa_t^2 - 0.66\;\kappa_W\kappa_t,
\end{eqnarray}
and the scalings of the fermionic couplings, $\kappa_f$, can be derived using
\begin{equation}
\kappa_f = \frac{\sqrt{2}v}{m_f}c_f^h
\end{equation}
for $f=t,b,\tau$ and $c_f^h$ couplings given in Appendix \ref{app:Hferm}. It should be noted that, in our work, we assume the same flavour structure in the up quark sector and in the lepton sector and, therefore, $\kappa_\tau = \kappa_t$. 

\begin{figure}[t!]
\begin{center}
\subfigure{\includegraphics[trim = 10mm 0mm 10mm 5mm, clip, width=.327\linewidth]{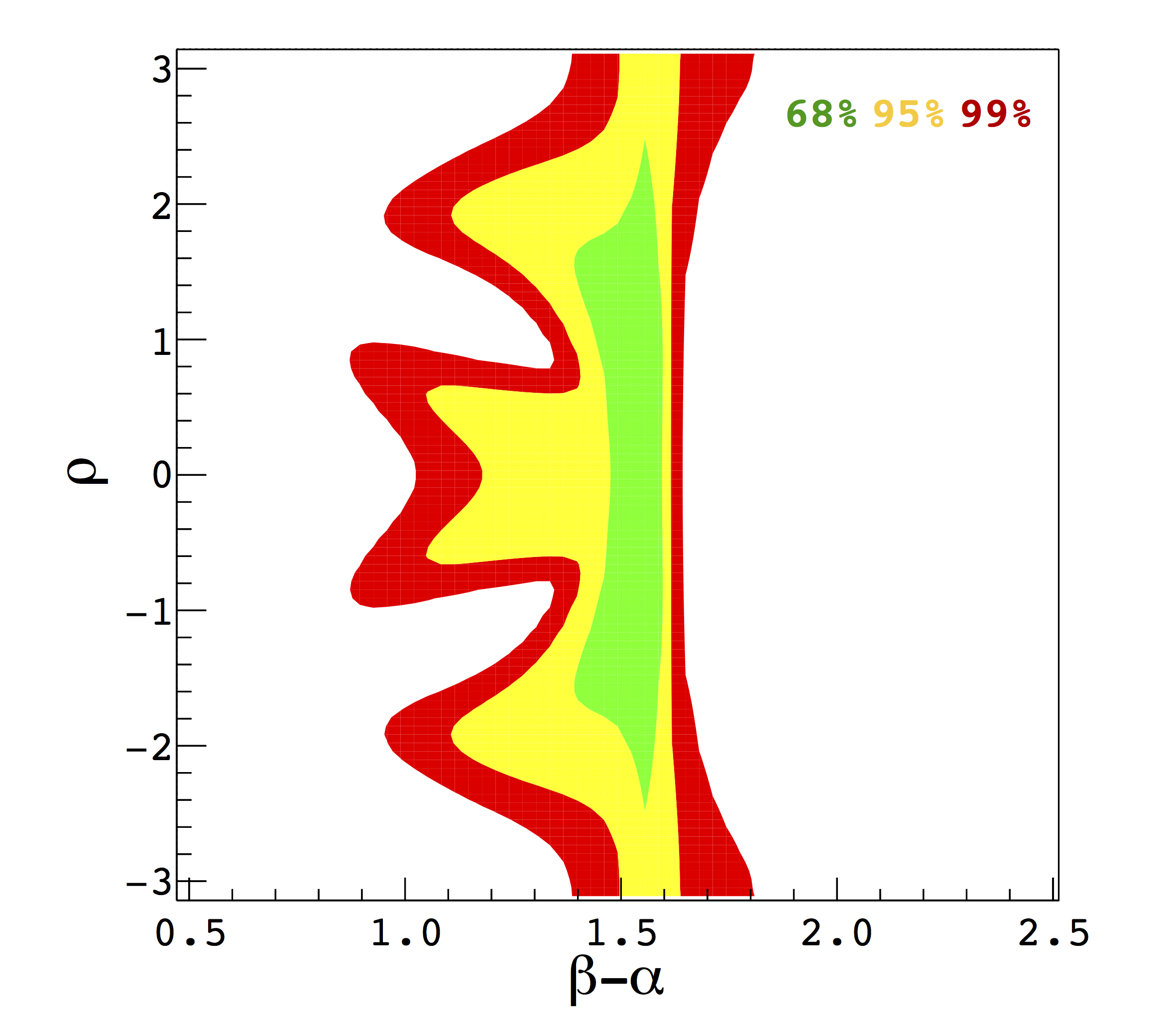}}
\subfigure{\includegraphics[trim = 10mm 0mm 10mm 5mm, clip, width=.327\textwidth]{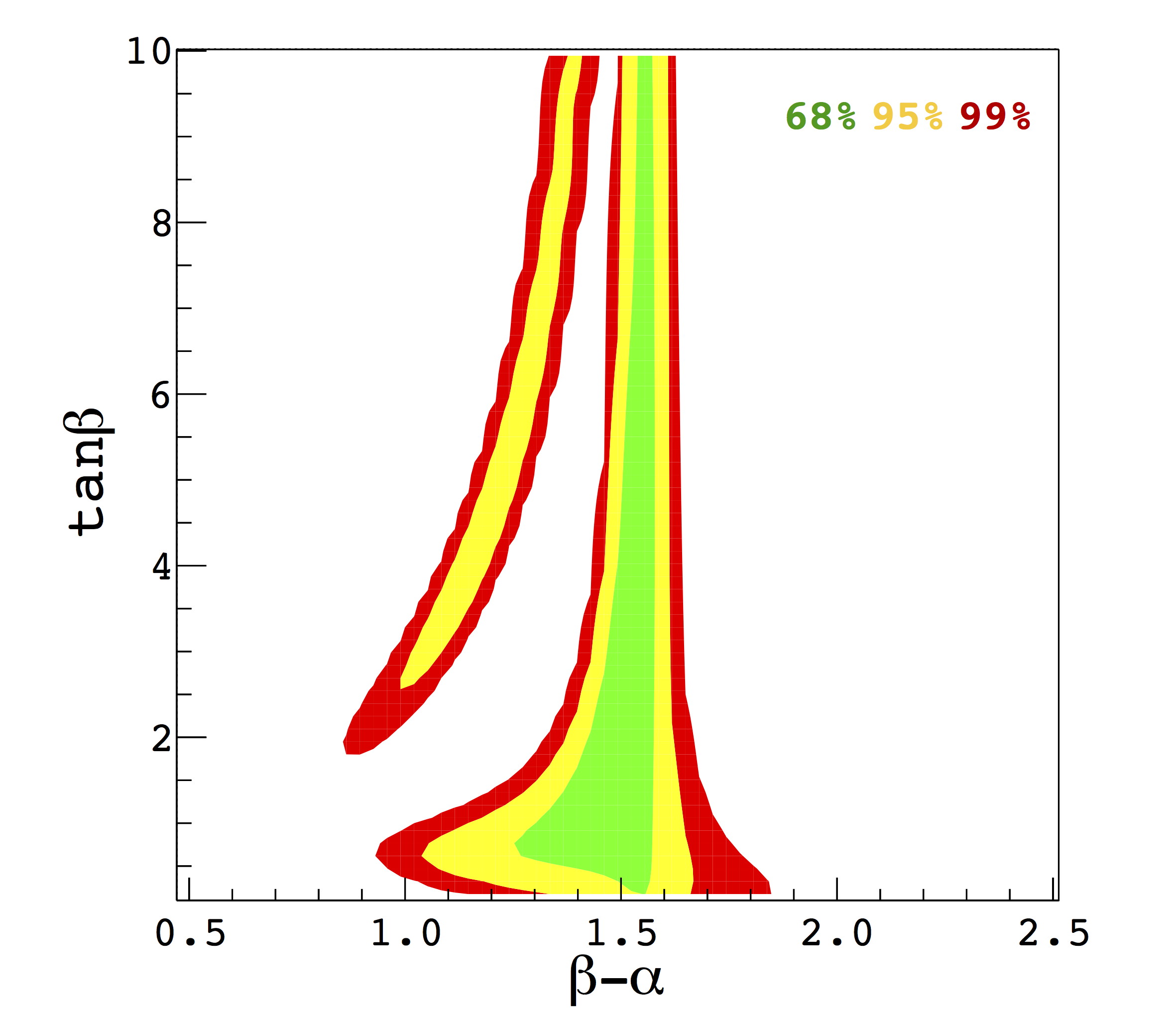}}
\subfigure{\includegraphics[trim = 10mm 0mm 10mm 5mm, clip, width=.327\textwidth]{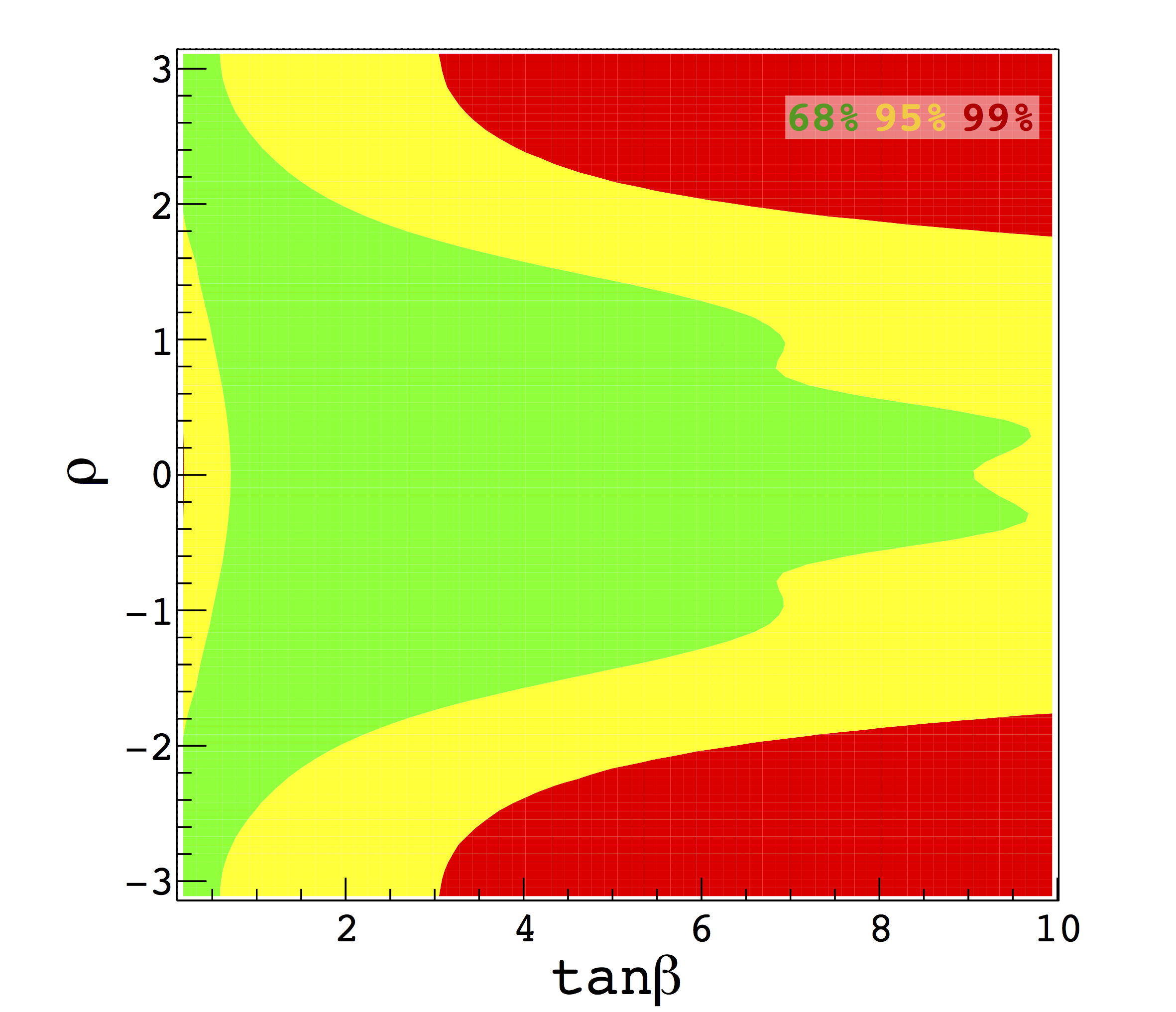}}
\caption{2D marginalized posterior distributions of the relevant model parameters using the Higgs coupling measurements listed in Table~\ref{tab:HC}. We vary the parameters in the range: $0.1\le\tan\beta\le15$, $-\pi\le\rho\le\pi$, and $0\lesssim (\beta - \alpha)\lesssim \pi$, with flat priors. The green, yellow, and red regions are the 68\%, 95\%, and  99\% favored regions, respectively.}
\label{fig:HC_C}
\end{center}
\end{figure}

In Fig.~\ref{fig:HC_C}  we present the 2D marginalized posterior distributions for the parameters $\rho$, $\tan\beta$, and $\beta-\alpha$, as constrained by the Higgs effective couplings data listed in Table~\ref{tab:HC}. As expected, the value of $|\beta-\alpha|$ is constrained to be not too large, since the 125\,GeV Higgs boson has SM-like properties. The values for $\rho$ and $\tan\beta$ are, instead, relatively unconstrained by the fit. Our analysis leads to results that are qualitatively similar to those of Ref.~\cite{Chiang:2015cba}, where the same model was used for the analysis. However, it should be noted that, in contrast to Ref.~\cite{Chiang:2015cba}, we use the final results from the Run I Higgs coupling combination, taking into account correlations. Furthermore, our statistical approach is different, as we obtain the 2D plots by marginalizing over the third parameter rather than assuming a fixed value for it.\footnote{The results in Ref.~\cite{Chiang:2015cba} are presented in terms of $\tan\beta$, $\rho$ and $a$ (defined in Eq.~(\ref{eq:def_a})).}

\section{Constraints from flavour-violating decays involving the Higgs boson}
\label{app:FVD}

The off-diagonal fermion couplings of the light neutral Higgs boson are, in general, non-zero, and can lead to a measurable value for branching ratios of $h\to\tau\mu$ and $t\to c h$.\footnote{A richer flavour structure including mixing of the third generations, not only with the second one, but also with the first generation would lead to additional interesting decay modes such as $h\to \tau e$.}
Allowing for deviations in the production cross section of the light Higgs boson, the experimental measurement of the $h\to\tau\mu$ branching ratio should be compared to~\cite{Chiang:2015cba}
\begin{eqnarray}
\textrm{BR}_{\rm exp}(h\to\tau\mu)&=&\frac{\sigma^{pp\to h}}{\sigma_{\rm SM}^{pp\to h}}\textrm{BR}_{\rm th}(h\to\tau\mu)\nonumber\\
&\simeq&\frac{\kappa_g^2 a^2\sin^2\rho}{36.5 \kappa_b^2 + 14.64\sin^2(\beta - \alpha) + 5.44\kappa_g^2 + 4\kappa_\tau^2},
\end{eqnarray}
where we denote with $\sigma^{pp\to h}$ and $\textrm{BR}_{\rm th}(h\to\tau\mu)$ the production cross section of the 125 GeV Higgs and the branching ratio of its decay to $\tau\mu$, as predicted in the scenario we study. For simplicity, we have only considered the gluon fusion production mechanism and the decay into bottom quarks, $WW$, $ZZ$, gluons, and tau leptons, leading to the terms in the denominator. The other decay modes will only bring small corrections that can be safely neglected. In the rest of the text we will refer to $\textrm{BR}_{\rm exp}(h\to\tau\mu)$ as $\textrm{BR}(h\to\tau\mu)$. The quantity $a$ is defined in Eq.~(\ref{eq:def_a}). 

In the quark sector, the branching ratio for the decay $t\to ch$ can also be sizable, and can be written as~\cite{Chiang:2015cba}\footnote{As shown by the structure of the Higgs couplings presented in Section~\ref{Sec:model}, the decay $t\to uh$ does not occur due to the absence of the corresponding flavour-violating couplings.}
\begin{equation}
\textrm{BR}(t\to ch)\simeq3.24\times 10^{-2}a^2\sin^2\rho.
\end{equation}
Other recent studies of the decay $t\to c h$ in a general 2HDM framework can be found in Refs.~\cite{Botella:2015hoa,Arhrib:2015maa,Fuyuto:2017ewj,Crivellin:2015hha} and those for $h\to\tau \mu$ can be found in Refs.~\cite{Sierra:2014nqa,Liu:2015oaa,Crivellin:2015hha,Benbrik:2015evd,Omura:2015xcg,Tobe:2016qhz,Chiang:2016vgf,Lee:2016dcb,Iguro:2017ysu}.

\begin{figure}[t!]
\begin{center}
\subfigure{\includegraphics[trim = 10mm 0mm 10mm 0mm, clip, width=.327\textwidth]{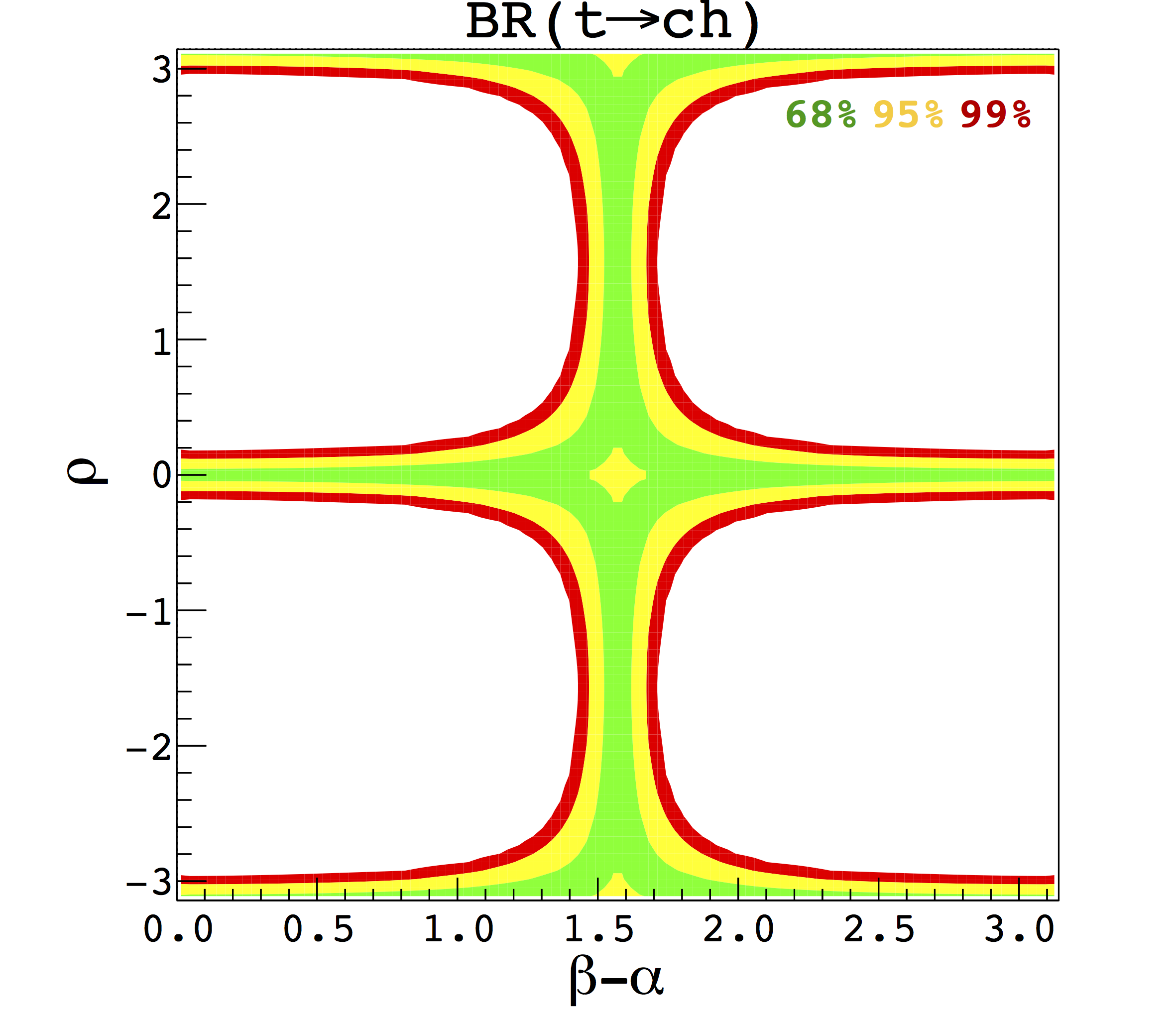}}
\subfigure{\includegraphics[trim = 10mm 0mm 10mm 0mm, clip, width=.327\textwidth]{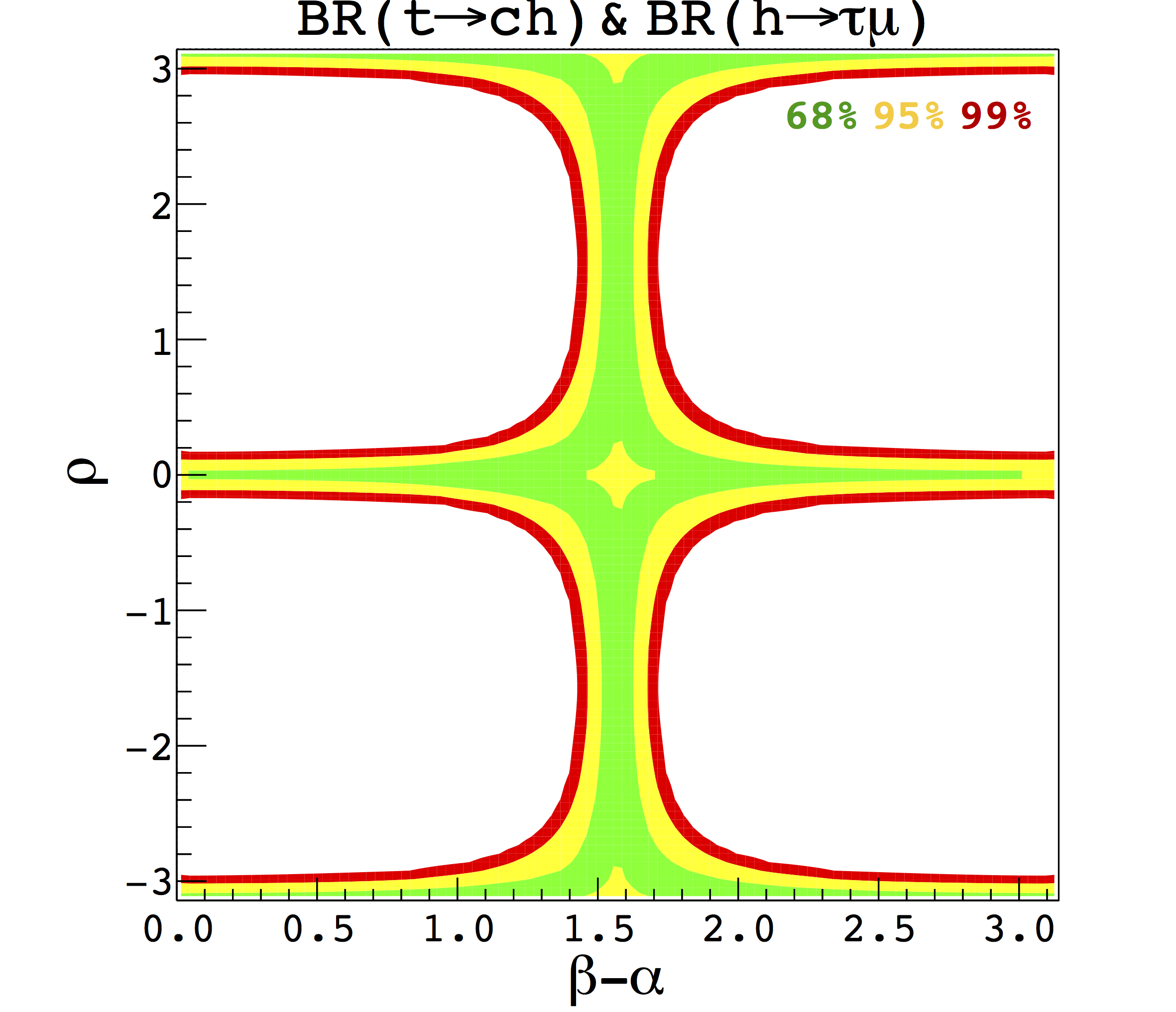}}
\subfigure{\includegraphics[trim = 10mm 0mm 10mm 0mm, clip, width=.327\textwidth]{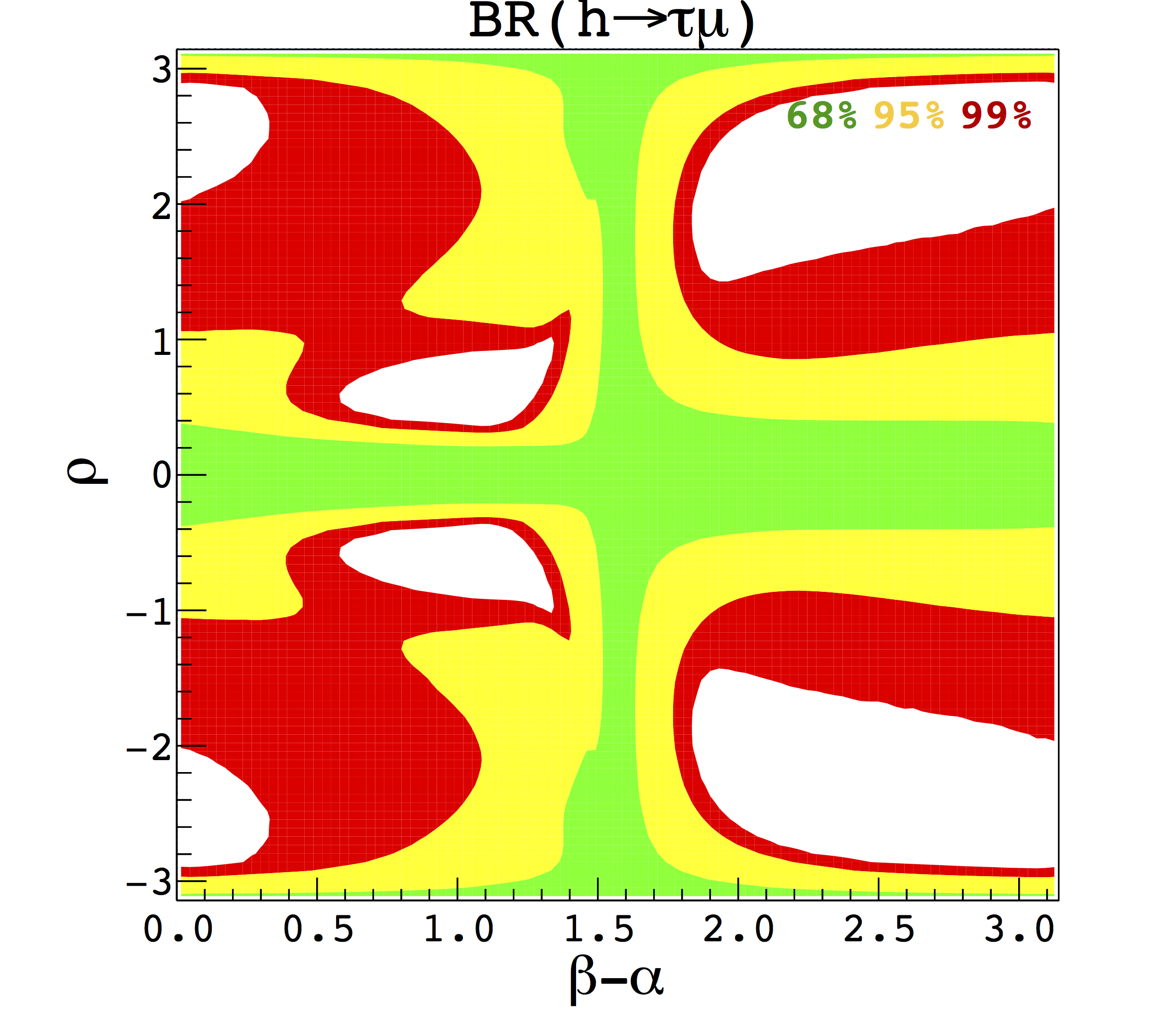}}
\caption{The constraints from $\textrm{BR}(t\to ch)$ in the left panel and from $\textrm{BR}(h \to\tau\mu)$ in the right panel. The middle panel shows the combination of the two constraints. The plots are produced by marginalizing over $\tan\beta$. The green, yellow, and red regions are the 68\%, 95\%, and  99\% favored regions, respectively.}
\label{fig:HiggsFlavDecays}
\end{center}
\end{figure}
\begin{figure}[t!]
\begin{center}
\subfigure{\includegraphics[trim = 10mm 0mm 10mm 5mm, clip, width=.327\textwidth]{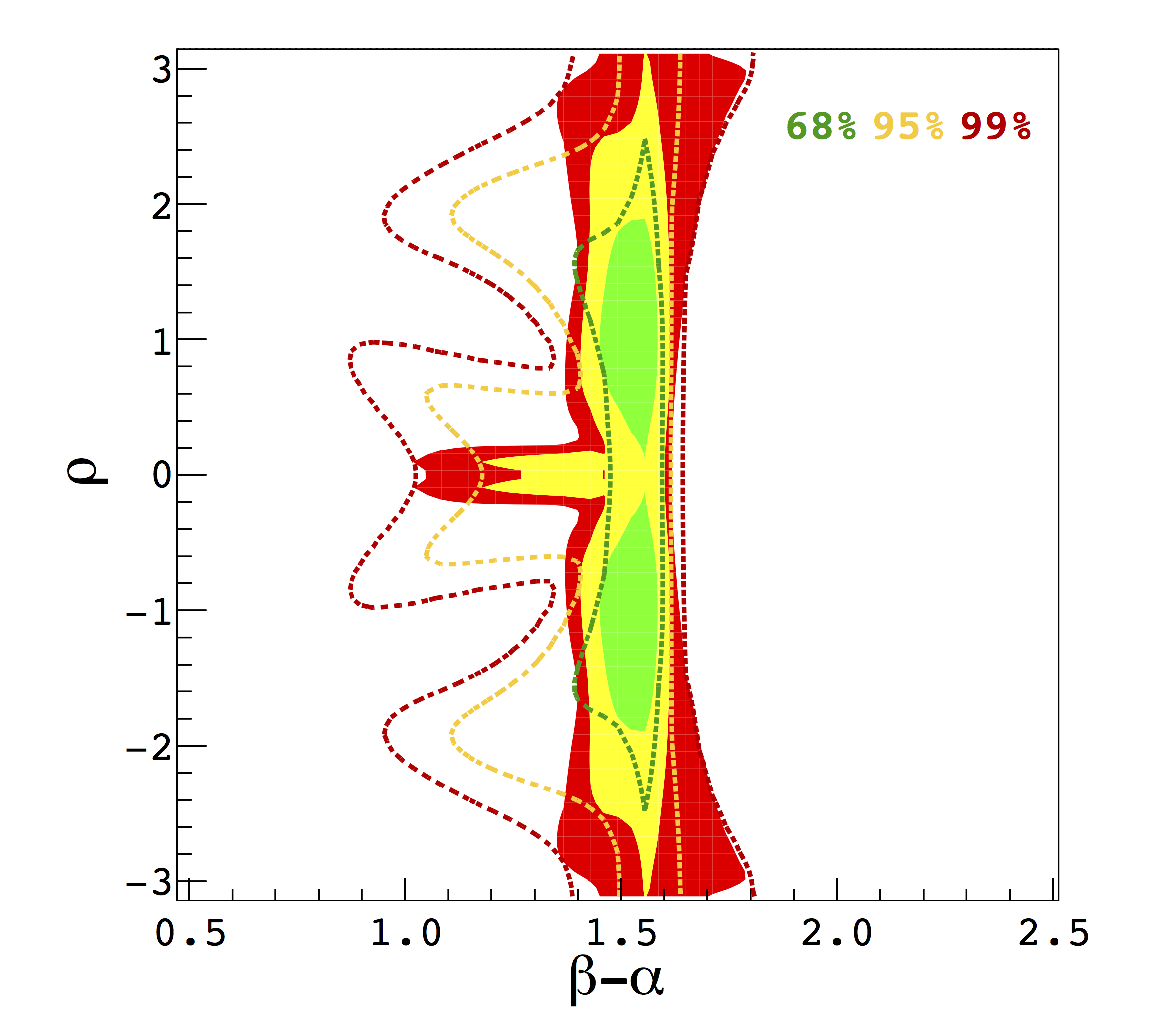}}
\subfigure{\includegraphics[trim = 10mm 0mm 10mm 5mm, clip, width=.327\textwidth]{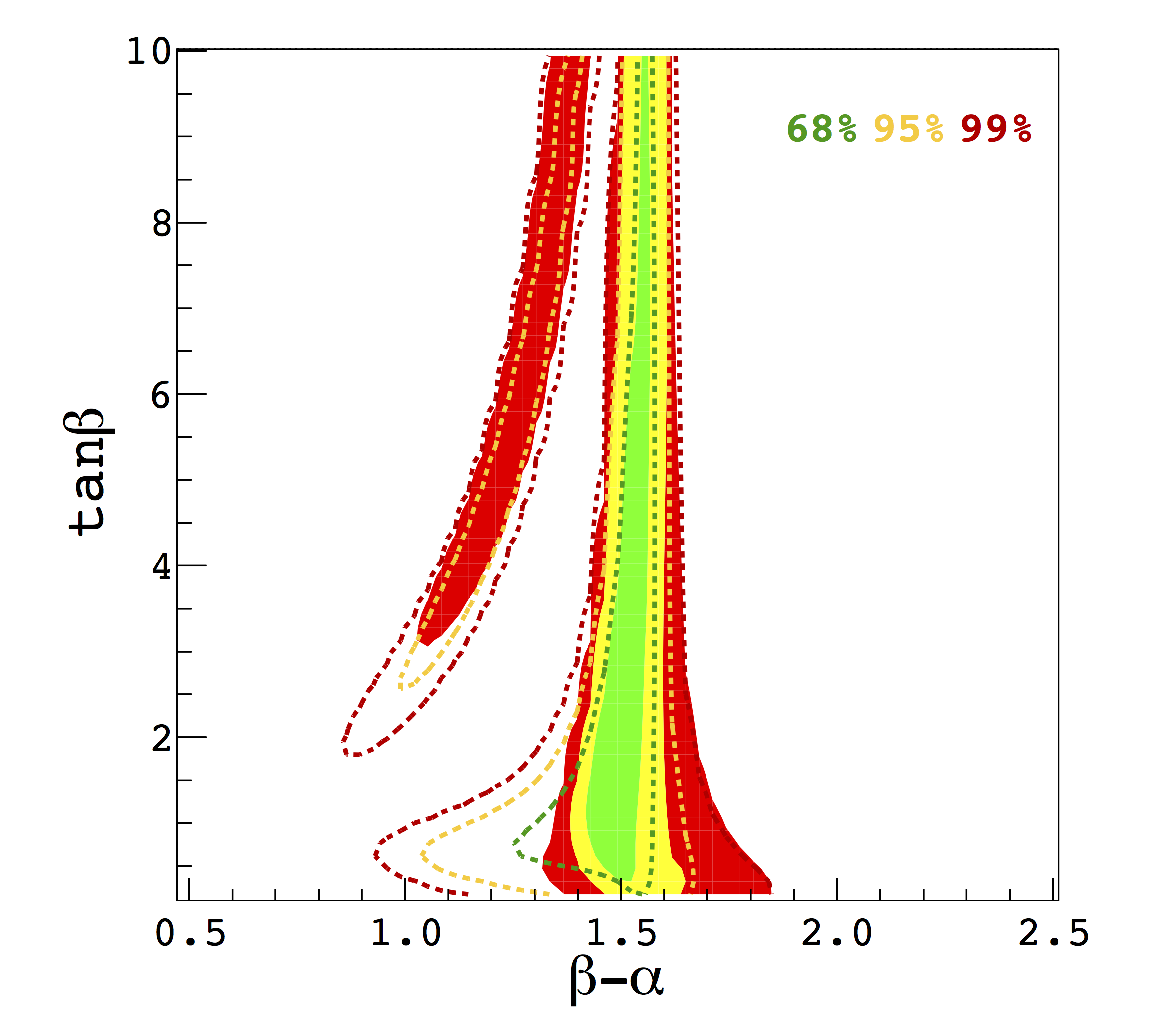}}
\subfigure{\includegraphics[trim = 10mm 0mm 10mm 5mm, clip, width=.327\textwidth]{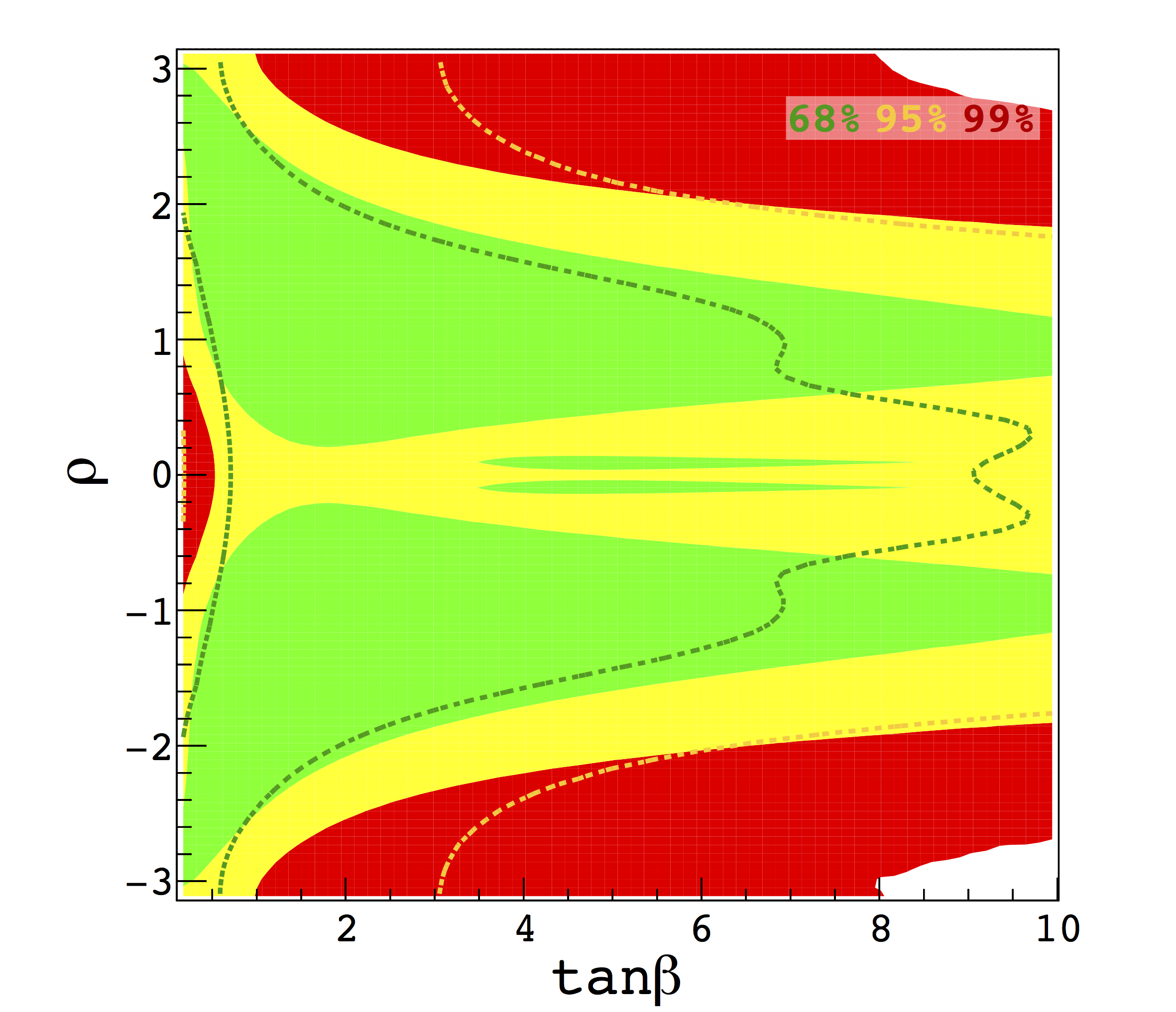}}
\caption{2D marginalized posterior distributions of the relevant model parameters for the combination of the Higgs couplings data (see Table~\ref{tab:HC}) and the flavour-violating $h\to\tau\mu$ and $t\to ch$ decays (see Table~\ref{tab:HFNC}). The dashed lines show the constraints from only the Higgs couplings data as shown in Fig.~\ref{fig:HC_C}. The green, yellow, and red regions (lines) are the 68\%, 95\%, and  99\% regions (contours), respectively.}
\label{fig:HiggsFlav}
\end{center}
\end{figure}

In Fig.~\ref{fig:HiggsFlavDecays}, we show the individual constraints from $\textrm{BR}(h\to\tau\mu)$ and $\textrm{BR}(t\to ch)$ in the $\rho$ vs. $\beta-\alpha$ plane, having marginalized over the other free parameter, $\tan\beta$. The green, yellow, and red regions are the 68\%, 95\%, and 99\% favored regions, respectively. The left panel corresponds to the bounds on the parameter space coming from $t\to ch$ data. The right panel of Fig.~\ref{fig:HiggsFlavDecays} shows the constraints coming from the $h\to\tau\mu$ decay. The combination of the two constraints is shown in the middle panel. As can be gauged from the individual panels, the constraints on the parameter space from $h \to \tau \mu$ and $t\to ch$ have some overlap at values of $\rho\ne 0$  and $\beta - \alpha\ne \pi/2$. The shape of the posterior distribution in the middle panel where the constraints from the two decays are combined is determined by the $\textrm{BR}(t\to ch)$ measurements, since they are more accurate than the corresponding measurements of $\textrm{BR}(h\to\tau\mu)$ as is evident from Table~\ref{tab:HFNC}.

We now proceed with the combination of these flavour-violating constraints with those from the flavour conserving Higgs couplings data, discussed in Appendix~\ref{app:HEC}. The 2D marginalized posterior distributions are presented in Fig.~\ref{fig:HiggsFlav}. The 68\%, 95\%, and  99\% contours from Fig.~\ref{fig:HC_C} are superimposed with dashed lines. Comparing the solid regions with the dashed lines, it is clear that the measurements of the $t \to c h$ and $h\to\tau\mu$ decays modify the constraints coming from the Higgs effective couplings significantly, reducing greatly the allowed degree of flavour violation encoded by the parameter $\rho$ away from the alignment limit $\beta-\alpha=\pi/2$. 

The combination of  constraints from Higgs data with the constraints from low-energy flavour observables that will be discussed in Appendix~\ref{app:FL} is shown in Fig.~\ref{fig:comb}.

\section{Constraints from low-energy flavour observables}
\label{app:FL}
\subsection{$b\to s \gamma$}\label{Sec:bsgamma}
The inclusive decay $b \to s \gamma$ is a very well measured flavour-violating process that is known to significantly constrain the parameter space of models that allow for sizable flavour-violating couplings of the Higgs boson(s) to the second and third generation quarks. Hence, it is important to understand how the parameter space of the model under consideration is affected by the measurement of this inclusive decay. The HFAG~\cite{Amhis:2016xyh} average for the several measurements of the branching ratio of $b\to s \gamma$ is:
\begin{equation}
\textrm{BR}(b \to s \gamma)_{\rm{exp}} = (3.32 \pm 0.15)\times 10^{-4},
\end{equation}
while the SM prediction at next-to-next-to-leading-order (NNLO) stands at~\cite{Misiak:2015xwa,Czakon:2015exa}:
\begin{equation}
\textrm{BR}(b \to s \gamma)_{\rm{SM}} = (3.36 \pm 0.23)\times 10^{-4},
\end{equation}
having a cutoff on the photon energy at 1.6\,GeV for both the average of the measurements and the theoretical prediction.

2HDMs generically predict a contribution to $b\to s \gamma$ from one loop charged and neutral Higgs boson exchange. In the framework we use, one loop neutral Higgs diagrams vanish since $\epsilon^d=0$. The charged Higgs loops will bring a new physics effect in the Wilson coefficients of the operators. 
\begin{equation}
O_7=\frac{e}{16\pi^2}m_b(\bar{s}\sigma^{\mu\nu}P_Rb)F_{\mu\nu}, \;{\rm and}\; O_8=\frac{g_s}{16\pi^2}m_b(\bar{s}\sigma^{\mu\nu}T^aP_Rb)G^a_{\mu\nu},
\end{equation}
The leading-order modification to the corresponding Wilson coefficients $C_7$ and $C_8$ are given by
\begin{eqnarray}
\delta C^{{\textrm{\tiny LO}}}_7 &=&\frac{v^2}{\lambda_t}\frac{1}{m_b}\sum^3_{j=1}\Gamma^{H^\pm*}_{u_R^j s_L}\Gamma^{H^\pm}_{u_L^j b_R} \frac{C^{{\textrm{\tiny LO}}}_{7,XY}(y_j)}{m_{u^j}}
+\frac{v^2}{\lambda_t}\sum^3_{j=1}\Gamma^{H^\pm*}_{u_R^j s_L}\Gamma^{H^\pm}_{u_R^j b_L} \frac{C^{{\textrm{\tiny LO}}}_{7,YY}(y_j)}{m^2_{u^j}},\nonumber\\
\delta C^{{\textrm{\tiny LO}}}_8 &=&\frac{v^2}{\lambda_t}\frac{1}{m_b}\sum^3_{j=1}\Gamma^{H^\pm*}_{u_R^j s_L}\Gamma^{H^\pm}_{u_L^j b_R} \frac{C^{{\textrm{\tiny LO}}}_{8,XY}(y_j)}{m_{u^j}}
+\frac{v^2}{\lambda_t}\sum^3_{j=1}\Gamma^{H^\pm*}_{u_R^j s_L}\Gamma^{H^\pm}_{u_R^j b_L} \frac{C^{{\textrm{\tiny LO}}}_{8,YY}(y_j)}{m^2_{u^j}},
\label{eq:dc}
\end{eqnarray}
where we have defined $\lambda_t\equiv V^*_{ts}V_{tb}$, $y_j\equiv m^2_{u^j}/m^2_{H^\pm}$ with $j=1,2,3$ denoting the flavour index, and the several charged Higgs couplings $\Gamma^{H^\pm}$  are given in Eqs.~(\ref{eq:HpmuLdR}) and (\ref{eq:HpmuRdL}). 
The loop functions are defined as
{\allowdisplaybreaks
\begin{eqnarray}
C^{{\textrm{\tiny LO}}}_{7,XY}(y_j)&=&\frac{y_j}{12}\left(\frac{-5y_j^2+8y_j-3+(6y_j-4)\ln y_j}{(y_j-1)^3}\right),\nonumber\\
C^{{\textrm{\tiny LO}}}_{7,YY}(y_j)&=&\frac{y_j}{4}\left(\frac{-y_j^2+4y_j-3-2\ln y_j}{(y_j-1)^3}\right),\nonumber\\
C^{{\textrm{\tiny LO}}}_{8,XY}(y_j)&=&\frac{y_j}{72}\left(\frac{-8y_j^3+3y_j^2+12y_j-7+(18y^2_j-12y_j)\ln y_j}{(y_j-1)^4}\right),\nonumber\\
C^{{\textrm{\tiny LO}}}_{8,YY}(y_j)&=&\frac{y_j}{24}\left(\frac{-y_j^3+6y_j^2-3y_j-2-6y_j\ln y_j}{(y_j-1)^4}\right).
\end{eqnarray}
}

The theoretical computation of the ${\rm BR}(b\to s \gamma)$ is quite involved in the SM. In the literature, there exist extensive studies of this decay within the Type II 2HDM taking into account NNLO corrections~\cite{Misiak:2015xwa}.
However, such a computation is not available for a generic Type III 2HDM. It is possible to write the branching ratio in terms of modifications to the SM Wilson coefficients assuming these NP contributions are real as is the case in this study.
This branching ratio with the SM contributions computed at NNLO along with the modifications to the Wilson coefficients generated by NP, $\delta C_7^{{\textrm{\tiny LO}}}$ and $\delta C_8^{{\textrm{\tiny LO}}}$, calculated at the LO, can be written as:
\begin{eqnarray}
10^4\times{\rm BR}(b \to s \gamma)^{{\textrm{\tiny NP,LO}}} &=& 3.36 - 8.22\; \delta C_7^{{\textrm{\tiny LO}}} + 5.36(\delta C_7^{{\textrm{\tiny LO}}})^2 -1.98\;\delta C_8^{{\textrm{\tiny LO}}}\nonumber\\
&&+2.43\;\delta C_7^{{\textrm{\tiny LO}}}\delta C_8^{{\textrm{\tiny LO}}}+0.431(\delta C_8^{{\textrm{\tiny LO}}})^2.
\label{eq:bsg}
\end{eqnarray}
This prediction is derived at the central values of the SM input parameters and has an uncertainty of $\mathcal O(10\%)$.

To scale the LO contribution from NP to NNLO we extract a k-factor from the branching ratios at different orders in Type II 2HDM. This factor is defined as:
\begin{equation}
k(m_{H^\pm},\tan\beta)=\frac{{\rm BR}(b \to s \gamma)^{{\textrm{\tiny Type II 2HDM,NNLO}}}}{{\rm BR}(b \to s \gamma)^{{\textrm{\tiny Type II 2HDM,LO}}}}
\end{equation}
The k-factor does not have a significant $\tan\beta$ dependence in the region of parameter space we are interested in and has a weak dependence at the level of few $\%$ on the mass of the charged Higgs boson. 
The k-factor is fit to a polynomial function of the charged Higgs boson mass only and we assume that this functional dependence represents the k-factor for the model we consider to a good approximation, since in the limit $\rho\to0$ the flavour off-diagonal couplings vanish rendering the relevant couplings of the Type II 2HDM kind. 
The k-factor, expressed as a fourth order polynomial in $m_{H^\pm}$, is given by
\begin{equation}
k(m_{H^\pm}) = 0.926 + 0.128\, m_{H^\pm} - 0.109\, m_{H^\pm}^2 + 0.0452\, m_{H^\pm}^3 - 0.00733\, m_{H^\pm}^4,
\end{equation}
with $m_{H^\pm}$ expressed in units of TeV. It should be noted here that this k-factor is only valid in the parameter space we consider in this study, specifically on the range of $\tan\beta$ and $m_{H^\pm}$ considered here. We also point out that while ${\rm BR}(b \to s \gamma)$ has a sizable $\tan\beta$ dependence at $\tan\beta<2$ in the Type II THDM, this does not translate into a $\tan\beta$ dependence of the k-factor that we extract.

\begin{figure}[h!]
\begin{center}
\subfigure{\includegraphics[width=.4\textwidth]{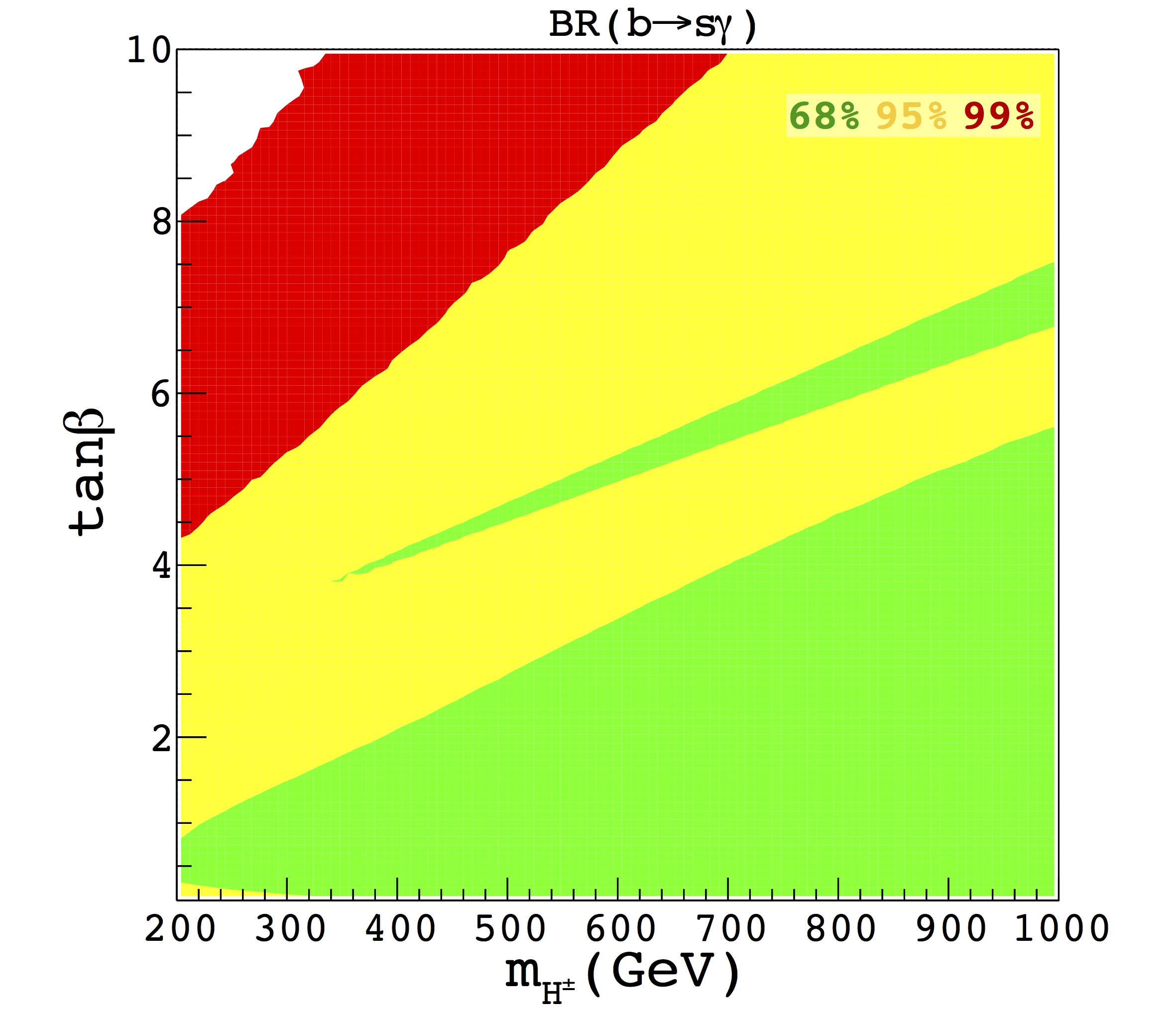}}
\subfigure{\includegraphics[width=.4\textwidth]{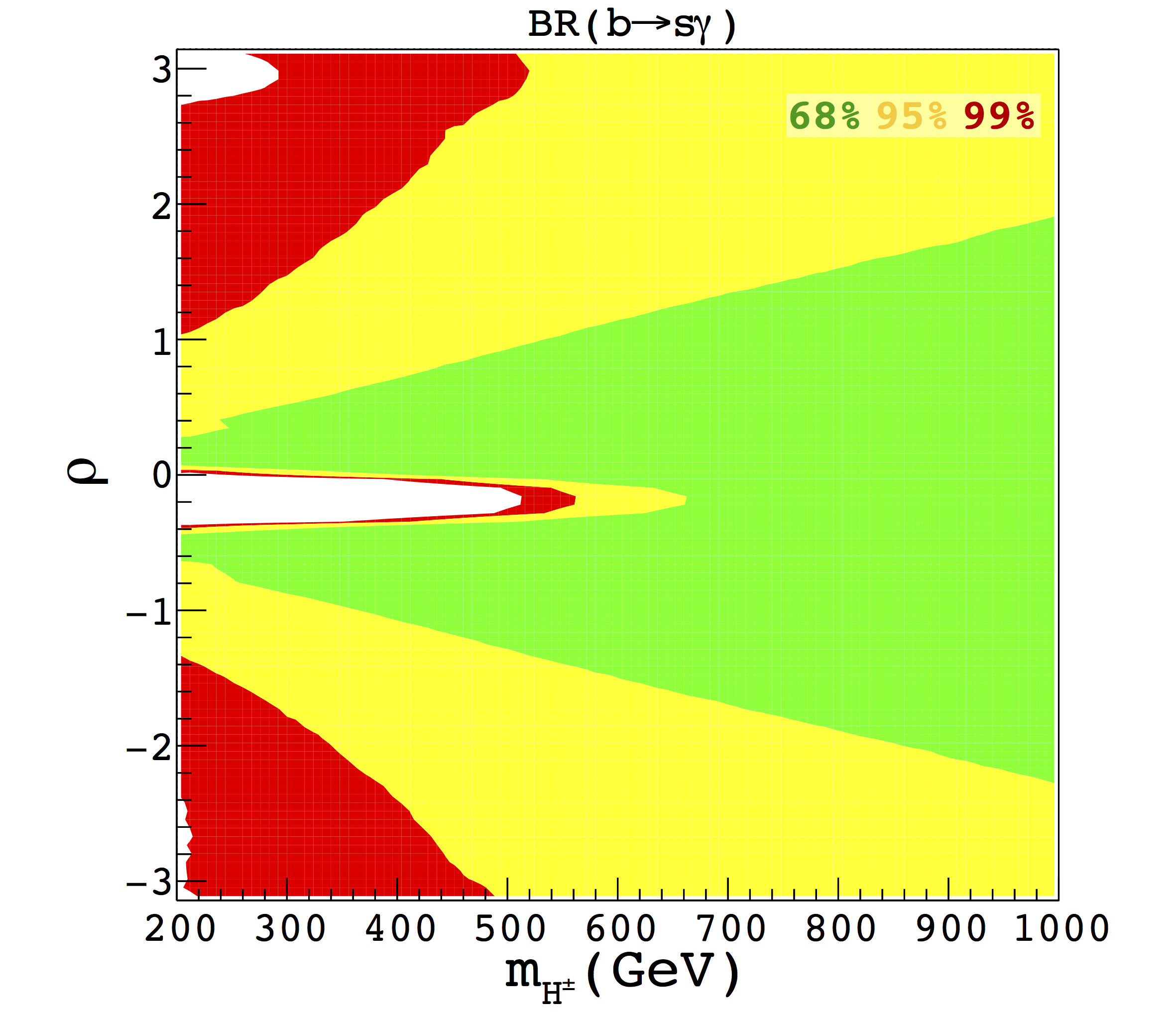}}
\caption{2D marginalized posterior distributions of the relevant model parameters for the constraint from $b \to s \gamma$ data. The green, yellow and red regions are the 68\%, 95\%, and  99\% favored regions, respectively.}%
\label{fig:bsg}
\end{center}
\end{figure}

In Fig.~\ref{fig:bsg} we show the constraints on the parameter space from the branching ratio of $b\to s \gamma$ at the 68\%, 95\%, and  99\% level in green, yellow and red, respectively. The posterior distributions are obtained marginalizing over all other parameters varied to give the 2D posterior distributions of $\tan\beta$ vs. $m_{H^\pm}$ and of $\rho$ vs. $m_{H^\pm}$ in the left and right panel, respectively. As can be appreciated, values of $m_{H^\pm}$ at around 200\,GeV are allowed at low values of $\tan\beta$. It is worthwhile to stress again here that the $\tan\beta$ dependence of the branching ratio of $b\to s \gamma$ in our model is quite different from the dependence in a Type II 2HDM. Generically, the bound in our model has a stronger $\tan\beta$ dependence and lower values of $\tan\beta$ are preferred by the experimental measurement. This can be seen by the $\tan\beta$ enhancement of the $\Gamma_{d^f_L u^i_R }^{H^ \pm }$ couplings in (\ref{eq:HpmuRdL}) ($ \epsilon^{u}_{ji}\left(\tan\beta + \cot\beta \right)$), in addition to the standard $\tan\beta$ enhancement of the $\Gamma_{u^f_L d^i_R }^{H^ \pm }$ couplings in (\ref{eq:HpmuLdR}) ($ \frac{m_{d_i }}{v_d} \delta_{ji}$). Also, as a result of the cancellations with the SM contributions at lower $\tan\beta$ for non-zero values of $\rho$, we get two separated $68\%$ probability regions in the $\tan\beta$ vs. $m_{H^\pm}$ plane as can be seen from the left panel of Fig.~\ref{fig:bsg}.

\subsection{$B \to \tau\nu$}
In the SM, the decay $B \to \tau \nu$ occurs through a tree-level exchange of a $W$ boson. In a 2HDM, this decay can also be mediated at tree-level by a charged Higgs boson, potentially giving rise to strong constraints on the parameter space.
The present HFAG~\cite{Amhis:2016xyh} average for the several measurements of the branching ratio is
\begin{equation}
\textrm{BR}(B \to \tau \nu)_{\rm{exp}} = (1.06 \pm 0.19)\times 10^{-4}.
\end{equation}
The SM prediction for the branching ratio from UTfit~\cite{Bona:2009cj} is rather consistent with this measurement and reads
\begin{equation}
\textrm{BR}(B \to \tau \nu)_{\rm{SM}} = (0.807 \pm 0.061)\times 10^{-4}.
\end{equation}
The branching ratio of $B \to \tau\nu$ including NP contributions coming from charged Higgs boson exchange is given by~\cite{Akeroyd:2003zr,Crivellin:2012ye}

\begin{equation}
\textrm{BR}(B \to \tau \nu) = \frac{G_F^2\left|V_{ub}\right|^2}{8\pi}m_\tau^2f_B^2m_B\left(1-\frac{m_\tau^2}{m_B^2}\right)^2\tau_B\left|1+\frac{m_B^2}{m_b m_\tau}\frac{C_R^{ub}-C_L^{ub}}{C_{\rm SM}^{ub}}\right|^2,
\end{equation}
 where, $\tau_B$ is the lifetime of the $B^+$ meson, $f_B$ its decay constant, and $m_B$ its mass. 
 $C_{\rm SM}^{ub}$, $C_R^{ub}$ and $C_L^{ub}$ are the Wilson coefficients of the operators
\begin{equation}
O^{ub}_{\rm SM} = (\bar{u}\gamma_\mu P_L b)(\bar{\tau}\gamma_\mu P_L\nu_\tau), \; O_R^{ub} = (\bar{u}P_Rb)(\bar{\tau}P_L\nu_{\tau}) \;\;{ \rm and}\;\; O_L^{ub}= (\bar{u}P_Lb)(\bar{\tau}P_L\nu_{\tau}),
\label{eq:opLR}
\end{equation}
respectively. $O^{ub}_{\rm SM}$ is generated in the SM by the $W$ boson exchange, while $O_R^{ub}$ and $O_L^{ub}$ are generated by the charged Higgs boson exchange. The SM tree-level contribution is given by $C_{\rm SM}^{ub}=4G_FV_{ub}/\sqrt{2}$, while in our model the contributions to $C_R^{ub}$ and $C_L^{ub}$ are
\begin{equation}
C_R^{ub} = -\frac{1}{m_{H^\pm}^2}\Gamma_{b_Ru_L}^{H^\pm}\Gamma_{\nu_L\tau_R}^{H^\pm}\;\;{\rm and}\;\; C_L^{ub} = -\frac{1}{m_{H^\pm}^2}\Gamma_{b_Lu_R}^{H^\pm}\Gamma_{\nu_L\tau_R}^{H^\pm},
\end{equation}
where the couplings $\Gamma_{ij}^{H^\pm}$ are given in Eqs.~(\ref{eq:HpmuLdR}),  (\ref{eq:HpmuRdL}), and (\ref{Higgs-leptons-vertices-decoupling}). Expanding the charged Higgs couplings, it can be shown that $C_L^{ub}$ is proportional to $m_u$ and is, therefore, vanishingly small. $C_R^{ub}$, on the other hand, is proportional to $m_b$ and is given by
\begin{equation}
C_R^{ub}\simeq V_{ub}\frac{m_b m_\tau}{2v^2}\frac{(1-\tan^2\beta) + (1+\tan^2\beta)\cos\rho}{m_{H^\pm}^2},
\label{eq:CR_ub}
\end{equation}
where we drop terms proportional to $m_\mu$. In the $\rho\to0$ limit $C_R^{ub}$ is approximately independent of $\tan\beta$. 
From Eq.~(\ref{eq:CR_ub}) we conclude that the contribution to $B\to\tau\nu$ for $\rho\ne 0$ and $\tan\beta\sim {\cal O}({\rm{few}})$ (as required by the constraint from $b\to s \gamma$) is relatively small even for low values of $m_{H^\pm}$. Hence, the measurement of $B\to\tau\nu$ does not bring a significant constraint on the parameter space of our model. Nevertheless, we will include this measurement in the global fit of the parameters discussed in Section~\ref{Sec:combination}.

\subsection{$R_D$ and $R_{D^*}$}

Beyond the $B\to\tau\nu$ decay, several $b\to c$ transitions have caught the attention in the last few years. In particular, there is a $\sim 4\sigma$ tension between the SM predictions for the ratios $R_D$ and $R_{D^*}$,
\begin{equation}
R_{D^{(*)}}=\frac{\textrm{BR}(B\to D^{(*)}\tau\nu)}{\textrm{BR}(B\to D^{(*)}\ell\nu)},
\end{equation}
and their measurement \cite{Amhis:2016xyh}.
The SM predictions for these lepton universality ratios are given by Ref.~\cite{Bernlochner:2017jka} (see also Ref.~\cite{Bigi:2016mdz})
\begin{eqnarray}
R_D^{\rm{SM}} &=&0.299 \pm 0.003,\nonumber\\
R_{D^*}^{\rm{SM}}&=&0.257 \pm 0.003,
\end{eqnarray}
while the several measurements of $R_D$ and $R_{D^*}$~\cite{Lees:2012xj,Lees:2013uzd,Huschle:2015rga,Aaij:2015yra,Abdesselam:2016cgx} have been combined by the HFAG collaboration~\cite{Amhis:2016xyh} to give:
\begin{eqnarray}
R_D^{\tiny\textrm{exp}} &=&0.403\pm 0.040\, (\rm{stat}) \pm 0.024\, (\rm{syst}),\nonumber\\
R_{D^*}^{\tiny\textrm{exp}} &=&0.310\pm 0.015\, (\rm{stat})  \pm 0.008\, (\rm{syst}),
\end{eqnarray}
with a correlation coefficient between the two measurements of $-0.23$.
In our 2HDM framework there can be significant contributions to both $R_D$ and $R_{D^*}$. A detailed analysis in a general Type III 2HDM with non-zero non-holomorphic contributions in the quark sector, but not in the lepton sector can be found in Ref.~\cite{Crivellin:2012ye}. Significantly large non-holomorphic contributions and light Higgs bosons are necessary to relieve this tension between the SM predictions and the corresponding experimental values. The contribution from charged Higgs boson exchange come through the same operators $O_L^{cb}$ and $O_R^{cb}$ as given in Eq.~(\ref{eq:opLR}) for $B\to\tau\nu$, exchanging $u\to c$. $R_D$ and $R_{D^*}$ receive the contributions \cite{Fajfer:2012vx,Sakaki:2012ft}
\begin{eqnarray}
R_D&=&R_D^{\rm SM}\left(1+1.5\Re\left(\frac{C_R^{cb}+C_L^{cb}}{C^{cb}_{\rm SM}}\right) + 1.0\left|\frac{C_R^{cb}+C_L^{cb}}{C^{cb}_{\rm SM}}\right|^2\right),\nonumber\\
R_{D^*}&=&R_{D^*}^{\rm SM}\left(1+0.12\Re\left(\frac{C_R^{cb}-C_L^{cb}}{C^{cb}_{\rm SM}}\right) + 0.05\left|\frac{C_R^{cb}-C_L^{cb}}{C^{cb}_{\rm SM}}\right|^2\right),
\end{eqnarray}
where $C_{\rm SM}^{cb}=4G_FV_{cb}/\sqrt{2}$, and $C_{R(L)}^{cb}$ are the Wilson coefficients of the operators $O_{R(L)}^{cb}$ given in (\ref{eq:opLR}) with the exchange $u\to c$. As in the case of the Wilson coefficient $C_R^{ub}$ given by Eq.~(\ref{eq:CR_ub}), $C_R^{cb}$ is proportional to $m_b$ and is given by
\begin{equation}
C_R^{cb}\simeq V_{cb}\frac{m_b m_\tau}{2v^2}\frac{(1-\tan^2\beta) + (1+\tan^2\beta)\cos\rho}{m_{H^\pm}^2}.
\end{equation}
However, contrary to $C_L^{ub}$, $C_L^{cb}$ is no longer vanishingly small and is given by
\begin{eqnarray}\label{eq:CLRDRDstar}
C_L^{cb}&\simeq &\frac{m_\tau m_t}{4v^2\tan^2\beta}\frac{(1-\tan^2\beta) + (1+\tan^2\beta)\cos\rho}{m_{H^\pm}^2}\times\nonumber\\
&&\left[V^*_{cb}\frac{m_c}{m_t}\left((1-\tan^2\beta) - (1+\tan^2\beta)\cos\rho\right)+V^*_{tb}(1+\tan^2\beta)\sin\rho\right],
\end{eqnarray}
where the last term derives from the $\epsilon_{32}^u$ contribution to the $\Gamma_{b_L c_R }^{H^ \pm }$ coupling in (\ref{eq:HpmuRdL}).
Terms proportional to $m_\mu$ have been dropped from both $C_R^{cb}$ and $C_L^{cb}$.
Despite the much larger Wilson coefficient, $C_L^{cb}$, if compared to $C_L^{ub}$, there is a necessity for a large $\tan\beta$ and/or quite low values for the charged Higgs boson mass, to get large enough values to explain the discrepancy between the SM values and the experimental measurements of $R_D$ and $R_{D^*}$~\cite{Crivellin:2012ye}. Furthermore, as shown in Ref.~\cite{Freytsis:2015qca}, the $R_D$ and $R_{D^*}$ anomalies prefer opposite-sign Wilson coefficients, $C_L^{cb}$ and $C_R^{cb}$. In the regime of large $\tan\beta$, $C_R^{cb}<0$, leading  to the requirement for a negative value of the $\rho$ parameter, in such a way as to obtain a positive $C_L^{cb}$ (see Eq. (\ref{eq:CLRDRDstar})). However, as presented in Appendix~\ref{Sec:bsgamma}, the measurement of $b\to s \gamma$ constrains sizable values of $\tan\beta$ and low values of the charged Higgs boson mass and, therefore, limits the size of the NP contribution to $R_D$ and $R_{D^*}$ to be rather small. In the combination of all constraints presented in Section~\ref{Sec:combination}, we include these measurements but do not attempt to explain the discrepancy between the SM prediction and the experimental results, which are still in their nascency.

\bibliographystyle{JHEP-CONF}
\bibliography{HiggsFCNC}
\end{document}